\documentclass{statsoc}
\usepackage{natbib}
\usepackage{amsmath,amssymb}
\usepackage{graphicx}
\usepackage{ulem}
\usepackage{times}
\usepackage{color}
\usepackage{anysize}
\usepackage[colorlinks,citecolor=blue,urlcolor=blue,linkcolor=blue]{hyperref}

\marginsize{1in}{1in}{0.8in}{0.65in}

\newtheorem{condition}{Condition}

\newtheorem{theorem}{Theorem}
\newtheorem{lemma}{Lemma}

\newtheorem{definition}{Definition}
\newtheorem{proposition}{Proposition}

\newcommand{\bX}{\mathbf{X}}
\newcommand{\tr}{\text{tr}}

\def\text#1{\mbox{\rm #1}}
\def\overset#1#2{\stackrel{#1}{#2} }

\let\oldsqrt\sqrt
\def\sqrt{\mathpalette\DHLhksqrt}
\def\DHLhksqrt#1#2{\setbox0=\hbox{$#1\oldsqrt{#2\,}$}\dimen0=\ht0
\advance\dimen0-0.2\ht0
\setbox2=\hbox{\vrule height\ht0 depth -\dimen0}{\box0\lower0.4pt\box2}}
\title[
Tuning-Free Heterogeneity Pursuit
]{
Tuning-free heterogeneity pursuit in massive networks
\thanks{This work was supported by NSF CAREER Awards DMS-0955316 and DMS-1150318 and a grant from the Simons Foundation. Part of this work was completed while the last two authors visited the Departments of Statistics at University of California, Berkeley and Stanford University. These authors sincerely thank both departments for their hospitality.}}
\author{Zhao Ren$^1$, Yongjian Kang$^2$, Yingying Fan$^2$ and Jinchi Lv$^2$}
\address{University of Pittsburgh$^1$ and University of Southern California$^2$}

\begin{document}

\begin{abstract}
Heterogeneity is often natural in many contemporary applications involving massive data. While posing new challenges to effective learning, it can play a crucial role in powering meaningful scientific discoveries through the understanding of important differences among subpopulations of interest. In this paper, we exploit multiple networks with Gaussian graphs to encode the connectivity patterns of a large number of features on the subpopulations. To uncover the heterogeneity of these structures across subpopulations, we suggest a new framework of tuning-free heterogeneity pursuit (THP) via large-scale inference, where the number of networks is allowed to diverge. In particular, two new tests, the chi-based test and the linear functional-based test, are introduced and their asymptotic null distributions are established. Under mild regularity conditions, we establish that both tests are optimal in achieving the testable region boundary and the sample size requirement for the latter test is minimal. Both theoretical guarantees and the tuning-free feature stem from efficient multiple-network estimation by our newly suggested approach of heterogeneous group square-root Lasso (HGSL) for high-dimensional multi-response regression with heterogeneous noises. To solve this convex program, we further introduce a tuning-free algorithm that is scalable and enjoys provable convergence to the global optimum. Both computational and theoretical advantages of our procedure are elucidated through simulation and real data examples.
\end{abstract} 


\section{Introduction} \label{sec:intro}
%
%
%
%
%
%
%
%

In the era of data deluge one can easily collect a massive amount of data from multiple sources, each of which may come from a certain subpopulation of a larger population of interest. For example, these subpopulations can represent different cancer types, brain disorders, or product choices. A large number of features are often associated with each subject. Understanding the heterogeneity in the association structures of these  features across subpopulations can be important in empowering meaningful scientific discoveries or effective personalized choices 
in our lives. Meanwhile heterogeneity in the data also poses new challenges to effective learning and calls for new developments of methods, theory, and algorithms with scalability and statistical efficiency.


Heterogeneity can take different forms in various applications such as the differences among the sparsity patterns or link strengths over multiple networks, and the differences in noise levels or distributions over multiple subpopulations. To avoid potential ambiguity, we would like to make it explicit that throughout this paper, we focus only on two particular types of heterogeneity which are the heterogeneity in sparsity patterns and the heterogeneity in noise levels. To approach the problem of heterogeneous learning in these contexts, we exploit the model setting of multiple networks with Gaussian graphs
each of which encodes the connectivity pattern among features for each subpopulation. The edges of these networks are characterized by the inverse covariances for each pair of nodes from a subpopulation. The focus on this particular type of network models enables us to present our main idea with technical brevity. 
See, for example,  \cite{Teng16} for an account of more general network models beyond graphical models. In fact, as a popular choice of network models Gaussian graphical models involving the inverse covariances have been used widely in applications to characterize the conditional dependency structure among variables. In such models, the joint distribution of $p$ variables $X_1, \cdots, X_p$ is modeled by a multivariate Gaussian distribution $N(0, \Omega^{-1})$, where the $p \times p$ matrix $\Omega$ is called the precision matrix or inverse covariance matrix of these $p$ variables. A basic fact is that each pair of variables, $X_a$ and $X_b$, are conditionally independent given all other variables if and only if the $(a,b)$th entry of the precision matrix $\Omega$ is zero. The conditional dependency structure in a Gaussian graph is therefore determined completely by the associated precision matrix $\Omega$. See, for instance, 
\cite{Lauritzen96} and \cite{WJ08} for more detailed accounts and applications of these models.

There is a growing literature on Gaussian graphical models. Much recent  attention has been given to the problem of support recovery and link strength estimation, which focuses on identifying the nonzero entries of the precision matrix and estimating their strengths. Among those endeavors, a majority of the work has focused primarily on the case of 
a single Gaussian graphical model;
see, for example, \cite{Meinshausen2006, Yuan2007, Friedman2008, FanFengWu09, Yuan2010inverse, cai2011constrained, Ravikumaretal11, liu2013gaussian, Zhang2014DTrace, ren2013asymptotic, Fan2016ISEE}, among many others. A common feature of this line of work is that the data is assumed to be homogeneous with all observations coming from a single population. More detailed discussions and comparisons of these methods can be found in, for instance, \cite{ren2013asymptotic} and \cite{Fan2016ISEE}. Yet as mentioned before  heterogeneity in the data can be prevalent in many contemporary applications involving massive data. The existing methods for analyzing data from each individual source become insufficient due to the assumption of homogeneity. Naively combining the results from these individual analyses may also yield suboptimal performance of statistical estimation and inference.

The setting of multiple networks with Gaussian graphical models has gained more recent attention. A lot of work assumes a time-varying graphical structure across
different graphs. 
In
particular, one assumes that there is a natural ordering of the graphs and the parameters of interest vary smoothly according to this order. For these developments, some smoothing techniques such as the kernel smoothing are key to the construction of the estimators as well as the analysis of their theoretical properties. While the time-varying graphical
model is not the focus of our current paper, one may refer to, for example, \cite{zhou2010time,kolar2010estimating,chen2013covariance,qiu2015joint,lu2015post}
for more details on this line of work.

In contrast, our setting of multiple networks with Gaussian graphical models is along another line that makes no assumption on the ordering of the graphs. In this line of work, the main assumption is a common sparsity structure across different graphs. 
In particular, the estimators proposed in \cite{guo2011joint}, \cite{jgl}, and \cite{zhu2014structural} employ the approach of penalized likelihood with different choices of the penalty function, while the MPE method introduced in \cite{mpe} takes a
weighted constrained $\ell_{\infty}$ and $\ell_{1}$ minimization approach, which can be seen as an extension of the CLIME estimator for a single graph \citep{cai2011constrained}. A common feature of such existing work is the focus on the problem of support recovery and link strength estimation. Moreover, by the nature of these methods their computational cost increases drastically with both the dimensionality and the number of graphs, which can limit their practical use in analyzing massive data sets.   
How to develop a scalable procedure for large-scale inference in the setting of multiple Gaussian graphical models still remains largely open.

To uncover the heterogeneity of the connectivity patterns among features  across subpopulations and address the aforementioned challenges, 
in this paper we suggest a new framework 
of tuning-free heterogeneity pursuit (THP) via large-scale inference, where the number of networks 
is allowed to diverge and the number of features can grow exponentially with the number of observations. Distinct from the existing methods, our procedure identifies the heterogeneity in sparsity patterns among a diverging number of graphs by testing the common sparsity structure of these $k$ Gaussian graphs. 
Specifically, we are interested in testing the null hypothesis \begin{equation}
H_{0,ab}: \ \omega _{a,b}^{0}=(\omega _{a,b}^{(1)}, \cdots
,\omega _{a,b}^{(k)})^{\prime }=\mathbf{0}  \label{eq:null}
\end{equation}%
associated with the joint link strength vector for each pair of variables $1 \leq a, b \leq p$ with $a\neq b$, where $\Omega^{(t)} = (\omega^{(t)}_{a,b})$ with $1 \leq t \leq k$ denotes the precision matrix associated with the $t$th graph. To approach the inference problem in (\ref{eq:null}), we propose two new tests, named the chi-based test and the linear functional-based test, for two different scenarios. The former test which is for the general scenario requires no extra information from the graphs and is shown to perform well as long as the $\ell_2$ norm of the joint link strength vector 
$\omega_{a,b}^0$ is large. The chi-based test is named after the property that the null distribution of this test statistic is shown to converge to the chi distribution. The latter one  
relies on some extra information on the signs of $\omega^{(t)}_{a,b}$. Such extra information is indeed available in some applications. For example, in some genome-wide association studies (GWAS) it was discovered that the association structures can be portable between certain subpopulations \citep{Marigorta2013GWAS}. In such scenario, the linear functional-based test  can be constructed and shown to perform well when the $\ell_1$ norm of the vector $\omega_{a,b}^0$ becomes large.

An interesting feature of both tests is that each of them is established under mild regularity conditions to be optimal in the sense of achieving the testable region boundary, where the testable region boundary is defined as the smallest signal strength below which no test is able to detect if the observations are from the null hypothesis against the alternative hypothesis and above which some test can distinguish successfully between the two hypotheses. We further show that for the linear functional-based test, the sample size requirement is in fact minimal. A natural question is whether naively combining the tests constructed from $k$ individual graphs might suffice. Our theoretical results provide insights into this question and demonstrate the advantages of our tests in terms of weaker sample size requirement than the naive combination approach. We also would like to mention that although the main focus of our paper is on hypothesis testing, our procedure can be modified easily by introducing an additional thresholding step for support recovery; see Section \ref{sec: comp} for detailed discussions and comparisons with existing approaches. In particular, our modified procedure achieves successful support recovery under milder sample size assumption than many existing methods. To the best of our knowledge, the testing of multiple networks with graphs 
and the optimality study are both new to the literature.

The challenges of heterogeneous learning in the setting of multiple networks  
are rooted on the inference with efficiency, the scalability, and the selection of tuning parameters which is often an implicit bottleneck of existing methods. Our THP framework 
addresses all these challenges in a harmonious fashion. Both theoretical guarantees and the tuning-free feature are enabled through efficient multiple-network 
estimation by our newly suggested approach of heterogeneous group square-root Lasso (HGSL) in the setting of high-dimensional multi-response regression with heterogeneous noises.
More specifically, we reduce the problem of estimating $k$ graphs simultaneously to that of running $p$ multi-response regressions with heterogeneous noises. 
This new formulation allows us to borrow information across graphs when estimating their structures, which results in improved rates of convergence.  
To solve the convex programs from these multi-response regressions, we
introduce a new tuning-free algorithm that is scalable and admits provable convergence to the global optimum.
Compared to existing methods in the literature, our new procedure enjoys four  main advantages. First, it is justified theoretically that our HGSL estimators have faster rates of convergence. Second, the HGSL method is capable of handling heterogeneous noises, the presence of which causes intrinsic difficulty for developing a tuning-free procedure. Third, our new algorithm is simple and tuning free, and scales up easily.
Fourth, we provide theoretical justification on the convergence of the tuning-free algorithm. 

The rest of the paper is organized as follows. Section \ref{sec:test} introduces the THP framework 
for heterogeneous learning in multiple networks via large-scale inference with the chi-based test and the linear functional-based test, and establishes their optimality properties.
We present the HGSL approach along with a tuning-free algorithm for fitting high-dimensional multi-response regression with heterogeneous noises, and provide the estimation and prediction bounds for the estimator as well as a convergence analysis for the algorithm in Section \ref{sec:initial}.
Section \ref{sec:data} details several numerical examples of simulation studies and real data analysis. We discuss some extensions of the suggested method to a few settings in Section \ref{sec:discussion}. The proofs of all the results and technical details are provided in the Supplementary Material.

\section{
Tuning-free heterogeneity pursuit in multiple networks via large-scale inference} \label{sec:test}

\subsection{Model setting} \label{Sec2.1}
As mentioned in the Introduction, we adopt the setting of multiple networks with Gaussian graphical models 
to encode the connectivity patterns among $p$ features $X_1, \cdots, X_p$ measured on $k$ subpopulations of a general population, which yields $k$ classes of data. In this model, for each class $1 \leq t \leq k$ the $p$-dimensional feature vector follows a multivariate Gaussian distribution
\begin{equation} \label{neweq016}
X^{(t)}=(X_{1}^{(t)},\cdots
,X_{p}^{(t)})^{\prime }\sim N(0,( \Omega ^{(t)}) ^{-1}),
\end{equation}
where the superscript $(t)$ means that these $p$ features are measured on the $t$th subpopulation and $%
\Omega ^{(t)}$ is the $p \times p$ precision matrix of the
$t$th class. In addition, the distributions of $X^{(1)}, \cdots ,X^{(k)}$
are assumed to be independent. Each of the $k$ precision matrices $%
\Omega ^{(t)}=(\omega _{a,b}^{(t)})_{p\times p}$ reflects the conditional
dependency structure among the $p$ features $X_{1}^{(t)},\cdots
,X_{p}^{(t)}$. 
In the high-dimensional setting where the dimensionality $p$ can be very large compared to the sample size, it is common in many applications such as genomic studies to
assume that each precision matrix $\Omega ^{(t)}$ has certain sparsity structure. The goals in these studies include the estimation of precision matrices $\Omega
^{(t)}$ and the inference on their entries $\omega _{a,b}^{(t)}$.

When there is only one class of data, that is, $k=1$, our setting coincides with that of single Gaussian graphical model. For the general case of multiple graphs with $k\geq 2$, it can be beneficial to borrow the strength across all $k$ classes
of data to achieve more accurate estimation of the $k$ precision matrices if the $k$ classes are related to each other. With this spirit, 
we assume that the $k$ classes
share some similar sparsity structure, and the heterogeneity in sparsity patterns captures the differences among these graphical structures. In particular, we are interested in the scenario where for each pair of
nodes $\left( a,b\right)$ with $1\leq a\neq b \leq p$,
either $\omega _{a,b}^{(t)}=0$ simultaneously for all $1\leq t \leq k$ or
alternatively the joint link strength vector 
$\omega _{a,b}^{0} = (\omega _{a,b}^{(1)},\cdots
,\omega _{a,b}^{(k)})^{\prime }$ is significantly different from the zero vector. Throughout the paper we denote by
\begin{equation}
\mathcal{E}=\left\{\left( a,b\right) : 1\leq a\neq b \leq p \text{ and } \omega _{a,b}^{0}\neq\mathbf{0}\right\}
\label{eq:edge set}
\end{equation}%
the edge set corresponding to the $k$ graphs given in model (\ref{neweq016}).

The main goal of our paper is to develop an effective and efficient procedure for testing the null hypothesis $H_{0,ab}$ defined in \eqref{eq:null} for multiple networks, 
which provides an inferential approach to uncovering the heterogeneity of the feature association structures across the $k$ subpopulations.
Depending on the type of the alternative hypothesis, we will introduce two different fully data-driven test statistics and establish their advantages over those obtained by naively combining the tests constructed from each individual graph.

\subsection{Chi-based test} \label{sec:test.l1}
We begin with introducing the first test for our THP framework in multiple networks. 
To ease the presentation, we introduce some compact notation. Denote by $a_{-j}$ the
subvector of a vector $a = (a_1, \cdots, a_p)^{\prime }$ with the $j$th component removed, and for any matrix $A = (a_{i,j})$ denote by $A_{*,j}$ its  $j$th column, $A_{-j,j}$ the
subvector of $A_{*,j}$ with the $j$th component removed, and $A_{*,-j}$ the submatrix of $A$ with the $j$th column removed. 
Our testing idea is based on a simple observation that for
each $1\leq j \leq p$, the conditional
distribution of $X_{j}^{(t)}$ given all remaining variables $X_{-j}^{(t)}$
in class $t$ follows the Gaussian distribution
\begin{equation}
X_{j}^{(t)}|X_{-j}^{(t)}\sim N(X_{-j}^{(t)\prime }C_{j}^{(t)},1/\omega
_{j,j}^{(t)})  \label{eq:cond_dist}
\end{equation}%
with the $(p - 1)$-dimensional coefficient vector $C_{j}^{(t)} =-\Omega _{-j,j}^{(t)}/\omega _{j,j}^{(t)}$. Based on the distributional representation in (\ref{eq:cond_dist}), one can see that the error random variables $\epsilon_j^{(t)} = X_j^{(t)} - X_{-j}^{(t)\prime}C_j^{(t)}$ are independent across $t$ and follow the distribution $N(0, 1/\omega^{(t)}_{j,j})$. Moreover, it holds for each pair of nodes $(a, b)$ with $1 \leq a,  b \leq p$ that
\begin{equation}
\mathrm{cov}(\epsilon _{a}^{(t)},\epsilon _{b}^{(t)})=\frac{\omega
	_{a,b}^{(t)}}{\omega _{a,a}^{(t)}\omega _{b,b}^{(t)}}\mbox{\rm .}
\label{eq:cov_e}
\end{equation}
The key representation in (\ref{eq:cov_e}) entails that accurate estimators of $\omega_{a,b}^{(t)}$ with $a\neq b$ can be constructed if one can estimate $\omega_{a,a}^{(t)}$, $\omega^{(t)}_{b,b}$, and $\mathrm{cov}(\epsilon_a^{(t)}, \epsilon_b^{(t)})$ well.

Another important observation is that under the null hypothesis $H_{0,ab}$ in \eqref{eq:null}, the conditional distributions of the $k$
classes $X_{j}^{(t)}|X_{-j}^{(t)}$ with $1 \leq t \leq k$ indeed share similar
sparsity structure on the coefficient vectors $C_{j}^{(t)}$ thanks to the representation $%
C_{j}^{(t)}=-\Omega _{-j,j}^{(t)}/\omega _{j,j}^{(t)}$. In fact, it is clear
that $C_{a,b}^{(t)}=0$ for all $1 \leq t \leq k$ under $H_{0,ab}$, where $C_{a,b}^{(t)} = -\omega_{a,b}^{(t)}/\omega_{a,a}^{(t)}$ is the component of  vector $C_{a}^{(t)}$ corresponding to variable $X_b^{(t)}$. This observation suggests that we can borrow information from different graphs when testing the joint sparsity structure of multiple graphs.
Motivated by such observation, we turn 
the problem of multiple-network 
estimation into 
that of high-dimensional multi-response linear regression
\begin{eqnarray}
\left(
\begin{array}{c}
X_{j}^{(1)} \\
X_{j}^{(2)} \\
\vdots  \\
X_{j}^{(k)}%
\end{array}%
\right) &=&\left(
\begin{array}{cccc}
X_{-j}^{(1)} &  &  &  \\
& X_{-j}^{(2)} &  &  \\
&  & \ddots  &  \\
&  &  & X_{-j}^{(k)}%
\end{array}%
\right) \left(
\begin{array}{c}
C_{j}^{(1)} \\
C_{j}^{(2)} \\
\vdots  \\
C_{j}^{(k)}%
\end{array}%
\right) +\left(
\begin{array}{c}
\epsilon_{j}^{(1)} \\
\epsilon_{j}^{(2)} \\
\vdots  \\
\epsilon_{j}^{(k)}%
\end{array}%
\right)  \label{eq:multi-reponse}
\end{eqnarray}%
for $1 \leq j \leq p$. A distinct feature of the above multi-response regression model (\ref{eq:multi-reponse}) is that it has heterogeneous noises since $\omega_{j,j}^{(t)}$ generally varies over $1 \leq t \leq k$.

As mentioned before, we also have the
group sparsity structure of the regression coefficient vector
$
C_{j}^{0}=\left( C_{j}^{(1)\prime },\cdots ,C_{j}^{(k)\prime }\right)' \in
\mathbb{R}^{(p-1)k}
$ in model (\ref{eq:multi-reponse}). More specifically, denote the $k$-dimensional subvector of $%
C_{j}^{0}$ corresponding to the $l$th group by
\begin{equation}
C_{j(l)}^{0}=\left( C_{j,l}^{(1)},\cdots ,C_{j,l}^{(k)}\right)^{\prime }. %
\label{eq:group subvector}
\end{equation}%
Then we see that $C_{j(l)}^{0}= \bf{0}$ for
all pairs $(j, l) \in \mathcal{E}^c$, the complement of $\mathcal{E}$ defined in \eqref{eq:edge set}. We will suggest in Section \ref{sec:initial} an efficient estimation procedure that utilizes the group sparsity structure in the regression coefficients and also accounts for the heterogeneity in the noises in model \eqref{eq:multi-reponse}.

From now on we work with a sample from model (\ref{neweq016}) that is comprised of $n^{(t)}$ independent and identically distributed (i.i.d.)
observations $X_{1,\ast }^{(t)},\cdots ,X_{n^{(t)},\ast }^{(t)}$ for each class $1 \leq t \leq k$, where $%
X_{i,\ast }^{(t)}=(X_{i,1}^{(t)},\cdots ,X_{i,p}^{(t)})^{\prime }\sim
N(0,(\Omega
^{(t)})^{-1})$ and the observations across
different classes are independent. Suppose that we have some initial estimator $\hat{C}_{j}^{0}=( \hat{C}%
_{j}^{(1)\prime },\cdots ,\hat{C}_{j}^{(k)\prime }) ^{\prime }$ for the $(p-1)k$-dimensional regression coefficient vector $%
C_{j}^{0}$, whose construction is detailed in Section \ref{sec:initial}. Then the random errors
for each $1 \leq t \leq k$ can be estimated by the residuals
\begin{equation}
\hat{E}_{i,j}^{(t)}=X_{i,j}^{(t)}-X_{i,-j}^{(t)\prime }\hat{C}_{j}^{(t)} \label{eq:residual all classes}
\end{equation}%
with $\ 1 \leq i \leq n^{(t)}$ and $1 \leq j \leq p$. In view of the representation in \eqref{eq:cov_e}, we can estimate $\omega
_{j,j}^{(t)}$ associated with the noise level  of class $t$ as $\hat{%
	\omega}_{j,j}^{(t)}= n^{(t)}/( \sum_{i=1}^{n^{(t)}}\hat{E}_{i,j}^{(t)}\hat{E}%
_{i,j}^{(t)})$. In contrast, the estimation of $\omega^{(t)}_{a,b}$ with $a \neq b$ is slightly more complicated. To estimate the negative covariance  $-\mathrm{cov}(\epsilon _{a}^{(t)},\epsilon _{b}^{(t)})=-\omega_{a,b}^{(t)}/(\omega_{a,a}^{(t)}\omega_{b,b}^{(t)})$, we exploit the following bias corrected statistic
\begin{equation} \label{neweq017}
T_{n,k,a,b}^{(t)}=\frac{1}{n^{(t)}}\left[ \sum_{i=1}^{n^{(t)}}\hat{E}%
_{i,a}^{(t)}\hat{E}_{i,b}^{(t)}+\sum_{i=1}^{n^{(t)}}\left( \hat{E}%
_{i,a}^{(t)}\right) ^{2}\hat{C}_{b,a}^{(t)}+\sum_{i=1}^{n^{(t)}}\left( \hat{E%
}_{i,b}^{(t)}\right) ^{2}\hat{C}_{a,b}^{(t)}\right].
\end{equation}%
Observe that the first term on the right-hand side of (\ref{neweq017}) corresponds to the sample covariance of the residuals from variables $X_a^{(t)}$ and $X_b^{(t)}$. When $a=b$, this sample covariance is asymptotically unbiased in estimating $\mathrm{var}(\epsilon _{a}^{(t)}) = 1/\omega_{a,a}^{(t)}$. Such sample covariance is, however, biased in the case of $a\neq b$ and thus two additional terms are introduced for $T_{n,k,a,b}^{(t)}$ in (\ref{neweq017}) to correct the bias. Indeed, we can show that after the bias correction the statistic $T_{n,k,a,b}^{(t)}$ is asymptotically close to the statistic 
\begin{equation}
J_{n,k,a,b}^{(t)}=\left[1-\omega _{a,a}^{(t)}( \hat{\omega}%
_{a,a}^{(t)}) ^{-1}-\omega _{b,b}^{(t)}( \hat{\omega}%
_{b,b}^{(t)}) ^{-1}\right]\frac{\omega _{a,b}^{(t)}}{\omega _{a,a}^{(t)}\omega
	_{b,b}^{(t)}},  \label{eq:J(t)}
\end{equation}
which is in turn asymptotically close to the negative covariance $-\mathrm{cov}(\epsilon _{a}^{(t)},\epsilon _{b}^{(t)})$. 

When there is only a single graph, that is, $k=1$, the above statistic $T^{(t)}_{n,k,a,b}$ in (\ref{neweq017}) reduces to the one introduced in \cite{liu2013gaussian} to address the bias issue in the testing for a single Gaussian graph. 
In the scenario of multiple graphs, we observe a similar phenomenon and provide in Theorem
\ref{thm:l2 testing} later a formal theoretical justification. It is worth mentioning that the key estimators $\hat{\omega}_{j,j}^{(t)}$ and $T_{n,k,a,b}^{(t)}$ introduced above are constructed using the residuals $\hat E_{i,j}^{(t)}$ instead of the estimated regression coefficients $\hat C_{a,b}^{(t)}$, though the regression coefficients $C_{a,b}^{(t)}$ are also closely related to the entries of the precision matrix $\Omega^{(t)}$. The main advantage of using residuals $\hat E_{i,j}^{(t)}$ over coefficients $\hat C_{a,b}^{(t)}$ is rooted on the fact that obtaining asymptotically unbiased estimates of the latter is much more challenging in high dimensions, largely due to the well-known bias issue associated with the regularization methods, than accurately estimating the former, which is closely 
related to the prediction problem.  


The new formulation in \eqref{eq:multi-reponse} not only allows us to solve the problem of multiple-graph estimation efficiently through $p$ multi-response regressions as detailed in Section \ref{sec:initial}, but also enables us to construct new tests that are more powerful than existing methods by borrowing information from different graphs. We are now ready to present the first such test. Due to the group sparsity structure and the target of our null hypothesis $H_{0,ab}: \omega _{a,b}^{0}=\mathbf{0}$ in (\ref{eq:null}), we naturally construct our test
statistics using certain functions of all statistics $T_{n,k,a,b}^{(t)}$ in (\ref{neweq017}) with  $1 \leq t \leq k$. Thanks to the joint estimation accuracy for the $(p-1)k$-dimensional regression coefficient vector $C_{j}^{0}$, 
we define our first test statistic, the chi-based test statistic $U_{n,k,a,b}$, as
\begin{equation}
U_{n,k,a,b}^{2}=\sum_{t=1}^{k}n^{(t)}\hat{\omega}_{b,b}^{(t)}\hat{\omega}%
_{a,a}^{(t)}\left( T_{n,k,a,b}^{(t)}\right) ^{2}  \label{eq:test stat chi}
\end{equation}%
for testing the null hypothesis $H_{0,ab}$ against the alternative
hypothesis for which the condition is imposed on the $\ell_2$ norm $%
\Vert \omega _{a,b}^{0}\Vert $. In other words, our test statistic
is powerful whenever the signal strength $\Vert \omega _{a,b}^{0}\Vert $ is
larger than some testable region boundary, which will be characterized later in Section \ref{sec:optimal}.

To characterize the limiting distribution of the chi-based test statistic $U_{n,k,a,b}$ in (\ref{eq:test stat chi}) under the null, we introduce two additional statistics $V_{n,k,a,b}^{\ast (t)}$ and
$U_{n,k,a,b}^{\ast }$ as
\begin{eqnarray}
V_{n,k,a,b}^{\ast (t)} &=&\sqrt{\frac{\omega _{b,b}^{(t)}\tilde{\omega}%
_{a,a}^{(t)}}{n^{(t)}}}\sum_{i=1}^{n^{(t)}}\left( E_{i,a}^{(t)}E_{i,b}^{(t)}-%
\mathbb{E}E_{i,a}^{(t)}E_{i,b}^{(t)}\right),  \label{eq:V star} \\
U_{n,k,a,b}^{\ast 2} &=&\sum_{t=1}^{k}\left( V_{n,k,a,b}^{\ast (t)}\right)
^{2}=\sum_{t=1}^{k}\frac{\omega _{b,b}^{(t)}\tilde{\omega}_{a,a}^{(t)}}{n^{(t)}%
}\left[ \sum_{i=1}^{n^{(t)}}\left( E_{i,a}^{(t)}E_{i,b}^{(t)}-\mathbb{E}%
E_{i,a}^{(t)}E_{i,b}^{(t)}\right) \right] ^{2},  \label{neweq018}
\end{eqnarray}%
where $E^{(t)}_{i,j} = X_{i,j}^{(t)} - X_{i,-j}^{(t)\prime}C_j^{(t)}$ is the random error and  $\tilde{\omega}_{j,j}^{(t)}=
n^{(t)}/(E_{*,j}^{(t)\prime}E_{*,j}^{(t)})$ is the
oracle estimator of $\omega _{jj}^{(t)}$ since the random error vector  $E_{\ast,j}^{(t)} = (E_{1,j}^{(t)}, \cdots, E_{n^{(t)},j}^{(t)})'$ is unobservable in practice. It is interesting to observe that under the null, the Gaussian vector $E_{\ast ,b}^{(t)}\sim
N(0,( \omega _{b,b}^{(t)}) ^{-1} I)$ is independent of $E_{\ast
,a}^{(t)}$, which entails that $V_{n,k,a,b}^{\ast (t)}\sim N(0,1)$ and they are
independent of each other over $1 \leq t \leq k$. Consequently, under the null
hypothesis $H_{0,ab}$ in (\ref{eq:null}) it holds that $U_{n,k,a,b}^{\ast 2}\sim \chi ^{2}(k)$. 


Before formally presenting our first main result, we introduce the following two regularity conditions on our model (\ref{neweq016}).

\begin{condition} \label{CondA1}
There exists some constant $M>0$ such that $1/M\leq \lambda
	_{\min }(\Omega ^{(t)})\leq \lambda _{\max }(\Omega ^{(t)})\leq M$ for each $%
	1 \leq t \leq k$, where $\lambda
	_{\min }$ and $\lambda
	_{\max }$ denote the smallest and largest eigenvalues of a matrix.
\end{condition}

\begin{condition} \label{CondA2}
It holds that $%
	n^{(1)} \asymp \cdots \asymp n^{(k)}$ with $\max_{1 \leq t \leq k}\{n^{(t)}\}/n^{(0)}\leq M_0$, where $\asymp$ means the same order, $%
	n^{(0)}=\min_{1 \leq t \leq k}\{n^{(t)}\}$, and $M_0$ is some positive constant.
\end{condition}

The well-conditionedness of the precision matrices $\Omega ^{(t)}$ assumed in Condition \ref{CondA1} simplifies our technical presentation. For simplicity, we also assume in Condition \ref{CondA2} that our sample is balanced with the sample sizes of each of the $k$ classes comparable to each other. With slight abuse of notation, we denote by $n^{(0)}$ this common level whenever the rate is involved. We proceed with introducing additional notation and technical conditions. Denote by $\Delta_{j}=\hat{C}_{j}^{0}-C_{j}^{0}$ and $\Delta_{j(l)}=\hat{C}_{j(l)}^{0}-C_{j(l)}^{0}$ the estimation errors of $\hat{C}_{j}^{0}$ and $\hat{C}_{j(l)}^{0}$, respectively, with the $k$-dimensional subvector $\hat{C}_{j(l)}^{0}$ defined in a similar way to $C_{j(l)}^{0}$ in (\ref{eq:group subvector}). To characterize the sparsity level, we define the joint sparsity of the $k$ networks 
as the maximum node
degree corresponding to the edge set $\mathcal{E}$ in (\ref{eq:edge set}),
\begin{equation}
s\equiv\max_{1 \leq a \leq p}\sum_{1 \leq b\neq a \leq p}1\{\omega _{a, b}^{0}\neq \mathbf{0}\}.
\label{eq:sparsity}
\end{equation}
We further assume that with high probability the initial estimator $\hat{C}_{j}^{0}$ satisfies  
\begin{eqnarray}
\frac{1}{\sqrt{k}}\left\Vert \Delta _{j}\right\Vert &\leq &C_{1}\left[s\frac{%
	1+(\log p)/k}{n^{(0)}}\right]^{1/2},  \label{eq:assumption1} \\
\sum_{l\neq j}\frac{1}{\sqrt{k}}\left\Vert \Delta _{j(l)}\right\Vert &\leq
&C_{2}s\left[\frac{1+(\log p)/k}{n^{(0)}}\right]^{1/2},  \label{eq:assumption2} \\
\frac{1}{k}\sum_{t=1}^{k}\frac{\left\Vert X_{\ast ,-j}^{(t)}\left( \hat{C}%
	_{j}^{(t)}-C_{j}^{(t)}\right) \right\Vert ^{2}}{n^{(t)}} &\leq &C_{3}s\frac{%
	1+(\log p)/k}{n^{(0)}},  \label{eq:assumption3}
\end{eqnarray}%
where $C_1, C_2$, and $ C_3$ are some positive constants and $\Vert \cdot \Vert$ denotes the $\ell_2$ norm. The properties (\ref{eq:assumption1})--(%
\ref{eq:assumption3}) are crucial working assumptions in our
testing for $k$ networks. 

Indeed, the new tuning-free approach of HGSL suggested in Section \ref{sec:initial} guarantees that we can obtain initial estimators  $\hat{C}_{j}^{0}$ each satisfying all these properties (\ref{eq:assumption1})--(%
\ref{eq:assumption3}) with probability at least $1-C_{0}p^{1-\delta }$ for some positive constants $C_{0}$ and $\delta > 1$. A distinct feature is that the analysis of our tuning-free estimator is new due to the heterogeneity of noises across different classes, which makes typical tuning-free procedures such as the scaled Lasso \citep%
{sun2012scaled} and the square-root Lasso \citep{belloni2011square} no longer work in the current setting; see Section \ref%
{sec:initial} for more detailed discussions.


\begin{theorem}
\label{thm:l2 testing}
Assume that Conditions \ref{CondA1}--\ref{CondA2} hold, the initial estimators $\hat{C}_{j}^{0}$ each satisfy properties (\ref{eq:assumption1})--(%
\ref{eq:assumption3}) with probability at least $1-C_{0}p^{1-\delta }$, $s\left( k+\log p\right)/n^{(0)}=o(1)$, and $\log (k/\delta
_{1})=O\{s[1+(\log p)/k]\}$ for some constants $C_{0} > 0, \delta > 1$ and $\delta
_{1} = o(1)$. Then for each pair $(a,b)$ with $1 \leq a\neq b \leq p$, it holds with probability at least $%
1-(12+C_{0})p^{1-\delta }-4\delta _{1}$ that 
\begin{eqnarray*}
\left\vert \left[ \sum_{t=1}^{k}n^{(t)}\hat{\omega}_{b,b}^{(t)}\hat{\omega}%
_{a,a}^{(t)}\left( T_{n,k,a,b}^{(t)}-J_{n,k,a,b}^{(t)}\right) ^{2}\right]
^{1/2}-U_{n,k,a,b}^{\ast }\right\vert &\leq &C\left( s\frac{k+\log p}{\sqrt{n^{(0)}}}\right) ,
\end{eqnarray*}%
where $C>0$ is some constant. 
Moreover,  under null hypothesis $H_{0,ab}$ in (\ref{eq:null}) 
we have $U_{n,k,a,b}^{\ast 2}\sim \chi ^{2}(k)$ and with the same probability bound that
$
\left\vert U_{n,k,a,b}-U_{n,k,a,b}^{\ast }\right\vert \leq C\left( s\frac{%
k+\log p}{\sqrt{n^{(0)}}}\right).
$
\end{theorem}

The coupling result in Theorem \ref{thm:l2 testing} motivates us to propose the chi-based test $\phi _{2}$ defined as
\begin{equation}
\phi _{2}=1\left\{ U_{n,k,a,b}>z_{k}^{l2}(1-\alpha )\right\}
\label{eq: l2 test}
\end{equation}%
for our THP framework in multiple networks which tests the null hypothesis $H_{0,ab}$ in (\ref{eq:null}) using the test statistic $U_{n,k,a,b}$ given in (\ref{eq:test stat chi}), where $\alpha \in
\left( 0,1\right) $ is a fixed significance level and $z_{k}^{l2}(1-\alpha )$ denotes the $100(1-\alpha)$th percentile of the
chi distribution with degrees of freedom $k$. The name of this test is from the property that the null distribution of the test statistic is asymptotically close to the chi distribution.

\begin{proposition}
\label{cor:l2 testing}
Assume that all the conditions of Theorem \ref{thm:l2 testing} hold and
$s^{2}(k+\log p)^{2}=o(n^{(0)})$. Then the chi-based test $\phi _{2}$ in (\ref{eq: l2 test}) has asymptotic significance level $\alpha$.
\end{proposition}

As formally justified in Proposition \ref{cor:l2 testing}, the chi-based test $\phi _{2}$ introduced in (\ref{eq: l2 test}) is indeed an asymptotic test with significance level $\alpha $ under the sample size requirement of $n^{(0)}\gg s^{2}(k+\log p)^{2}$, in the asymptotic setting in
which the number of nodes $p$, the number of networks $k$, and the joint sparsity of the networks $s$
can diverge simultaneously as the common level of sample sizes $n^{(0)}\rightarrow \infty$.

\subsection{Linear functional-based test} \label{Sec2.3}
The chi-based test $\phi _{2}$ introduced in Section \ref{sec:test.l1} serves as a general procedure to test whether the joint link strength vector $\omega _{a,b}^{0}$ is zero when there is no additional information assumed on the $k$ networks. In some scenarios when certain extra knowledge is available, it is possible to design more powerful testing procedures.
In this spirit, we now present an alternative test for our THP framework in multiple networks based on a linear functional of $\omega _{a,b}^{0}$, which
is closely related to its $\ell_1$ norm. The main motivation is that in some applications such as the GWAS example mentioned in the Introduction \citep{Marigorta2013GWAS}, the sign
relationship of some target edge across $k$ graphs is provided implicitly or
explicitly. For example, one may expect that all the $\omega _{a,b}^{(t)}$  with $1\leq  t \leq k$ share the same sign, that is, they are either all nonpositive or all nonnegative. In such scenario, testing the null hypothesis $H_{0,ab}:\omega
_{a,b}^{0}=\mathbf{0}$ is equivalent to testing $\Vert \omega
_{a,b}^{0}\Vert _{1}=\vert \sum_{t=1}^{k}\omega
_{a,b}^{(t)}\vert =0$. In a more general setting, the sign relationship
can be represented by a unique sign vector $\xi =(\xi _{1},\cdots ,\xi
_{k})^{\prime }\in \{1,-1\}^{k}$, up to a single sign, such that
$\Vert \omega _{a, b}^{0}\Vert _{1}=\sum_{t=1}^{k}\xi _{t}\omega
_{a, b}^{(t)}$ or $\Vert \omega _{a, b}^{0}\Vert _{1}=\vert
\sum_{t=1}^{k}\xi _{t}\omega _{a, b}^{(t)}\vert$, and thus the null hypothesis $H_{0,ab}:\omega
_{a,b}^{0}=\mathbf{0}$ takes an equivalent form of $\Vert \omega _{a, b}^{0}\Vert _{1}=\vert
\sum_{t=1}^{k}\xi _{t}\omega _{a, b}^{(t)}\vert =0$.

Given the above sign vector $\xi$, we define our second test statistic, the linear functional-based test statistic $V_{n,k,a,b}(\xi )$, as
\begin{equation}
V_{n,k,a,b}(\xi )=\sum_{t=1}^{k}\xi _{t}\sqrt{n^{(t)}\hat{\omega}_{a, a}^{(t)}%
\hat{\omega}_{b, b}^{(t)}}T_{n,k,a,b}^{(t)}  \label{eq:test stat l1}
\end{equation}%
with the bias corrected statistic $T_{n,k,a,b}^{(t)}$ given in (\ref{neweq017}). To characterize the limiting distribution of the linear functional-based test statistic $V_{n,k,a,b}$ under the null, we introduce another statistic $V_{n,k,a,b}^{\ast }(\xi )$ as
\begin{equation*}
V_{n,k,a,b}^{\ast }(\xi )=\sum_{t=1}^{k}\xi _{t} V_{n,k,a,b}^{\ast
(t)}=\sum_{t=1}^{k}\xi _{t} \sqrt{\frac{\omega _{b,b}^{(t)}\tilde{\omega}%
_{a,a}^{(t)}}{n^{(t)}}}\sum_{i=1}^{n^{(t)}}\left( E_{i,a}^{(t)}E_{i,b}^{(t)}-%
\mathbb{E}E_{i,a}^{(t)}E_{i,b}^{(t)}\right) ,
\end{equation*}%
where the statistic $V_{n,k,a,b}^{\ast (t)}$ is given in (\ref{eq:V star}). With the extra sign information, our new test statistic is powerful whenever
the signal strength $\Vert \omega _{a, b}^{0}\Vert _{1}$ becomes
large; see Section \ref{sec:optimal} for the characterization of the testable region boundary under the alternative hypothesis for which the condition is imposed
on the $\ell_1$ norm $\Vert \omega _{a, b}^{0}\Vert _{1}$. It is easy to see that
under the null, $V_{n,k,a,b}^{\ast (t)}\sim N(0,1)$ are independent of each
other over $1 \leq t \leq k$, and consequently $V_{n,k,a,b}^{\ast
}(\xi)\sim N(0,k)$ for any given sign vector $\xi$.


\begin{theorem}
\label{thm:l1 testing}
Assume that all the conditions of Theorem \ref{thm:l2 testing} hold. 
Then for each pair $(a,b)$ with $1 \leq a\neq b \leq p$, it holds with probability at least $%
1-(12+C_{0})p^{1-\delta }-4\delta _{1}$ that 
\begin{eqnarray}
\left\vert\sum_{t=1}^{k}\xi _{t} \left[ \sqrt{n^{(t)}\hat{\omega}_{b, b}^{(t)}%
\hat{\omega}_{a, a}^{(t)}}\left( T_{n,k,a,b}^{(t)}-J_{n,k,a,b}^{(t)}\right)
-V_{n,k,a,b}^{\ast (t)}\right]\right\vert
&\leq &C\left( s\frac{k+\log p}{\sqrt{n^{(0)}}}\right) ,
\label{eq:thm l1}
\end{eqnarray}%
where $C>0$ is some constant. 
Moreover,  under null hypothesis $H_{0,ab}$ in (\ref{eq:null})
we have $J_{n,k,a,b}^{(t)}=0$, $V_{n,k,a,b}^{\ast
}(\xi)\sim N(0,k)$ 
and with the same probability bound,
$
\left\vert V_{n,k,a,b}(\xi)-V_{n,k,a,b}^{\ast }(\xi)\right\vert \leq C\left( s\frac{%
k+\log p}{\sqrt{n^{(0)}}}\right).
$
\end{theorem}

Theorem \ref{thm:l1 testing} quantifies the asymptotic behavior of the linear functional-based test statistic $V_{n,k,a,b}(\xi )$ under the null hypothesis $H_{0,ab}$ in (\ref{eq:null}). Assume further that the sign vector $\xi $ is given uniquely such that $\Vert \omega _{a,b}^{0}\Vert _{1}=\sum_{t=1}^{k}\xi
_{t}\omega _{a,b}^{(t)}$ under the alternative hypothesis. Then Theorem \ref{thm:l1 testing}
and the definition of the statistic $J_{n,k,a,b}^{(t)}$ in (\ref{eq:J(t)}) motivate us to propose a one-sided test, the linear functional-based test $\phi _{1}$, defined as
\begin{equation}
\phi _{1}=1\left\{ \frac{V_{n,k,a,b}(\xi)}{\sqrt{k}}<z(\alpha )\right\}
\label{eq: l1 test}
\end{equation}%
for our THP framework in multiple networks, where $\alpha \in
\left( 0,1\right) $ is a fixed significance level and $z(\alpha )$ stands for  the $100 \alpha$th percentile of the standard
Gaussian distribution. When the sign vector $\xi $ is given up to a single sign, for example, 
when we know only that all the signs $\xi _{t}$ with $1 \leq t \leq k$ are identical, it is
more natural to define a two-sided test. We omit the details of such
two-sided test for simplicity.


\begin{proposition}
\label{cor:l1 testing}
Assume that all the conditions of Theorem \ref{thm:l1 testing}
hold 
and
$s^{2}k^{-1}(k+\log p)^{2}=o(n^{(0)})$. Then the linear functional-based test $\phi _{1}$ in (\ref{eq: l1 test}) has asymptotic significance level $\alpha$.
\end{proposition}

Proposition \ref{cor:l1 testing} which is based on Theorem \ref{thm:l1 testing} shows that the linear functional-based test $\phi _{1}$ introduced in (\ref{eq: l1 test}) is indeed an asymptotic test with
significance level $\alpha $ under the sample size requirement of $n^{(0)}\gg
s^{2}k^{-1}(k+\log p)^{2}$. It is worth mentioning that most existing results in the literature either focus on testing procedures for a
single graph or develop estimation procedures for multiple graphs without
statistical inference in high dimensions. In contrast, our developments in Theorems \ref{thm:l2 testing}--\ref{thm:l1 testing} and Propositions \ref{cor:l2 testing}--\ref{cor:l1 testing} 
provide procedures of large-scale 
inference in multiple graphs for the first time. For the case of a single graph with $k = 1$, our test statistics essentially reduce to the one introduced in \cite%
{liu2013gaussian}. This suggests an alternative way of constructing test statistics, which is to construct a test statistic for each individual graph $1 \leq t \leq k$ as in \cite%
{liu2013gaussian} and then naively pool them together in the same way as for our tests $\phi_2$ and $\phi_1$.

Let us gain some insights into our tests with a comparison to the above naive combination procedure.
The advantage of our linear functional-based test $\phi _{1}$ is
reflected on the sample size requirement of $s^{2}k^{-1}(k+\log
p)^{2}=o(n^{(0)})$ established in Proposition \ref{cor:l1 testing}, thanks to the information of structural similarity across the $k$ graphs which makes the working assumptions (\ref{eq:assumption1})--(\ref{eq:assumption3}) possible. In
comparison, to test the null hypothesis $H_{0,ab}:\omega _{a, b}^{0}=\mathbf{0}$ one can also apply
the procedure in \cite{liu2013gaussian} to each of the $k$ graphs and
then construct a similar linear functional-based test as in (\ref{eq: l1 test}).
For such naive combination procedure, it can be shown that a stronger sample size assumption $%
s^{2}k\left( \log p\right) ^{2}=o(n^{(0)})$ is required. In fact, we further establish in Section \ref%
{sec:optimal} that the sample size requirement $%
s^{2}k^{-1}(k+\log p)^{2}=o(n^{(0)})$ for our linear functional-based test $\phi _{1}$ is minimal in a decision theoretic framework.

Similarly the advantage of our chi-based test $\phi _{2}$ is rooted on the sample size requirement of $s^{2}(k+\log p)^{2}=o(n^{(0)})$ obtained in Proposition \ref{cor:l2 testing}. In contrast,
one can also construct a similar chi-based test as in (\ref{eq: l2 test}) based on the
residuals $\hat{E}_{i,j}^{(t)}$ which are obtained through an application of the
procedure in \cite{liu2013gaussian} to each individual graph. For such naive combination testing procedure, it can be shown that the sample size assumption $s^{2}k\left( \log p\right) ^{2}=o(n^{(0)})$ is
required. This demonstrates that in a range of typical scenarios when the number of networks does not grow excessively fast with $k=o\{\left( \log p\right) ^{2}\}$, our
chi-based test $\phi _{2}$ indeed has a weaker sample size requirement.

\subsection{Optimality of tests and minimum sample size requirement} \label{sec:optimal}
So far we have introduced our THP framework in multiple networks with two different types of tests for testing the null hypothesis $H_{0,ab}: \omega _{a,b}^{0}=\mathbf{0}$ in (\ref{eq:null}). The constructions of our test statistics are motivated by the possible alternative hypothesis. In particular, the chi-based test $\phi _{2}$ should be powerful as long as the joint link strength 
$\Vert \omega
_{a,b}^{0}\Vert $ is away from zero, while the linear functional-based test $%
\phi _{1}$ will be powerful when the signs of $\omega _{a,b}^{0}$ are
known and $\Vert \omega _{a,b}^{0}\Vert _{1}$ becomes 
large. Along this direction, we now further investigate two types of
composite alternative hypotheses. We define the set of all $s$-sparse multiple networks 
as
\begin{equation} \label{neweq019}
\mathcal{F}(s)=\mathcal{F}(s,M)=\left\{ \Omega
^{0}:\max_{1 \leq a \leq p}\sum_{1 \leq b\neq a \leq p}1\{\omega _{a, b}^{0}\neq \mathbf{0}\}\leq s%
\mbox{ and
Condition \ref{CondA1} holds}\right\},
\end{equation}%
where 
$\Omega
^{0}=\{\Omega ^{(t)}\}_{t=1}^{k}$ stands for the set of $k$ precision matrices with slight abuse of notation and $s$ is some positive integer. Then the null hypothesis $H_{0,ab}$ in (\ref{eq:null}) can be rewritten as
\begin{equation} \label{neweq020}
H_{0,ab}=H_{0,ab}(s):\Omega ^{0}\in \mathcal{N}(s)\equiv\left\{ \Omega
^{0}:\Omega ^{0}\in \mathcal{F}(s), \, \omega _{a,b}^{0}=\mathbf{0}\right\} .
\end{equation}%
In particular, we consider the following two alternative hypotheses
\begin{eqnarray}
H_{1,ab}^{l2}(s,\epsilon ) &:&\Omega ^{0}\in \mathcal{A}^{l2}(s,\epsilon
)\equiv\left\{ \Omega ^{0}:\Omega ^{0}\in \mathcal{F}(s), \, \left\Vert \omega
_{a,b}^{0}\right\Vert \geq \epsilon \right\} ,  \label{eq:alternative l2} \\
H_{1,ab}^{l1}(s,\epsilon ,\xi ) &:&\Omega ^{0}\in \mathcal{A}%
^{l1}(s,\epsilon ,\xi )\equiv\left\{ \Omega ^{0}:\Omega ^{0}\in \mathcal{F}%
(s), \, \xi ^{\prime }\omega _{a,b}^{0}=\left\Vert \omega _{a,b}^{0}\right\Vert
_{1}\geq \epsilon \right\},  \label{eq:alternative l1}
\end{eqnarray}
where the former is introduced to investigate the chi-based test $\phi _{2}$, the latter is for the linear functional-based test $\phi _{1}$, and $\epsilon > 0$.

It is clear that the difficulty of testing the null $H_{0,ab}$ in (\ref{neweq020}) against the alternative $%
H_{1,ab}^{l2}(s,\epsilon )$ in (\ref{eq:alternative l2}) or against the alternative $H_{1,ab}^{l1}(s,\epsilon
,\xi )$ in (\ref{eq:alternative l1}) depends critically on the quantity $\epsilon $. The smaller $%
\epsilon $ is, the more difficult to distinguish between the null and alternative
hypotheses. A natural and fundamental question is what the boundary of
the testable region is. Such a boundary means that it is impossible to detect whether the
observations are from the null against the alternative as long as $\epsilon $
is smaller than it, while there exists some test which can
distinguish between the two hypotheses whenever $\epsilon $ is far larger than it.

To characterize the testable region boundary, we introduce the separating rate $\epsilon _{n}$ of null $%
H_{0,ab}$ against alternative $H_{1,ab}^{l2}(s,\epsilon )$ or $%
H_{1,ab}^{l1}(s,\epsilon ,\xi )$. For any fixed
significance level $\alpha \in (0,1)$ and power $\alpha < \beta < 1 $, the
\textit{separating rate} for alternative $H_{1}=H_{1,ab}^{l2}(s,\epsilon )$ or $H_{1,ab}^{l1}(s,\epsilon ,\xi )$ is said to be $\epsilon _{n}$ if
there exist some test $\psi _{0}$ of asymptotic significance level $\alpha $
and some absolute large constant $c>0$ such that 
\begin{equation}
\lim_{n^{(0)}\rightarrow \infty }\inf_{v\in \mathcal{A}(c)}\mathbb{P}_{v}(\psi
_{0}\mbox{ \rm  rejects }H_{0,ab})\geq \beta ,  \label{eq:test upper}
\end{equation}%
while there exists some absolute small constant $c^{\prime }>0$ such that for
any test $\psi $ of asymptotic significance level $\alpha $, it holds that
\begin{equation}
\lim_{n^{(0)}\rightarrow \infty }\inf_{v\in \mathcal{A}(c^{\prime })}\mathbb{P}_{v}(\psi
\mbox{ \rm
rejects }H_{0,ab})<\beta ,  \label{eq:test lower}
\end{equation}%
where $\mathcal{A}(c)$ represents $\mathcal{A}^{l2}(s,c\epsilon_{n} )$ or $\mathcal{A}%
^{l1}(s,c\epsilon _{n},\xi )$. By symmetry, it is easy to see that the
separating rate $\epsilon _{n}$ for alternative $H_{1,ab}^{l1}(s,\epsilon
,\xi )$ defined above is free of 
the sign vector $\xi $.

Our major goals in this section are twofold. First,
we identify the separating rates $\epsilon
_{n}$ for alternative $H_{1,ab}^{l2}(s,\epsilon )$ under the sample size
assumption $s^{2}(k+\log p)^{2}=o(n^{(0)})$ and for alternative $%
H_{1,ab}^{l1}(s,\epsilon ,\xi )$ under the sample size assumption $%
s^{2}k^{-1}(k+\log p)^{2}=o(n^{(0)})$. 
In particular, we show later in
Theorem \ref{thm:lower rate} that  $\epsilon _{n}\asymp \sqrt{%
k^{1/2}/n^{(0)}}$ for alternative $H_{1,ab}^{l2}(s,\epsilon )$ and $\epsilon
_{n}\asymp \sqrt{k/n^{(0)}}$ for alternative $H_{1,ab}^{l1}(s,\epsilon ,\xi )
$. Moreover, our newly suggested chi-based test $\phi _{2}$ and linear
functional-based test $\phi _{1}$ achieve these two separating rates, respectively, and hence are optimal in this sense.
Second,
we investigate the optimality of the sample size assumption $s^{2}k^{-1}(k+\log
p)^{2}=o(n^{(0)})$ for the $\ell_1$ type alternative $%
H_{1,ab}^{l1}(s,\epsilon ,\xi )$ in (\ref{eq:alternative l1}).
Specifically, we establish later in Theorem \ref{thm:lower sample size} that in
order to have separating rate $\epsilon _{n}\asymp \sqrt{k/n^{(0)}}$, this
sample size requirement is necessary under the setting of $k=O(\log p)$.
Therefore, we conclude that the linear functional-based test $\phi _{1}$ is
optimal to test null $H_{0,ab}$ from alternative $H_{1,ab}^{l1}(s,\epsilon ,\xi )$ under the minimum sample size requirement. It is worth mentioning that the novelty and major contributions of our second goal lie in a new construction of a related minimax lower bound argument.

\begin{theorem}
\label{thm:lower rate} (1) Under the conditions of Proposition \ref{cor:l2
testing}, 
the separating rate for testing $H_{0,ab}$ against $%
H_{1,ab}^{l2}(s,\epsilon )$ is $\epsilon _{n}=\sqrt{k^{1/2}/n^{(0)}}$ and the chi-based test $\phi _{2}$ in (\ref{eq: l2 test}) achieves this rate, that is,{\ for any given $\beta >\alpha $, (\ref%
{eq:test upper}) is valid with $\psi _{0}=\phi _{2}$} and $\mathcal{A}(c)=\mathcal{A}^{l2}(s,c\epsilon_{n} )$ for some sufficiently large constant $c>0$.

(2) Under the conditions of Proposition \ref{cor:l1 testing},  
the
separating rate for testing $H_{0,ab}$ against $H_{1,ab}^{l1}(s,\epsilon
,\xi )$ is $\epsilon _{n}=\sqrt{k/n^{(0)}}$ and the linear
functional-based test $\phi _{1}$ in (\ref{eq: l1 test})
achieves this rate. 
\end{theorem}

In fact, the detection problems of the separating rates for $H_{1,ab}^{l2}(s,\epsilon )$ and $%
H_{1,ab}^{l1}(s,\epsilon ,\xi )$ investigated in Theorem %
\ref{thm:lower rate} are closely related to those of optimal quadratic
functional and linear functional estimation for Gaussian sequence models, respectively. See, for example, \cite%
{baraud2002non,ingster2012nonparametric,collier2015minimax} for more
details. Yet Gaussian graphical models are much more complicated than
Gaussian sequence models. Even for the simple setting of $k=1$, it was shown in
\cite{ren2013asymptotic} that minimax estimation of each single edge $\omega
_{a,b}$ can be different from the parametric rate $\sqrt{n}$. This subtle
difference is reflected in the sample size requirements stated in Theorem %
\ref{thm:lower rate} for the setting of multiple networks. 

\begin{theorem}
\label{thm:lower sample size}
Assume that $k\leq M_{1}\log p$, $s>2$, $s^{2}k^{-1}(k+\log p)^{2}>Cn^{(0)}$, $p>s^{\mu }$, and $s[1+(\log p)/k]/n^{(0)}=o(1)$ for some 
large
constants $M_{1}, C>0$ and some 
$\mu >2$. Then given any $\alpha <\beta < 1$ and some constant $c>0$, 
there exists no test of asymptotic significance level $%
\alpha $ satisfying  (\ref{eq:test upper}) with $\mathcal{A}(c)=\mathcal{A}^{l1}(s,c\epsilon _{n},\xi )$ and $\epsilon _{n}=\sqrt{k/n^{(0)}}$. 
\end{theorem}

Theorem \ref{thm:lower sample size} further justifies that the sample size requirement of $%
s^{2}k^{-1}(k+\log p)^{2}=o(n^{(0)})$ for the $\ell_1$ type alternative $%
H_{1,ab}^{l1}(s,\epsilon ,\xi )$ in (\ref{eq:alternative l1}) is indeed sharp. To obtain such result, one needs to construct a lower
bound involving the sample size requirement and the separating rate. For the
single graph setting of $k=1$, this is related to the minimax lower bound of
estimating each single edge $\omega _{a,b}$, which was explored in \cite{ren2013asymptotic}. The lower bound argument in \cite%
{ren2013asymptotic} is, however, not applicable in the current setting even for the case of $k=1$, since the
construction of the least favorable subset of the parameter space in \cite%
{ren2013asymptotic} does not allow $\omega _{a,b}$ to be 
close to zero,
which is in fact the focus of the testing problem. To overcome such difficulty, we propose a very different least favorable subset in our analysis of Theorem %
\ref{thm:lower sample size}.

\subsection{
Comparisons with existing methods
} \label{sec: comp}

As mentioned in the Introduction, there is a rich and growing line of research on multiple networks in the setting of Gaussian graphical
models. 
Due to the space constraint, we compare our procedure with some most relevant ones in the literature. Our work makes no assumption on the ordering for the $k$ networks. 
Existing work along this line includes, for instance,  \cite{guo2011joint,jgl,zhu2014structural,mpe}.
The main advantages of our proposed THP method over these existing approaches are threefold.
First, our THP framework with the two specific testing procedures provides statistical inference for each joint link strength vector 
$%
\omega _{a,b}^{0}$ over $k$ networks 
to reflect its statistical significance.
This is of crucial importance for model interpretation, 
false discovery rate control, and global multiple precision matrices estimation in applications.
In contrast, none of these previous attempts along this line goes beyond point estimation to investigate statistical inference.

Second, our theoretically optimal procedure is tuning free and data driven.
This is mainly due to a novel approach of HGSL as a convex program as well as a computationally fast algorithm with convergence guarantees suggested in Section \ref{sec:initial} for the setting of high-dimensional multi-response regression with heterogeneous noises, which may be of independent interest.  
Different from ours, all existing methods typically involve one or more tuning parameters. Moreover, some of these methods rely on nonconvex optimization problems whose global solutions cannot always be
guaranteed to be computable. In contrast, our procedure not only enjoys the computational efficiency but also avoids the additional practical and theoretical issues caused by the use of the cross-validation; see the simulation studies in Section \ref{sec:simulationmodelsetting} for a detailed comparison on the computational cost of our algorithm with competitors 
which demonstrates the computational advantage of our procedure. Third, our procedure admits the optimality properties established for two different types of tests in terms of the separating rates, which follow from three new lower bound arguments introduced in Sections \ref{SecA.3} and \ref{SecA.4} of the Supplementary Material. To
the best of our knowledge, there are no such immediate results available in the
literature of multiple Gaussian graphical models. The obtained
optimality results ensure that our testing procedures are optimal.

More thorough theoretical comparisons of our
method with competitor ones are possible but involved, particularly given that no inference results are provided for these existing methods. For a fair comparison, we now focus on the requirements for support recovery results of different methods under the
assumption that all $k$ graphs share a common sparsity structure. To this end,
we need to go a little further based on our chi-based test $\phi _{2}$ by
replacing $\alpha $ in (\ref{eq: l2 test}) by $p^{-2-\rho }$ with some $\rho
>0$. Specifically, for any given $\rho
>0$ we define the THP estimator $\mathcal{\hat{E}}$ for the support or edge set $%
\mathcal{E}$ corresponding to the $k$ graphs in (\ref{eq:edge set}) as%
\begin{equation} \label{neweq015}
(a,b)\in \mathcal{\hat{E}}\text{ \ \ \ \ when  \ \ }%
U_{n,k,a,b}>z_{k}^{l2}(1-p^{-2-\rho }),
\end{equation}%
where all the notation is the same as in (\ref{eq: l2 test}).
The following proposition establishes that the THP estimator $\mathcal{\hat{E}}$ introduced in (\ref{neweq015}) is indeed capable of recovering the network structure exactly with large probability as long as the minimum signal strength is above a certain threshold.

\begin{proposition} \label{prop:suppreco}
Assume that all the conditions of Proposition \ref{cor:l2 testing} hold and
$\min_{(a,b)\in \mathcal{E}%
}\Vert \omega _{a,b}^{0}\Vert >C\sqrt{[(k\log p)^{1/2}+\log p]/n^{(0)}}$
for some sufficiently large constant $C>0$. Then the THP estimator
$\mathcal{\hat{E}}$ given in (\ref{neweq015}) satisfies 
$\mathcal{\hat{E}} = \mathcal{E}$ with probability at least $1-O(p^{-\rho })$.
\end{proposition}

In view of the separating rate $C\sqrt{k^{1/2}/n^{(0)}}$ obtained in Theorem \ref{thm:lower rate} (1) for a single joint link strength vector, 
we see that the lower bound on the minimum signal strength $\min_{(a,b)\in \mathcal{E}%
}\Vert \omega _{a,b}^{0}\Vert$ in Proposition \ref{prop:suppreco} for support recovery comes with an extra factor of $(\log p)^{1/4}$ for the case of $\log p=O(k)$, or with the factor $k^{1/4}$ replaced by $(\log p)^{1/2}$ for the case of  $k=O(\log p)$. We would like to point out that such increased minimum signal strength generally cannot be avoided and stems from the union bound argument taken over all pairs of nodes $(a, b)$ in the edge set $\mathcal{E}$.  

Let us gain some insights into the advantage of our THP procedure on support recovery in comparison to some existing approaches.
To recover
the support successfully, at least the minimum signal strength requirement of $\min_{(a,b)\in \mathcal{E%
}}\Vert \omega _{a,b}^{0}\Vert \geq C\sqrt{k}$ is needed in \cite{guo2011joint},
and the assumption of $\min_{(a,b)\in \mathcal{E}}\Vert \omega _{a,b}^{0}\Vert \geq
CM_{n}\sqrt{(k\log p)/n}$ is needed in \cite{mpe}, where $M_{n}\equiv\max_{1 \leq t \leq k}\max_{1 \leq b \leq p}\Sigma
_{a=1}^{p}\vert \omega _{a, b}^{(t)}\vert $ denoting the largest matrix $1$%
-norm among $k$ graphs can diverge 
with $n^{(0)}$ under our setting, and $C$ is some positive constant. In addition, no theoretical
justification is provided for the method in \cite{jgl}, and the support recovery result in \cite{zhu2014structural} cannot
be easily compared due to an extra clustering structural assumption. In
summary, compared with existing methods our optimal THP approach yields a sharper minimum signal strength requirement for recovering the support of the networks 
with common structure, thanks to our optimal testing procedures.

\section{Tuning-free 
heterogeneous group square-root Lasso} \label{sec:initial}

\subsection{
Heterogeneous group square-root Lasso: a convex program} \label{sec:initial_1}
Our THP framework suggested in Section \ref{sec:test} for uncovering the heterogeneity in sparsity patterns among multiple networks via large-scale inference relies critically on an efficient procedure for fitting the high-dimensional multi-response linear regression model (\ref{eq:multi-reponse}) for each node $1 \leq j \leq p$. We now introduce such an approach HGSL that can be of independent interest when one is in need of a tuning-free method for the general  setting of high-dimensional multi-response regression with heterogeneous noises.
Specifically, we need to construct some initial estimators $\hat{C}_{j}^{0}=( \hat{C}%
_{j}^{(1)\prime },\cdots ,\hat{C}_{j}^{(k)\prime }) ^{\prime }$ for the $(p-1)k$-dimensional regression coefficient vectors $%
C_{j}^{0}=\left( C_{j}^{(1)\prime },\cdots ,C_{j}^{(k)\prime }\right)'$ in model (\ref{eq:multi-reponse}) with $1 \leq j \leq p$ that each satisfy properties (\ref{eq:assumption1})--(%
\ref{eq:assumption3}) with significant probability, say, at least $1-C_{0}p^{1-\delta }$ for some positive constants $C_{0}$ and $\delta > 1$.

By symmetry, we can focus only on the case of $j=1$ hereafter without loss of generality. Recall that in our model (\ref{neweq016}), for each graph $1 \leq t \leq k$ we have an $n^{(t)} \times p$ data matrix $%
\mathbf{X}^{(t)}\mathbf{=(}X_{1,\ast }^{(t)},\cdots
,X_{n^{(t)},\ast }^{(t)}\mathbf{)}^{\prime }$ with i.i.d. rows $X_{i,\ast
}^{(t)}=(X_{i,1}^{(t)},\cdots ,X_{i,p}^{(t)})^{\prime }\sim N(0,(\Omega
^{(t)})^{-1})$ for $1 \leq i \leq n^{(t)}$. Using the matrix notation, the multi-response linear regression model \eqref{eq:multi-reponse} can be rewritten as%
\begin{eqnarray}
\left(
\begin{array}{c}
X_{\ast ,1}^{(1)} \\
X_{\ast ,1}^{(2)} \\
\vdots  \\
X_{\ast ,1}^{(k)}%
\end{array}%
\right) &=&\left(
\begin{array}{cccc}
\mathbf{X}_{\ast ,-1}^{(1)} &  &  &  \\
& \mathbf{X}_{\ast ,-1}^{(2)} &  &  \\
&  & \ddots  &  \\
&  &  & \mathbf{X}_{\ast ,-1}^{(k)}%
\end{array}%
\right) \left(
\begin{array}{c}
C_{1}^{(1)} \\
C_{1}^{(2)} \\
\vdots  \\
C_{1}^{(k)}%
\end{array}%
\right) +\left(
\begin{array}{c}
E_{\ast ,1}^{(1)} \\
E_{\ast ,1}^{(2)} \\
\vdots  \\
E_{\ast ,1}^{(k)}%
\end{array}%
\right) \nonumber \\
 &\equiv&\mathbf{X}_{\ast ,-1}^{0}C_{1}^{0}+E_{\ast ,1}^{0}
\label{eq:HeterGroupLasso}
\end{eqnarray}%
lying in the $N$-dimensional Euclidean space, where $X_{*,1}^{(t)} = (X_{1,1}^{(t)}, \cdots, X_{n^{(t)}, 1}^{(t)})'$, $N=\sum_{t=1}^{k}n^{(t)}$ denotes the total sample size, $E_{\ast ,1}^{(t)}=(E_{1,1}^{(t)},\cdots
,E_{n^{(t)},1}^{(t)})^{\prime }$ is the same as in \eqref{eq:V star} with
i.i.d. components from distribution $N(0,(\omega _{1,1}^{(t)})^{-1})$, and we adopt the compact notation introduced in Section \ref{sec:test.l1}. In addition, we have the
group sparsity structure for the regression coefficient vector $C_{1}^{0}$, which means that all but at most $s$ subvectors $C_{1(l)}^{0}\in \mathbb{R}^{k}$ are zero with $C_{1(l)}^{0}$ and $s$ defined in (\ref{eq:group subvector}) and (\ref{eq:sparsity}), respectively.

The joint group structure and sparsity structure in the multi-response linear regression model (\ref{eq:HeterGroupLasso}) naturally motivate us to exploit some variant of the group Lasso method \citep{YL06} to estimate the coefficient vector $C_{1}^{0}$. The asymptotic
properties of the standard group Lasso are well understood and imply faster
rates of convergence in estimating $C_{1}^{0}$ and $\mathbf{X}_{\ast
,-1}^{0}C_{1}^{0}$, compared to the standard Lasso approach \citep{Tibshirani96}. See, for
instance, \cite{huang2010benefit} and \cite{lounici2011oracle} for more  
details. The optimal choice of an important tuning parameter, the regularization 
parameter $\lambda \geq 0$, in these methods, 
however,
depends critically on the common noise level $\sigma $ and is thus typically
unknown in practice. Hence one needs a practical and data-driven choice of $\lambda$ that can lead to optimal estimation. Such important issue
has been investigated recently
in \cite{bunea2014group} and \cite{mitra2014benefit} by extending the tuning-free methods of the
square-root Lasso \citep{belloni2011square} and the scaled Lasso \citep%
{sun2012scaled} to the group setting, 
respectively.

Yet the aforementioned existing tuning-free approaches in the standard group Lasso setting are not applicable in the model setting (\ref{eq:HeterGroupLasso}), which is due to the distinct feature of  heterogeneity of the noise level in our model.
Indeed,
instead of a common noise level for all components of the error vector $E_{\ast ,1}^{0} = (E_{\ast ,1}^{(1) \prime}, \cdots, E_{\ast ,1}^{(k) \prime})^{\prime }$, we
allow each class to have its own noise level, say, $(\omega _{1,1}^{(t)})^{-1}$ for $1 \leq t \leq k$. The strategy used in the square-root Lasso and the scaled Lasso,
which essentially includes an additional parameter for the noise level, can
handle only the homogeneous noises. To deal with such heterogeneity, 
we extend the group square-root Lasso one step further to allow for  heterogeneous noises. We would like to point out that such
extension for achieving the tuning-free feature is generally 
never trivial, and the novelty of our analysis is due to
an intrinsic constant level upper bound obtained on  
the fitted residual level for each class;
see Lemma \ref{lem:constant upper} in Section \ref{SecB.7} of the Supplementary Material for more details.

To ease the presentation, we first introduce some notation. Define a function $%
Q_{t}(\beta ^{(t)})=\Vert X_{\ast ,1}^{(t)}-\mathbf{X}%
_{\ast ,-1}^{(t)}\beta ^{(t)}\Vert ^{2}/n^{(0)}$ with $\beta ^{(t)}=(\beta _{2}^{(t)},\cdots ,\beta _{p}^{(t)})^{\prime
}\in \mathbb{R}^{p-1}$ matching the index set of $C_{1}^{(t)}$ and $1 \leq t \leq k$. Denote by $\beta
^{0}=(\beta ^{(1)\prime },\cdots ,\beta ^{(k)\prime
})^{\prime }$ a $(p-1)k$-dimensional vector and $\beta
_{(l)}^{0}=(\beta _{l}^{(1)},\cdots ,\beta _{l}^{(k)})^{\prime }\in \mathbb{R%
}^{k}$  the $l$th group of $\beta ^{0}$ with $1 \leq l \leq p$ in the same way as we defined $C_{1(l)}^{0}$ in (\ref{eq:group subvector}).
We
further introduce a diagonal matrix $\bar{D}_{1}^{(t)}=%
\mathrm{diag}(\mathbf{X}_{\ast ,-1}^{(t)\prime }\mathbf{X}_{\ast
,-1}^{(t)}/n^{(t)})$ of order $p-1$ and then put them together to form a new diagonal scaling matrix $\bar{D}_{1}$ of order $(p-1)k$, with the submatrix of $\bar{D}_{1}$ corresponding
to the $l$th group denoted by $\bar{D}_{1(l)}$ and the $t$th
entry on the diagonal of $\bar{D}_{1(l)}$ given by $\mathbf{X}_{\ast
,l}^{(t)\prime }\mathbf{X}_{\ast ,l}^{(t)}/n^{(t)}$.

Our new approach of the heterogeneous group square-root Lasso (HGSL) 
is defined as the one given by
the following optimization problem
\begin{equation}
\hat{C}_{1}^{0}={\arg \min}_{\beta ^{0}\in \mathbb{R}^{(p-1)k}}%
\left\{\sum_{t=1}^{k}Q_{t}^{1/2}(\beta ^{(t)})+\lambda \sum_{l=2}^{p}\left\Vert
\bar{D}_{1(l)}^{1/2}\beta _{(l)}^{0}\right\Vert \right\},  \label{eq:GSRLH}
\end{equation}%
where the regularization parameter $\lambda > 0$ which is chosen to be  independent of the noise levels $(\omega _{1,1}^{(t)})^{-1}$ for $1 \leq t \leq k$ will be provided explicitly later. Clearly, our HGSL procedure
defined in (\ref{eq:GSRLH}) is a convex program and yields an estimator for the $(p-1)k$-dimensional regression coefficient vectors $C_{1}^{0}$. For the estimation of general $C_{j}^{0}$ with $1 \leq j \leq p$, one can simply replace the corresponding subscript $1$ by $j$ in the above method (\ref{eq:GSRLH}). The optimization problem in (\ref{eq:GSRLH}) coincides with the standard
square-root Lasso in \cite{belloni2011square} for the case of $k=1$, and  differs from the standard
group square-root Lasso in \cite{bunea2014group} which is defined with the loss function $(\sum_{t=1}^{k}Q_{t}(\beta ^{(t)}))^{1/2}$ in place of ours $\sum_{t=1}^{k}Q_{t}^{1/2}(%
\beta ^{(t)})$ when $k \geq 2$.
Without such new feature in the formulation, the standard
group square-root Lasso, however, cannot carry over to take into account the heterogeneity issue when the noise level varies across different classes.

As revealed in the analysis of Theorem \ref{thm:GSRLH} to be presented, a key ingredient for the success of our HGSL estimators is an event $\mathcal{B}_{1}$ defined as
\begin{equation}
\mathcal{B}_{1}=\left\{ \frac{\max_{2\leq l\leq p}\left\Vert \bar{D}_{E1}^{-1/2}%
\bar{D}_{1(l)}^{-1/2}\mathbf{X}_{\ast ,(l)}^{0\prime }E_{\ast
,1}^{0}\right\Vert }{\sqrt{n^{(0)}}}\leq \lambda \frac{\xi -1}{\xi +1}%
\right\}   \label{eq: lambda condition}
\end{equation}%
for any fixed scalar $\xi >1$, where $\mathbf{X}_{\ast ,(l)}^{0}$ is an $N\times k$
submatrix of $\mathbf{X}_{\ast ,-1}^{0}$ given by columns 
corresponding to the $l$th group and $\bar{D}_{E1}$ is a $k \times k$ diagonal matrix
with $t$th diagonal entry the squared $\ell_2$ norm of the error 
vector
$E_{\ast ,1}^{(t)}$, that is, $(\bar{D}_{E1})_{t,t}=\Vert E_{\ast
,1}^{(t)}\Vert ^{2}$ for $1 \leq t \leq k$. Similarly we can define the event $\mathcal{B}_{j}$ as in (\ref{eq: lambda condition})
for each node $1 \leq j \leq p$. Each event $\mathcal{B}_{j}$ represents the one
that the pure
noise incurred is dominated by the penalty level. In order to ensure that event $\mathcal{B}_{j}$ holds with high probability, we need to carefully
pick a sharp choice of the regularization parameter $\lambda $ that is free of the heterogeneous noise levels. 


\begin{theorem}
\label{thm:GSRLH}
Assume that Conditions \ref{CondA1}--\ref{CondA2} hold, 
$s\leq
C_{\xi }n^{(0)}/\log p$ for some constant $C_{\xi }>0$, 
and 
let $\hat{C}_{j}^{0}$ be the solution as in (\ref{eq:GSRLH}) for $1 \leq j \leq p$
with 
$
\lambda =\frac{\xi +1}{\xi -1}\left[ \frac{k+2\delta \log p+2\sqrt{\delta
k\log p}}{n^{(0)}(1-\tau )}\right] ^{1/2},
$
$\tau ^{2}=8(\delta \log p+\log k)/n^{(0)}=o(1)$, and $\delta > 1$ some constant. Then the event $\mathcal{B}_{j}$ holds 
with
probability at least $1-3p^{1-\delta}$, and it holds with probability at least $1-4p^{1-\delta}$ that 
\begin{eqnarray}
\sum_{1 \leq l \leq p, \, l \neq j}\frac{1}{\sqrt{k}}\left\Vert \hat{C}%
_{j(l)}^{0}-C_{j(l)}^{0}\right\Vert  &\leq &Cs\left[\frac{1+(\log p)/k}{n^{(0)}}%
\right]^{1/2},  \label{eq:L1 bound} \\
\left\Vert \hat{C}_{j}^{0}-C_{j}^{0}\right\Vert  &\leq &C\left[s\frac{1+(\log p)/k%
}{n^{(0)}}\right]^{1/2},  \label{eq:L2 bound} \\
\frac{1}{k}\sum_{t=1}^{k}\frac{\left\Vert \mathbf{X}_{\ast ,-1}^{(t)}\left(
\hat{C}_{j}^{(t)}-C_{j}^{(t)}\right) \right\Vert ^{2}}{n^{(0)}} &\leq &Cs%
\frac{1+(\log p)/k}{n^{(0)}},  \label{eq: weigted prediction}
\end{eqnarray}%
where $C>0$ is some constant. 
\end{theorem}


Theorem \ref{thm:GSRLH} establishes the estimation and prediction bounds for  our HGSL estimators. The novelty of our technical analysis comes from an intrinsic upper bound on the fitted residual level for each class. It is worth mentioning that with the knowledge of such quantity, we can also apply the regular group Lasso with a tuning parameter depending on this quantity and obtain a corresponding justifiable theorem.
The intrinsic upper bound in our analysis, however, does not appear in the HGSL optimization problem in (\ref{eq:GSRLH}) and provides only theoretical support, while the regular group Lasso implemented in the above way has to apply it in the tuning parameter explicitly. Consequently, this possibly
loose intrinsic upper bound can yield large bias for the regular group Lasso,  but still sharp results for our HGSL method; see the proofs of Theorem \ref{thm:GSRLH} and
Lemma \ref{lem:constant upper} in Sections \ref{SecA.5} and \ref{SecB.7} of the Supplementary Material, respectively, for more details. 

Let us gain some further insights into our tuning-free HGSL method by comparing the sharpness of our regularization parameter $\lambda $ specified in Theorem \ref{thm:GSRLH} with
the one used in \cite{bunea2014group} for the setting of homogeneous noises.
One advantage of our choice of $\lambda $ comes from the use of the scaling
matrix $\bar{D}_{1}$, which makes the noise per column of $\mathbf{X}_{\ast
,(l)}^{0}$ homogeneous and sharpens $\lambda $ by a factor given by 
the ratio of the largest and the smallest $\ell_2$ norms among all columns.
Moreover, thanks to the simple block diagonal structure of matrices $\mathbf{X}_{\ast
,(l)}^{0}$ a direct and sharp chi-square tail probability \citep{laurent2000adaptive} provides us sharper constant
factors for both $k$ and $\log p$.

In addition to the choice of parameter $\lambda$ for HGSL established in Theorem \ref{thm:GSRLH}, in practice we can also calculate the sharp parameter $\lambda $ using simulation. For instance, we can simulate the value of $\|\bar{D}%
_{E1}^{-1/2}\bar{D}_{1(2)}^{-1/2}\mathbf{X}_{\ast ,(2)}^{0\prime }E_{\ast
,1}^{0}\|/(n^{(0)})^{1/2}$ for $10,000$ times and pick the $100(1-1/p^{\delta})$th percentile of
its empirical distribution as our choice of $\lambda (\xi -1)/(\xi +1)$ with
some constant $\delta >1$. Here we take $\delta >1$ because of the union bound argument given
that only the setting of $l=2$ is simulated. It is important to note that the
components of $\bar{D}_{E1}^{-1/2}\bar{D}_{1(2)}^{-1/2}\mathbf{X}_{\ast
,(2)}^{0\prime }E_{\ast ,1}^{0}$ are independent and their distributions can
be characterized easily since they do not depend on the variances of $\mathbf{X}_{\ast
,(2)}^{0\prime }$ and $E_{\ast ,1}^{0}$. More specifically, for each
replication $1 \leq T \leq 10,000$ we simulate the $t$th component of $\bar{D%
}_{E1}^{-1/2}\bar{D}_{1(2)}^{-1/2}\mathbf{X}_{\ast ,(2)}^{0\prime }E_{\ast
,1}^{0}$ independently by first generating $Z_{1,t,T}, Z_{2,t,T} \sim N(0,I)\in \mathbb{R}%
^{n^{(t)}}$  
independently
and then calculating $Z_{t,T}=(n^{(t)})^{1/2}Z_{1,t,T}^{\prime
}Z_{2,t,T}/(\|Z_{1,t,T}\| \|Z_{1,t,T}\|)^{1/2}$. The simulated
value of $\|\bar{D}_{E1}^{-1/2}\bar{D}_{1(2)}^{-1/2}\mathbf{X}_{\ast
,(2)}^{0\prime }E_{\ast ,1}^{0}\|$ can then be written as $%
(\sum_{t=1}^{k}Z_{t,T}^{2})^{1/2}$. Thus our simulation strategy provides a specific choice of the parameter $\lambda$ given by
\begin{equation}
\lambda_{\textit{sim}}=\frac{1}{\sqrt{n^{(0)}}}\frac{\xi +1}{\xi -1}\inf \left\{
v:\sum_{T=1}^{10000}1\left\{ \Big(\sum_{t=1}^{k}Z_{t,T}^{2}\Big)^{1/2}<v\right\}
/10000\geq 1-1/p^{\delta }\right\} .  \label{eq: sim lambda}
\end{equation}
We will further discuss the choices of $\delta$ and $\xi$ in Section \ref{sec:simulationmodelsetting} when implementing our proposed procedure THP with the HGSL.


\subsection{
Scalable HGSL algorithm with provable convergence
} \label{Sec3.2}

The tuning-free feature of HGSL established in Section \ref{sec:initial_1} provides a crucial 
step toward the scalability of our THP framework when one needs to analyze a large number of networks with massive number of nodes jointly. To further boost the scalability, we now introduce a new computational algorithm to solve the convex program of HGSL problem in (\ref{eq:GSRLH}) in a simple yet efficient fashion, which will be referred to as the HGSL algorithm hereafter for simplicity. As is common in regularization problems, we rescale  
each column of $\mathbf{X}_{\ast ,-1}^0$ to have $\ell_2$ norm $(n^{(t)})^{1/2}$ and denote by $\bar{\bX}_{\ast,-1}^{0} = \text{diag}\{\bar{\bX}_{\ast,-1}^{(1)}, \cdots, \bar{\bX}_{\ast,-1}^{(k)}\}$ the resulting new design matrix; that is, $\bar{\bX}_{\ast,-1}^{0} = \mathbf{X}_{\ast ,-1}^0\bar D_1^{-1/2}$ with the scaling matrix $\bar D_1$ given in Section \ref{sec:initial_1}. Let us consider another HGSL optimization problem
\begin{equation}
\hat{\bar{C}}_{1}^{0}={\arg\min}_{\beta^0\in \mathbb{R}^{(p-1)k}}%
\left\{\sum_{t=1}^{k}\bar{Q}_{t}^{1/2}(\beta ^{(t)})+\lambda \sum_{l=2}^{p}\left\Vert
\beta _{(l)}^0\right\Vert\right\}, \label{eq: group square-root Lasso1}
\end{equation}
where $\bar{Q}_{t}(\beta ^{(t)})=\Vert X_{\ast ,1}^{(t)}-\bar{\bX}_{\ast,-1}^{(t)}\beta ^{(t)}\Vert ^{2}/n^{(0)}$ for $1 \leq t \leq k$ and 
the rest of the notation is defined similarly as in \eqref{eq:GSRLH}. In fact, the new HGSL optimization problem in (\ref{eq: group square-root Lasso1}) is closely related to the original HGSL optimization problem in (\ref{eq:GSRLH}), through a simple equation $\hat{\bar{C}}_{1}^{0} = \bar D_1^{1/2}\hat C_1^{0}$ linking the minimizers of these two problems.
Thus the problem of solving \eqref{eq:GSRLH} reduces to that of solving \eqref{eq: group square-root Lasso1}.

To ease the presentation, we slightly abuse the notation and rewrite the new HGSL optimization problem \eqref{eq: group square-root Lasso1} in a general form 
\begin{equation}
\hat\beta={\arg \min}_{\beta\in \mathbb{R}^{pk}}%
\left\{(n^{(0)})^{-1/2}\sum_{t=1}^{k}\Vert Y^{(t)}-\bX^{(t)}\beta ^{(t)}\Vert+\lambda \sum_{l=1}^{p}\left\Vert
\beta _{(l)}\right\Vert\right\}, \label{eq: group square-root Lasso2}
\end{equation}
where $Y^{(t)} \in \mathbb{R}^{n^{(t)}}$, $\bX^{(t)} \in \mathbb{R}^{n^{(t)}\times p}$, and $\beta^{(t)} \in \mathbb{R}^{p}$ are the response vector, the design matrix, and the regression coefficient vector, respectively, corresponding to the $t$th network 
for $1 \leq t \leq k$ with the $pk$-dimensional vector $\beta=((\beta^{(1)})', \cdots, (\beta^{(k)})')'$  
and $\beta_{(l)}$ a $k$-dimensional subvector of $\beta$ formed by 
each $l$th component of $\beta^{ (t)}$ with $1 \leq t \leq k$. Similarly we define the $p$-dimensional subvectors $\hat\beta^{ (t)}$ of $\hat\beta$ with $1 \leq t \leq k$, and its $k$-dimensional subvectors $\hat\beta_{(l)}$ with $1 \leq l \leq p$.

So far our original HGSL optimization problem in (\ref{eq:GSRLH}) has been reduced to the general HGSL optimization problem in (\ref{eq: group square-root Lasso2}) with the same tuning-free choice of the parameter $\lambda$ as discussed in Section \ref{sec:initial_1} and the relationship between the two minimizers elucidated above. To solve the convex  optimization problem in ($\ref{eq: group square-root Lasso2}$), we suggest a new scaled iterative thresholding algorithm.
Our HGSL algorithm is designed specifically for the HGSL problem with convergence guarantees, motivated by the algorithm for the group square-root Lasso with homogeneous noises in \cite{bunea2014group} as well as a more general algorithm developed in \cite{she2012}. In practice, to 
reduce the bias of the estimator $\hat{\beta}$ incurred by the regularization  
in ($\ref{eq: group square-root Lasso2}$) one can obtain the final estimate by a refit on the support of the computed sparse $\hat{\beta}$ using the ordinary least-squares estimator.

Our HGSL algorithm consists of two main steps, with the first step for rescaling and the second one for iteration. In the first step, we rescale the response vector, the design matrix, and the regularization parameter 
as
\begin{equation}\label{eq: scale}
Y^{(t)}/K_0\rightarrow Y^{(t)}, \ \bX^{(t)}/K_0\rightarrow \bX^{(t)}, \ \lambda/K_0\rightarrow \lambda \  \text{ for } 1 \leq t \leq k,
\end{equation}
where $K_0 > 0$ is some preselected sufficiently large scalar. Clearly the solution to the optimization problem (\ref{eq: group square-root Lasso2}) remains the same after the rescaling specified in (\ref{eq: scale}). Such step, however, reduces the norm of the design matrix, which can guarantee the convergence of the iterative algorithm as shown in 
Theorem \ref{com theorem} later. We again slightly abuse the notation and still use $Y^{(t)}$, $\bX^{(t)}$, and $\lambda$ to denote the response vector, the design matrix, and the regularization 
parameter after rescaling hereafter. 
In particular, the choice of $K_0=\max_{1 \leq t \leq k} \| \bX^{(t)} \|_{\ell_2}$ with $\|\cdot\|_{\ell_2}$ denoting the spectral norm of a matrix, which is suggested by inequality (\ref{ine}) in the proof of Theorem \ref{com theorem} in Section \ref{SecA.6} of the Supplementary Material, works well in our simulation studies.

In the second step, we solve iteratively the general HGSL optimization problem in (\ref{eq: group square-root Lasso2}) 
with the rescaled data matrix from the first step, and let $\beta(m)$ be the solution returned by the $m$th iteration for each integer $m \geq 0$. 
For the initial value $\beta(0)$, we set it as the zero vector in our numerical studies, which works well.
Denote by $\beta(m)^{(t)}$ and $\beta(m)_{(l)}$ the subvectors of $\beta(m)$ similarly as in \eqref{eq: group square-root Lasso2}. For the $(m+1)$th iteration with input $\beta(m)$, we define  $R(m)=((R(m)^{(1)})',\cdots,(R(m)^{(k)})')' \in \mathbb{R}^{pk}$ with
\[ R(m)^{(t)}=  (\bX^{(t)})'\left(\bX^{(t)}\beta(m)^{(t)}-Y^{(t)}\right)/\left[(n^{(0)})^{1/2}\left \| \bX^{(t)}\beta(m)^{(t)}-Y^{(t)} \right \|\right] \]
for $1 \leq t \leq k$, denote by $R(m)_{(l)}$ a $k$-dimensional subvector of $R(m)$ corresponding to the $l$th group for $1 \leq l \leq p$, 
and introduce a scaling factor
$A(m)=\sum_{t=1}^{k}\left[(n^{(0)})^{1/2}\left \| \bX^{(t)}\beta(m)^{(t)}-Y^{(t)} \right \|\right]^{-1}$. 
Then we compute $\beta(m+1)$ as
\begin{equation} \label{algorithm update}
\beta (m+1)_{(l)} =\overrightarrow{ \Theta }\Big(\beta (m)_{(l)} -\frac{R(m)_{(l)}}{A(m)};\frac{\lambda }{A(m)}\Big) \quad \text{ for }  1 \leq l \leq p,
\end{equation}
where $\overrightarrow{ \Theta }$ is the
multivariate soft-thresholding operator defined as
\begin{equation} \label{neweq001}
\overrightarrow{ \Theta }(0;\lambda )= 0 \text{ \quad and \quad } \overrightarrow{ \Theta }(a;\lambda )= a\Theta(\left \| a \right \|;\lambda )/\left \| a \right \| \  \text{ for } a \neq \mathbf{0}
\end{equation}
with $ \Theta(t;\lambda )=\text{sgn}(t)(\left |  t\right |-\lambda )_{+}$ representing the soft-thresholding rule. 
In practice, we stop the iteration when the difference between the solutions from two consecutive iterates falls below a prespecified small threshold for convergence.


\begin{theorem}\label{com theorem}
Assume that $\lambda > 0$ and 
$\min_{1 \leq t\leq k}\inf_{\xi \in A^t}\left \| \bX^{(t)}\xi -Y^{(t)}  \right \|>c_0$ 
with $A^t=\{v\beta(m)^{(t)}\\+(1-v)\beta(m+1)^{(t)}:v\in[0,1], m=0,1,\cdots\}$ and $c_0 > 0$ some constant. Then 
for large enough $K_0$, the sequence of computed solutions $\beta(m)$ converges to the global optimum of the HGSL 
problem \eqref{eq: group square-root Lasso1}.
\end{theorem}


Theorem \ref{com theorem} justifies formally that our suggested scalable HGSL algorithm indeed enjoys provable convergence to the global optimum of our convex HGSL optimization problem. The scalability of the HGSL algorithm is rooted on both the tuning-free feature and the simple iterative thresholding nature. It is also worth mentioning that a similar regularity condition to the one assumed in Theorem \ref{com theorem} was imposed in \cite{bunea2014group} to prove the convergence of their algorithm for the group square-root Lasso with homogeneous noises. As mentioned before, in the end one can further apply a refit using the support of the computed sparse solution to obtain a final estimate with possibly reduced bias. 



\section{Numerical studies} \label{sec:data}

\subsection{Simulation studies} \label{sec:simulationmodelsetting}
We now proceed with investigating the finite-sample performance of our proposed framework THP with the chi-based test $\phi _{2}$ and the linear functional-based test $\phi _{1}$, which are referred to as procedures THP-$\phi _{2}$ and THP-$\phi _{1}$, respectively, for simplicity, in some simulation examples.
In particular, Section \ref{Sec4.1.1} presents the hypothesis testing results of our methods. As discussed in the Introduction and Section \ref{sec: comp},  
the existing methods on multiple graphs have focused on the estimation problem instead of statistical inference.  
As such, we modify our procedures correspondingly to obtain estimates for the precision matrix and then compare them with some popularly used 
approaches such as the MPE \citep{mpe} and the GGL and FGL \citep{jgl} in Section \ref{subsec: comparison}.  
Section \ref{subsec: heavy-tailed} further examines the robustness of our methods in the presence of heavy-tailed distributions. 


We consider two different model settings, Models I and II, for generating the $k$ networks with Gaussian graphical models 
given by precision matrices $\Omega^{(t)}=(\omega^{(t)}_{a,b})$ with $1 \leq t \leq k$. In both models, the block diagonal structure is used to introduce sparsity in the precision matrices in the sense that all the entries outside 
the diagonal blocks are equal to zero. More specifically, our Model I assumes that all $k$ precision matrices share the same block diagonal  structure and all diagonal blocks have the same size. For each pair $(a,b)$ with $1 \leq a\neq b \leq p$, if the $(a,b)$th entry belongs to a diagonal block, then we draw the values for $\omega_{a,b}^{(1)}, \cdots, \omega_{a,b}^{(k)}$ independently from the uniform distribution $U[0.2,0.4]$ or $U[0.6, 1.2]$, depending on whether it belongs to the upper half diagonal blocks or the lower half diagonal blocks, respectively. All the off-diagonal entries within the diagonal blocks are generated 
independently. Finally we set the diagonal entries as $1$ for the upper half diagonal blocks and $3$ for the lower half ones. Observe that in Model I, each joint link strength vector $\omega _{a,b}^{0}=(\omega _{a,b}^{(1)}, \cdots
,\omega _{a,b}^{(k)})^{\prime }$ with $a \neq b$
is either a zero vector or of $k$ nonzero components. 

To make the sparsity pattern more flexible compared to Model I, our Model II 
employs a different data generating scheme for entries inside the diagonal blocks with the rest of the setting the same as in Model I.
Specifically, for each entry $(a,b)$ with $a\neq b$ inside a diagonal block  we first flip a fair coin. If it is heads, then the joint link strength vector $\omega _{a,b}^{0}$  
is generated in the same way as in Model I. If it is tails, we randomly draw an integer $k_0$ from the uniform distribution over $\{1,\cdots, k\}$, and then set $\omega_{a,b}^{(t)}=0$ for each $1 \leq t\neq k_0 \leq k$ and generate $\omega_{a,b}^{(k_0)}$ from the uniform distribution $U[0.2, 0.4]$ or $U[0.6, 1.2]$, depending on whether the pair $(a,b)$ falls in the upper half diagonal blocks or the lower half diagonal blocks, respectively. Clearly, Model II is sparser than Model I.

For each of the two models introduced above, we further consider three different settings of parameters by varying the number of networks 
$k$ and the number of nodes $p$, while fixing the sample sizes $n^{(t)}=n^{(0)}$ at $100$ for Model I and at $200$  
for Model II with $1 \leq t \leq k$.  
We also fix the block size to be $8$ and set the number of repetitions as $100$ in each simulation setting. The tuning-free regularization parameter $\lambda$ is chosen as $\lambda_{\textit{sim}}$ in (\ref{eq: sim lambda}) using our simulation strategy with $\delta=1$ and $\xi=\infty$. Alternatively one can also use the choice of parameter $\lambda$ given in Theorem \ref{thm:GSRLH}, which results in similar but slightly worse performance  compared to the use of $\lambda_{\textit{sim}}$.

\subsubsection{Testing results} \label{Sec4.1.1}
To see how our proposed methods THP-$\phi _{2}$ and THP-$\phi _{1}$ perform in finite samples, let us start with the hypothesis testing results in Models I and II.
For each simulated data set, we apply the THP procedure with the chi-based test $\phi_2$ and the linear functional-based test $\phi_1$ with sign vector $\xi=(1, \cdots, 1)'$ to each pair of nodes $(a,b)$ with $a\neq b$ to detect whether some edges exist between nodes $a$ and $b$ for any of the $k$ networks. 
We set the significance level $\alpha$ to be $0.05$ and employ two different methods to calculate the critical values. The first method computes the critical values using the asymptotic null distributions established in Theorems \ref{thm:l2 testing} and \ref{thm:l1 testing}, with the corresponding critical values named as ``Theoretical" in Tables \ref{testing model 1} and \ref{testing model 2}. The second method, called ``Empirical" in Tables \ref{testing model 1} and \ref{testing model 2}, computes the critical values empirically based on the values of the test statistic $U_{n,k,a,b}$ for the chi-based test $\phi_2$, or the test statistic $V_{n,k,a,b}(\xi)$ for the linear functional-based test $\phi_1$, for the entries outside the diagonal blocks. 
Since the entries outside the diagonal blocks are all equal to zero across the $k$ networks,
the 5\% critical value can be calculated as the 95th percentile of the pooled test statistics for all such null entries.

It is worth pointing out that the ``Empirical" critical value mentioned above relies on the knowledge of true nulls and thus can only be calculated in simulation studies. The main purpose of using both methods for determining the critical values is to compare the ``Theoretical" values with the ``Empirical" ones to justify our findings on the null distributions of our tests $\phi_2$ and $\phi_1$ in Theorems \ref{thm:l2 testing} and \ref{thm:l1 testing}, respectively. With these critical values, we can calculate the false positive rate (FPR) and the false negative rate (FNR). Clearly, with the ``Empirical" critical value the FPR should be exactly 5\%, and thus we omit its values and include only the FPR based on the ``Theoretical" critical value in Tables \ref{testing model 1} and \ref{testing model 2}, which present the means and standard errors of testing results in Models I and II, respectively. The FNRs based on both critical values are reported. In fact, we see from Tables \ref{testing model 1} and \ref{testing model 2} that the ``Theoretical" values for both FPR and FNR are very close to the ``Empirical" ones, indicating that the asymptotic null distributions obtained in Theorems \ref{thm:l2 testing} and \ref{thm:l1 testing} indeed match the empirical distributions very closely. 
To better evaluate these methods, we also vary the critical value and generate a full receiver operating characteristic (ROC) curve. The areas under the ROC curves are summarized in 
Tables \ref{testing model 1} and \ref{testing model 2}. It is seen that both methods THP-$\phi _{2}$ and THP-$\phi _{1}$ have areas under the ROC curve close to 1 across all settings.

\begin{table}
\caption{\label{testing model 1}
Means and standard errors (in parentheses) of testing results for THP methods in Model I with $\alpha=0.05$.}
{\centering
\begin{tabular}{|l|lll|ll|l|l|} \hline
Method &      & $k$     & $p$     & \multicolumn{2}{|c|}{FNR ($\times 10^{-2}$)}    & \multicolumn{1}{|l|}{FPR }    & \multicolumn{1}{|l|}{ROC Area}\\ & &       &       &  Empirical & Theoretical &  ($\times 10^{-2}$)     & ($\times 10^{-2}$) \\ \hline &Setting 1   & 5     & 50    & 0.375 (0.484) & 0.369 (0.454) & 5.044 (0.656) & 99.90 (0.078)\\
THP-$\phi_1$ &Setting 2    & 10    & 50    & 0 (0)  & 0 (0)  & 4.945 (0.752) & 1 (0) \\
&Setting 3   & 10    & 200   &0.001 (0.014) & 0.001 (0.014) & 5.005 (0.170) & 1 (0) \\
\hline & Setting 1    & 5     & 50    &3.268 (1.568) & 3.161 (1.422) & 5.123 (0.722) & 99.26 (0.319)\\ 
THP-$\phi_2$ & Setting 2    & 10    & 50    & 0.006 (0.060) & 0.006 (0.060) & 5.352 (0.751) & 1 (0.010) \\
& Setting 3    & 10    & 200   & 0.077 (0.100) & 0.077 (0.098)& 4.896 (0.177)& 99.97 (0.019) \\ 
\hline
\end{tabular}}
\end{table}%

\begin{table}
\caption{\label{testing model 2}
Means and standard errors (in parentheses) of testing results for THP methods in Model II with $\alpha=0.05$.}
{\centering	
\begin{tabular}{|l|lll|ll|l|l|}
\hline
Method &      & $k$     & $p$     & \multicolumn{2}{|c|}{FNR ($\times 10^{0}$)}       & \multicolumn{1}{|l|}{FPR}    & \multicolumn{1}{|l|}{ROC Area}  \\
& &       &       &  Empirical & Theoretical & ($\times 10^{-2}$)      & ($\times 10^{-2}$) \\
\hline
&Setting 1   & 5     & 50    & 0.226 (0.043) & 0.224 (0.038) & 5.151 (0.821)& 94.54 (1.346) \\
THP-$\phi_1$&Setting 2    & 10    & 50    & 0.327 (0.041) & 0.327 (0.038) & 5.046 (0.932) & 90.26 (2.07) \\
&Setting 3    & 10    & 200   &0.306 (0.017) & 0.305 (0.016) & 5.04 (0.233) & 91.12 (0.771) \\
\hline
&Setting 1    & 5     & 50    & 0.066 (0.019) & 0.064 (0.017)& 5.125 (0.747) & 98.42 (0.520)\\
THP-$\phi_2$&Setting 2    & 10    & 50    & 0.099 (0.021) & 0.094 (0.020) & 5.416 (0.750) & 97.66 (0.560)  \\
&Setting 3    & 10    & 200   &   0.090 (0.010) & 0.090 (0.010) & 5.017 (0.149) & 97.79 (0.302)\\
\hline
\end{tabular}}
\end{table}%
		
		
In particular, we see from Table \ref{testing model 1} that the linear functional-based test $\phi_1$ is significantly better than the chi-based test $\phi_2$ over all settings of Model I. From setting 1 to setting 2, both testing procedures become better, while both procedures perform worse from setting 2 to setting 3. These are consistent with our theoretical results. To understand this, let us take the entry $(1,2)$ as an example. In view of Theorem \ref{thm:lower rate}, the separating rate for alternative $H_{1,12}^{l1}(s,\epsilon,\xi)$ with the corresponding optimal test $\phi_1$ is $\|\omega^{0}_{1,2}\|_1 \geq \epsilon_n \asymp \sqrt{k/n^{(0)}}$. Since the components of the joint link strength vector $\omega^{0}_{1,2}$ are i.i.d. from the uniform distribution $U[0.2,0.4]$, as the number of networks  $k$ increases the separating rate condition becomes weaker because  $\|\omega^{0}_{1,2}\|_1$ grows linearly with $k$, while the right-hand side $\epsilon_n \asymp \sqrt{k/n^{(0)}}$ grows at a slower rate of $\sqrt{k}$. Thus the growth of $k$ makes the separating rate condition easier to be satisfied.
The results for the chi-based test $\phi_2$ can be understood similarly.
		
Comparing Table \ref{testing model 2} with Table \ref{testing model 1}, we see that the performance of both testing procedures $\phi_2$ and $\phi_1$ becomes worse. This is reasonable since Model II is sparser than Model I and thus the separating rate conditions indicated in Theorem \ref{thm:lower rate} are harder to be satisfied for these sparser entries with only one nonzero component across $k$ networks, 
because this nonzero entry needs to have magnitude much larger than $\epsilon_n \asymp \sqrt{k/n^{(0)}}$ for test $\phi_1$ or $\epsilon_n \asymp \sqrt{k^{1/2}/n^{(0)}}$ for test $\phi_2$. As a consequence, different from Table \ref{testing model 1} in which the separating rate conditions become easier for denser entries with all $k$ nonzero components as $k$ increases, these conditions become more stringent for sparser entries with only one nonzero component as $k$ increases. Such increased difficulty for sparser entries is more severe for the linear functional-based test $\phi_1$ than for the chi-based test $\phi_2$ in light of the separating rates $\epsilon_n$ in Theorem \ref{thm:lower rate}.
		
		%
		
\subsubsection{Precision matrix estimation}\label{subsec: comparison}

As mentioned before, almost all existing methods on multiple graphs focus on the estimation part. To compare with these existing methods, we modify our THP procedure to generate sparse estimates of the precision matrices. More specifically, we suggest a two-step procedure. In the first step, for each entry $(a,b)$ with $a\neq b$, we conduct hypothesis testing at significance level $\alpha$ to see whether the null hypothesis $H_{0,ab}$ in (\ref{eq:null}) is rejected or not. The critical values at significance level $\alpha$ are calculated using the asymptotic distributions established in Theorems \ref{thm:l2 testing} and \ref{thm:l1 testing}. In the second step, for each $1 \leq a \leq p$ we estimate the $(a,a)$th entry of the $t$th graph as $\hat{\omega}_{a,a}^{(t)}$, and for each rejected null hypothesis $H_{0,ab}$ we estimate the $(a,b)$th entry of the $t$th graph as $-\hat{\omega}_{a,a}^{(t)}\hat{\omega}_{b,b}^{(t)}T^{(t)}_{n,k,a,b}$  in view of (\ref{eq:J(t)}), where all the notation is the same as in Section \ref{sec:test.l1}.

In our two-step procedure suggested above, there is one tuning parameter which is the significance level $\alpha$. To tune such parameter, we generate an independent validation set with the same sample sizes $n^{(t)}=n^{(0)} = 100$ for Model I and $200$  
for Model II with $1 \leq t \leq k$. 
Then for each given value of $\alpha$, we obtain a set of sparse precision matrix estimates $\hat{\Omega}^{0} = (\hat{\Omega}^{(1)},\cdots, \hat{\Omega}^{(k)})$ for the $k$ graphs using the training data, and calculate the value of the loss function
\begin{equation}\label{eq:loss}
L(\hat{\Omega}^{0}) = \sum_{t=1}^k\left\{\log[\det(\hat\Omega^{(t)})]-\tr(\hat\Sigma^{(t)}\hat\Omega^{(t)})\right\},
\end{equation}
where $\hat{\Sigma}^{(1)},\cdots, \hat{\Sigma}^{(k)}$ are the sample covariance matrix estimators for the $k$ graphs constructed based on the validation data. The parameter $\alpha$ is then chosen by minimizing the loss function in (\ref{eq:loss}) over a grid of 10 values for $\alpha$.
We compare our THP approach with three commonly used competitor methods MPE, GGL, and FGL, each with one regularization parameter to tune. For a fair comparison, for each method we use the same validation set to tune the regularization parameter and choose the one minimizing the loss function in \eqref{eq:loss} over a grid of 10 values.

\begin{table}
	\caption{\label{estimation 1}
Means and standard errors (in parentheses) of precision matrix estimation results for different methods in Model I.}
	{\centering
		\begin{tabular}{|ccc|c|ccc|}
			\hline
& $k$     & $p$     & Method & $\ell_1$    & $\ell_2$    & $\ell_F$ \\
			\hline
			Setting 1    & 5     & 50    & THP-$\phi_1$    &4.968 (0.041) & 3.417 (0.036) & 6.657 (0.036)  \\
			&              &       & THP-$\phi_2$    & 5.68 (0.070) & 3.894 (0.081) & 7.578 (0.131) \\
			&             &       & MPE   & 7.556 (0.024) & 6.347 (0.056) & 11.53 (0.083) \\
			&            &       & GGL   & 8.331 (0.009) & 7.289 (0.005) & 13.05 (0.005) \\
			&             &       & FGL   & 7.989 (0.046) & 7.247 (0.044) & 13.13 (0.069) \\
			\hline
			Setting 2    & 10    & 50    & THP-$\phi_1$    &5.117 (0.102) & 3.281 (0.103) & 6.416 (0.194) \\
			&              &       & THP-$\phi_2$    & 5.191 (0.104) & 3.333 (0.108) & 6.542 (0.202) \\
			&              &       & MPE   & 7.075 (0.022) & 5.618 (0.048) & 10.44 (0.070) \\
			&             &       & GGL   & 8.193 (0.006) & 7.241 (0.005) & 12.98 (0.010) \\
			&              &       & FGL   & 8.132 (0.003) & 7.461 (0.003) & 13.36 (0.004) \\
			\hline
			Setting 3   & 10    & 200   & THP-$\phi_1$    &  5.84 (0.096) & 3.997 (0.116) & 14.3 (0.474) \\
			&           &       & THP-$\phi_2$    &6.466 (0.111) & 4.674 (0.142) & 16.79 (0.594) \\
			&&& MPE & -- & -- & --\\
			&           &       & GGL   & 8.467 (0.006) & 7.489 (0.003) & 27.01 (0.003) \\
			&&&FGL & -- & -- & --\\
			\hline
		\end{tabular}}
	\end{table}%

	\begin{table}
		\caption{\label{estimation 2}
Means and standard errors (in parentheses) of precision matrix estimation results for different methods in Model II.}
		{\centering
			\begin{tabular}{|ccc|cccc|}
				\hline
 & $k$     & $p$     & Method & $\ell_1$    & $\ell_2$    & $\ell_F$\\
				\hline
				Setting 1    & 5     & 50    & THP-$\phi_1$    &  3.651 (0.035) & 2.091 (0.018) & 4.723 (0.023) \\
				&       &       & THP-$\phi_2$    & 3.368 (0.045) & 2.042 (0.023) & 4.392 (0.043) \\
				&       &       & MPE   &  4.909 (0.020) & 3.289 (0.015) & 6.668 (0.018)  \\
				&       &       & GGL   & 7.087 (0.009) & 5.155 (0.004) & 9.653 (0.005) \\
				&       &       & FGL   & 6.748 (0.007) & 4.942 (0.004) & 9.563 (0.006) \\
				\hline
				Setting 2   & 10    & 50    & THP-$\phi_1$    & 3.095 (0.018) & 1.898 (0.009) & 4.213 (0.011)  \\
				&              &       & THP-$\phi_2$    & 3.019 (0.020) & 1.878 (0.011) & 4.099 (0.013) \\
				&              &       & MPE   & 3.613 (0.013) & 2.264 (0.010) & 4.325 (0.014)  \\
				&              &       & GGL   & 5.708 (0.006) & 4.325 (0.003) & 8.238 (0.004)  \\
				&             &       & FGL   & 5.606 (0.005) & 4.27 (0.003) & 8.228 (0.004)  \\
				\hline
				Setting 3   & 10    & 200   & THP-$\phi_1$    &6.035 (0.077) & 2.7 (0.018) & 11.18 (0.078) \\
				&              &       & THP-$\phi_2$    & 5.595 (0.085) & 3.448 (0.061) & 15.19 (0.306)\\
				&&& MPE & -- & -- & --\\
				&              &       & GGL   & 6.976 (0.005) & 5.195 (0.004) & 18.23 (0.004) \\
				&&&FGL & -- & -- & --\\
				\hline
			\end{tabular}}%
		\end{table}%

To evaluate the performance of different methods, we calculate three loss functions of the matrix $1$-norm, the spectral norm, and the Frobenius norm for the estimation errors, which are denoted as $\ell_1$, $\ell_2$, and $\ell_F$, respectively. The precision matrix estimation results for different methods in Models I and II are summarized in Tables \ref{estimation 1} and \ref{estimation 2}, respectively.
In particular, for setting 3 of both models the results of MPE and FGL are not reported because the results cannot be obtained within a reasonable amount of time due to their excessively high computational costs.
To gain some insights into the computational costs of various methods, we record in Table \ref{cost} the average computational cost measured as the CPU time in seconds for each method. 
Since the computational cost of THP-$\phi_1$ is almost identical to that of THP-$\phi_2$, only the results for the latter 
are reported.

We see from Table \ref{estimation 1} that across all three settings, both methods THP-$\phi_2$ and THP-$\phi_1$ outperform the MPE, FGL, and GGL significantly. Similar phenomenon can be observed from Table \ref{estimation 2}.  In light of the computational cost presented in Table \ref{cost}, our methods are much faster than MPE and FGL over all the settings. Thus the overall performance of our methods is superior to that of all three competing methods. Observe that setting 1 differs from setting 2 only in the number of networks 
$k$. Therefore, it is fair to conclude that compared to other approaches, our methods have greater advantages in estimating a large number of graphs
simultaneously, which is in line with our theoretical findings that our methods allow the number of networks $k$ to diverge with the sample size $n^{(0)}$ 
at a faster rate.

\begin{table}
\caption{\label{cost} Average computational costs 
 of different methods in seconds. 
}
{\centering	
\begin{tabular}{|l|rrrr|rrrr|rrrr|}
\hline
&  \multicolumn{4}{|c|}{Setting 1 ($\times 10^0$)} &\multicolumn{4}{|c|}{Setting 2 ($\times 10^{1}$)}&\multicolumn{4}{|c|}{Setting 3 ($\times 10^{2}$)}\\
\hline
& THP    & MPE   & GGL   & FGL   & THP    & MPE   & GGL   & FGL   & THP    & MPE   & GGL   & FGL \\
\hline
Model I & 7.2  & 57.7 & 9.2   & 64.8 & 2.1 & 8.7 & 2.6  & 13.5 & 3.9 & 36.7 & 3.7 & 18.2 \\
Model II & 18.1 & 69.8 & 18.2  & 44.4 & 3.0 & 10.0 & 3.5  & 28.7 & 6.8 & 38.6 & 5.9 & 23.1 \\
\hline
\end{tabular}}
\end{table}%

\subsubsection{Heavy-tailed distributions} \label{subsec: heavy-tailed}
Model misspecification \citep{CSS10} can often occur in 
applications. Thus it is important to examine the robustness of proposed methods. With this in mind, we now investigate the finite-sample performance of our THP procedure in the presence of heavy-tailed distributions such as the Laplace distribution, as opposed to the Gaussianity assumed in our theoretical developments.
For each previous setting in Models I and II, after generating the precision matrix $\Omega^{(t)}$, instead of sampling the data matrix $\bX^{(t)}$ from the 
Gaussian distribution with mean zero and covariance matrix $(\Omega^{(t)})^{-1}$ we draw $\bX^{(t)}$ from the multivariate Laplace distribution with covariance matrix $(\Omega^{(t)})^{-1}$. More specifically, we first generate a random vector whose components are i.i.d. Laplace random variables with location parameter zero and scale parameter $1/\sqrt{2}$, and then multiply this vector by $(\Omega^{(t)})^{-1/2}$ to obtain the desired Laplace random vector. All the rest of the settings are the same as before.
					
Table \ref{testing model 3} presents the testing results of our methods THP-$\phi_2$ and THP-$\phi_1$ in the setting of heavy-tailedness. Compared to the results in Tables \ref{testing model 1} and \ref{testing model 2}, we observe that across all settings of Models I and II, the performance of our methods stays almost the same when the Gaussian distribution is replaced by the Laplace distribution, demonstrating the robustness of our methods to the heavy-tailed distributions. 
We have also explored other heavy-tailed distributions such as the $t$-distribution with 5 degrees of freedom and the results are very similar. To save the space, these additional results are not presented here but are available upon request.	
			
\begin{table}
	\caption{\label{testing model 3}
Means and standard errors (in parentheses) of testing results for THP methods in Models I and II with the Laplace distribution and $\alpha=0.05$.}
	{\centering
		\begin{tabular}{|c|ccc|cc|c|c|}
			\hline
			\multicolumn{8}{|c|}{Model I} \\
			\hline
Method			&    & $k$     & $p$     & \multicolumn{2}{c|}{FNR($\times 10^{-2}$)}       & FPR    & ROC Area \\
			& &       &       &  Empirical & Theoretical &     ($\times 10^{-2}$)  &($\times 10^{-2}$)  \\
			\hline
			& Setting 1 & 5     & 50    & 0.345 (0.480)& 0.357 (0.440) & 4.986 (0.723) & 99.91 (0.068) \\
			THP-$\phi_1$& Setting 2 & 10    & 50    & 0 (0)  & 0 (0) & 5.089 (0.991)& 100 (0) \\
			&Setting 3 & 10    & 200   & 0 (0)  & 0 (0)  & 5.03 (0.172)& 100 (0) \\
			\hline
			& Setting 1 & 5     & 50    &3.012 (1.555) & 2.810 (1.438) & 5.293 (0.669)& 99.32 (0.287)\\
			THP-$\phi_2$& Setting 2 & 10    & 50    & 0 (0)  & 0 (0) & 5.701 (0.824) & 100 (0.004) \\
			& Setting 3 & 10    & 200   &  0.066 (0.094)& 0.063 (0.094) & 5.073 (0.171) & 99.98 (0.016)\\
			\hline
			\multicolumn{8}{|c|}{Model II} \\
			\hline
Method			&    & $k$     & $p$     & \multicolumn{2}{c|}{FNR ($\times 10^{0}$)}        & FPR    & ROC Area \\
			&&       &       &  Empirical & Theoretical &     ($\times 10^{-2}$)  &($\times 10^{-2}$)  \\
			\hline
			&Setting 1 & 5     & 50    &  0.226 (3.594) & 0.226 (3.414) & 5.046 (0.973)& 94.49 (1.311)\\
			THP-$\phi_1$&Setting 2 & 10    & 50    &0.317 (3.765) & 0.319 (3.497) & 5.011 (0.908)& 90.88 (1.806)\\
			&Setting 3 & 10    & 200   &0.309 (1.574) & 0.308 (1.567) & 5.048 (0.219) & 91.03 (0.766) \\
			\hline
			&Setting 1 & 5     & 50    &0.069 (0.020) & 0.066 (0.019) & 5.388 (0.854) & 98.43 (0.512)\\
			THP-$\phi_2$&Setting 2 & 10    & 50    & 0.093 (0.020) & 0.090 (0.019)& 5.375 (0.725) & 97.66 (0.629)\\
			&Setting 3 & 10    & 200   & 0.089 (0.010) & 0.088 (0.010)& 5.083 (0.177) & 97.83 (0.320)\\
			\hline
		\end{tabular}}
	\end{table}%

\subsection{Real data analysis} \label{Sec4.2}
In addition to the simulation examples, we also demonstrate the performance of our suggested methods THP-$\phi_2$ and THP-$\phi_1$ on a real data example  of the epithelial ovarian cancer. As introduced in
\cite{Tothill2008data}, the ovarian cancer has six molecular subtypes, which are referred to as C1 through C6 following the notation in \cite{Tothill2008data}.
They discovered that there is a significant difference in expression levels of genes associated with stromal
and immune cell types between C1 and other subtypes. It was also discovered that C1 patients suffer from a lower survival rate. We consider the RNA expression data measured on $n^{(1)}= 78$ patients from C1 subtype and $n^{(2)} = 113$ patients from all other subtypes combined. 
The number of genes in this study 
is $p=87$. Our goal is to recover the networks 
of genes related to the apoptosis pathway from the KEGG database \citep{Orgata1999,Kanehisa2012}  for disease subtype C1 and other subtypes combined such that we can identify which genes are crucial in both disease subtype C1 and all other  subtypes combined. Thus the number of graphs in our setting 
is $k=2$.
						
\begin{figure}
\centering
\begin{minipage}{0.4\textwidth}
\includegraphics[width=\textwidth]{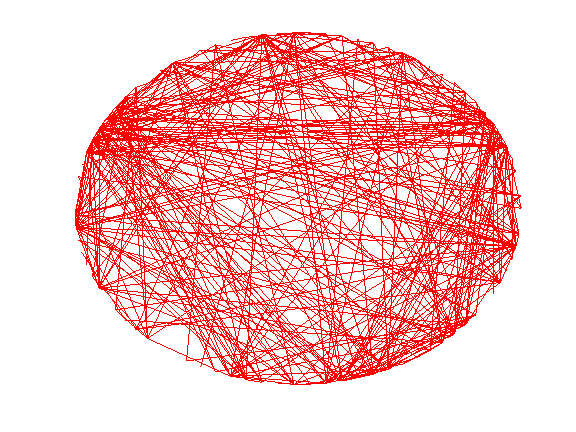}
\end{minipage}
\hfill
\begin{minipage}{0.4\textwidth}
\includegraphics[width=\textwidth]{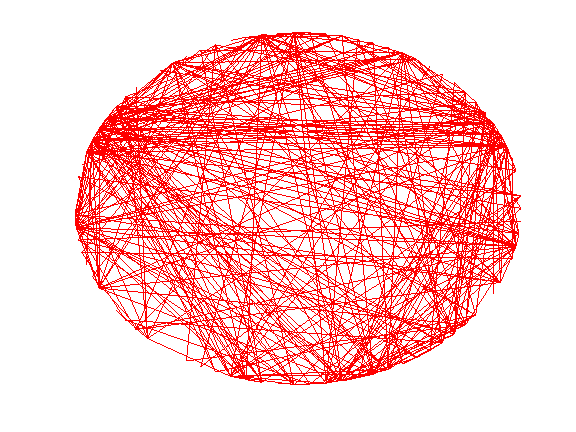}
\end{minipage}
\caption{\label{realdataplot}Common edges between C1 and other types identified by methods THP-$\phi_1$ (left panel) and THP-$\phi_2$ (right panel)}
\end{figure}

We apply our proposed methods to this 
data set with significance level $\alpha=0.001$. For each entry $(a,b)$ with $a\neq b$, if the corresponding null hypothesis $H_{0,ab}$ in (\ref{eq:null}) is rejected then  we posit that there is an edge connecting node $a$ and node $b$ in at least one of the two graphs. Figure \ref{realdataplot} presents the connectivity structures identified by methods THP-$\phi_2$ and THP-$\phi_1$. We further would like to find out which nodes are crucial in defining the connectivity structures identified in Figure \ref{realdataplot}. Motivated by the definition of central nodes introduced in \cite{mpe}, we define important nodes as the ones with the largest degrees in the graphs depicted in Figure \ref{realdataplot}. Table \ref{tab: nodes} lists the top 10 nodes with the highest degrees identified by methods THP-$\phi_2$ and THP-$\phi_1$. Since two graphs are considered, there are two possible sign vectors $(1,1)'$ and $(1,-1)'$ up to a single sign for our linear functional-based test $\phi_1$. Without the knowledge of the sign vector, we test both relationships, that is, the sum and the subtraction, and conduct the corresponding two-sided tests. The results for the subtraction are, however, not convincing since the corresponding graph is too sparse, where the largest degree among all nodes is 4, the second largest degree is 2, and all the other degrees are less or equal to 1. Thus we present only the results for the sum.

Let us gain some insights into the genes revealed in Table \ref{tab: nodes}.  Among these genes, 1L1B, MYD88, NFKB1, and PIK3R5 have been identified as key genes and been
implicated in the ovarian cancer risk or progression \citep{mpe,giudice2013determinants}. Moreover, BIRC3 and FAS have been proved to function importantly in ovarian cancer. In particular, it has been discovered that upregulation of FAS reverses the development of resistance to Cisplatin in epithelial ovarian cancer \citep{yang2015upregulation,jonsson2014distinct}, which demonstrates the importance of the nodes identified by our methods in ovarian cancer.
						
\begin{table}
\caption{\label{tab: nodes} Top 10 nodes with highest degrees identified by THP methods 
in descending order.}
\centering
\begin{tabular}{|c|l|}
\hline
Method & Node\\
\hline
THP-$\phi_1$    & MYD88,      NFKB1,    CSF2RB,   PIK3R5,       FAS, PIK3CG,    TRADD,    BIRC3,  \\
&  IL1B,  NFKBIA \\
THP-$\phi_2$    &  NFKB1,    MYD88,    CSF2RB,   PIK3R5,    BIRC3, PIK3CG,      FAS,    IL1B,\\
&   CAPN1,  NFKBIA  \\
\hline
\end{tabular}
\end{table}%

\section{Discussions} \label{sec:discussion}
In this paper we have introduced the tuning-free heterogeneity pursuit (THP) framework with the chi-based test and the linear functional-based test to detect the heterogeneity in sparsity patterns of multiple networks in the setting of Gaussian graphical models. Such a framework is not only scalable to large scales, but also enjoys optimality properties in the scenario where the number of networks is allowed to diverge and the number of features can be much larger than the sample size. Our theoretical justifications show that under mild regularity conditions, the linear functional-based test has the minimum requirement on the sample size.

Yet the optimality of the sample size requirement for the chi-based test, that is, the minimum sample size requirement with the optimal separating rate $\epsilon _{n}=\sqrt{k^{1/2}/n^{(0)}}$ for testing null $H_{0,ab}$ against alternative $H^{l2}_{1,ab}$, still remains as an open problem for future invesigation.  
The main challenges lie in the need of constructing a new lower bound as in Theorem \ref{thm:lower sample size} for alternative $H^{l1}_{1,ab}$, which involves both the sample size requirement and the separating rate. Moreover, the technical analysis in the proof of Theorem \ref{thm:l2 testing} contains a relatively loose bound between the $\ell_1$ and $\ell_2$ norms, which implies that the sample size requirement imposed in Proposition \ref{cor:l2 testing}  
may not be sharp, though sharper than that for the naive combination testing procedure discussed in Section \ref{Sec2.3}.

As mentioned in the Introduction, our paper has focused only
on two particular aspects of heterogeneity which are the heterogeneity in sparsity patterns over multiple networks 
and the heterogeneity in noise levels over multiple subpopulations. The appealing features of our THP framework for addressing these 
issues are empowered by our newly suggested convex approach of heterogeneous group square-root Lasso (HGSL) for the setting of  high-dimensional multi-response regression with heterogeneous noises.
Other aspects of heterogeneous learning and inference can certainly be interesting as well. For example, in practice one might be interested in studying whether the entries across different graphs are identical or not, that is, the heterogeneity in link strengths. This is a more general yet more challenging problem that deserves further study. Some efforts along this direction have been made in the literature. For instance, \cite{jgl} proposed a penalized likelihood method using the fused Lasso to estimate the common link strength among multiple Gaussian graphs. This method, however, focuses only on the estimation of common link strength and lacks theoretical justification for its performance. Moreover, their proposed algorithm is not scalable due to the complicated form of the likelihood function. Thus it would be interesting to extend the methods developed in our paper to the problem of testing for   heterogeneity in link strengths.

Our studies are only among the first attempts to address the challenging issues of heterogeneity in multiple networks in the setting of 
Gaussian graphical models. It would be interesting to extend our inferential approach to the settings of  
multiple matrix graphical models, multiple tensor graphical models, and multiple non-Gaussian graphical models, as well as other network models beyond graphical models. Furthermore, the false discovery rate (FDR) control \citep{BH95, BC15} is often an important issue in practice. It would also be interesting to further extend the THP framework to provide tools that can control the FDR in multiple networks effectively. In some applications, it is possible that a fraction of the class labels for the subpopulations or even all the class labels can be unavailable, in which clustering techniques can play a crucial role. In addition, there can exist some latent features which would require a broader class of network structures. 
The developments on heterogeneity identification in multiple networks can also motivate new approaches for regression and classification problems that have networks as an input. The possible extensions addressing these issues are beyond the scope of the current paper and will be interesting topics for future research.

%
%
%
%
%

\bibliographystyle{rss}
\bibliography{MGGM}


\newpage
\quad \vspace{0.05in}

\begin{center}
\textbf{\Large Supplementary material to ``
Tuning-free heterogeneity pursuit in massive networks''}

\bigskip
\author{Zhao Ren$^1$, Yongjian Kang$^2$, Yingying Fan$^2$ and Jinchi Lv$^2$}

\smallskip
\textit{University of Pittsburgh$^1$ and University of Southern California$^2$}
\end{center}

\medskip

\noindent This Supplementary Material contains the proofs of Theorems \ref{thm:l2 testing}--\ref{com theorem} and Propositions \ref{cor:l2 testing}--\ref{prop:suppreco} in Section \ref{sec:proof}, as well as the proofs of key lemmas and additional technical details in Sections \ref{SecB} and \ref{SecC}, respectively.

\smallskip

\appendix
\setcounter{page}{1}
\setcounter{section}{0}
\renewcommand{\theequation}{A.\arabic{equation}}
\setcounter{equation}{0}

\section{Proofs of main results} \label{sec:proof}

\subsection{Proofs of Theorem \ref{thm:l2 testing} and Proposition \ref{cor:l2 testing}} \label{SecA.1}

The proofs of Theorems \ref{thm:l2 testing}--\ref{thm:l1 testing} and Propositions \ref{cor:l2 testing}--\ref{cor:l1 testing} rely on two key sets of results in Lemmas \ref{lem:diagnoal} and \ref{lem:lin comb} in Sections \ref{SecB.1} and \ref{SecB.2}, respectively, where we use the compact notation $[\ell]$ to denote the set $\{1, \cdots, \ell\}$ for any positive integer $\ell$ whenever there is no confusion. Our results are important consequences of Lemmas \ref{lem:diagnoal} and \ref{lem:lin comb}. Indeed, it holds that
\begin{equation*}
\sum_{t=1}^{k}\left\vert \sqrt{n^{(t)}\hat{\omega}_{2,2}^{(t)}\hat{\omega}%
_{1,1}^{(t)}}\left( T_{n,k,1,2}^{(t)}-J_{n,k,1,2}^{(t)}\right)
-V_{n,k,1,2}^{\ast (t)}\right\vert \leq T_{1}+T_{2},
\end{equation*}%
where
\begin{eqnarray*}
T_{1} &=&\sum_{t=1}^{k}\sqrt{n^{(t)}\hat{\omega}_{2,2}^{(t)}\hat{\omega}%
_{1,1}^{(t)}}\left\vert T_{n,k,1,2}^{(t)}-J_{n,k,1,2}^{(t)}-\frac{1}{n^{(t)}}%
\sum_{i=1}^{n^{(t)}}\left( E_{i,1}^{(t)}E_{i,2}^{(t)}-\mathbb{E}%
E_{i,1}^{(t)}E_{i,2}^{(t)}\right) \right\vert , \\
T_{2} &=&\sum_{t=1}^{k}\left\vert 1-\sqrt{\frac{\hat{\omega}_{2,2}^{(t)}\hat{%
\omega}_{1,1}^{(t)}}{\omega _{2,2}^{(t)}\tilde{\omega}_{1,1}^{(t)}}}\right\vert
\left\vert \sqrt{\frac{\omega _{2,2}^{(t)}\tilde{\omega}_{1,1}^{(t)}}{n^{(t)}}}%
\sum_{i=1}^{n^{(t)}}\left( E_{i,1}^{(t)}E_{i,2}^{(t)}-\mathbb{E}%
E_{i,1}^{(t)}E_{i,2}^{(t)}\right) \right\vert .
\end{eqnarray*}

According to Lemma \ref{lem:diagnoal}, we have $\vert \hat{\omega}%
_{j,j}^{(t)}-\omega _{j,j}^{(t)}\vert \leq C^{\prime }(\sqrt{\frac{\log
(k/\delta _{1})}{n^{(0)}}}+s\frac{\left( k+\log p\right) }{n^{(0)}})=o(1)$
with probability at least $1-6p^{1-\delta }-2\delta _{1}$ uniformly for all $%
t\in [k]$ and $j=1,2$. Therefore, Condition \ref{CondA1} implies that all $\hat{\omega}%
_{j,j}^{(t)}$ are bounded from both below and above, which together with Lemma \ref%
{lem:lin comb} and $s\left( k+\log p\right) /n^{(0)}=o(1)$ leads to
\begin{equation*}
T_{1}\leq C\left( s\frac{k+(\log p)}{\sqrt{n^{(0)}}}\right)
\end{equation*}%
with probability at least $1-12p^{1-\delta }-2\delta _{1}$, where positive constant $C$
depends on constants $M,M_{0},\delta ,C_{1},\\C_{2}$, and $C_{3}$.

It remains to upper bound term $T_{2}$. Note that Lemma \ref{lem:diagnoal} together with Condition \ref{CondA1} implies that $\tilde{\omega}_{1,1}^{(t)}$ is bounded. In addition, Condition \ref{CondA1} also implies
that $E_{i,1}^{(t)}E_{i,2}^{(t)}$, $i\in [n^{(t)}]$ are i.i.d. sub-exponential with bounded constant parameter. Consequently, Bernstein's inequality (see, e.g., Proposition 5.16,
\cite{vershynin2010introduction}) entails immediately that $\max_{k}\vert
V_{n,k,1,2}^{\ast (t)}\vert <\sqrt{C'\log (k/\delta _{1})}$ with
probability at least $1-2\delta _{1}$, where positive constant $C'$ depends on $M$ only. Therefore, this fact and Lemma \ref%
{lem:diagnoal} along with the union bound further yield with probability at least $%
1-6p^{1-\delta }-4\delta _{1}$ that
\begin{eqnarray*}
T_{2} &\leq &\sqrt{C'\log (k/\delta _{1})}\sum_{t=1}^{k}\left\vert 1-\sqrt{%
\frac{\hat{\omega}_{2,2}^{(t)}\hat{\omega}_{1,1}^{(t)}}{\omega _{2,2}^{(t)}%
\tilde{\omega}_{1,1}^{(t)}}}\right\vert  \\
&\leq &C\sqrt{\log (k/\delta _{1})}\left( \sum_{t=1}^{k}\left\vert \tilde{%
\omega}_{1,1}^{(t)}-\hat{\omega}_{1,1}^{(t)}\right\vert
+\sum_{t=1}^{k}\left\vert \omega _{2,2}^{(t)}-\hat{\omega}_{2,2}^{(t)}\right%
\vert \right)  \\
&\leq &C\left( k\sqrt{\frac{\log (k/\delta _{1})}{n^{(0)}}}+s\frac{\left(
k+(\log p)\right) }{n^{(0)}}\right) \sqrt{\log (k/\delta _{1})} \\
&\leq &C(s\frac{k+(\log p)}{\sqrt{n^{(0)}}}),
\end{eqnarray*}%
where the second inequality follows from the fact that all $\hat{\omega}_{j,j}^{(t)},$ $%
\tilde{\omega}_{j,j}^{(t)}$, and $\omega _{j,j}^{(t)}$ are bounded from both below and
above, the third inequality is due to Lemma \ref{lem:diagnoal}, and the last
inequality follows from our sample size assumptions $\log (k/\delta
_{1})=O(s(1+(\log p)/k))$ as well as $\log (k/\delta _{1})=o(n^{(0)})$. The positive
constant $C$ above depends on constants $M,\delta ,C_{1},C_{2}$, and $C_{3}$.

Combining the bounds of $T_{1}$ and $T_{2}$ above, we deduce that the following inequality holds with probability at least $%
1-12p^{1-\delta }-4\delta _{1}$,
\begin{equation}
\sum_{t=1}^{k}\left\vert \sqrt{n^{(t)}\hat{\omega}_{2,2}^{(t)}%
\hat{\omega}_{1,1}^{(t)}}\left( T_{n,k,1,2}^{(t)}-J_{n,k,1,2}^{(t)}\right)
-V_{n,k,1,2}^{\ast (t)}\right\vert \leq C\left( s\frac{k+\log p}{\sqrt{n^{(0)}}}\right),
\label{eq:thm l1 abs}
\end{equation}%
where constant $C>0$ depends only on $M,M_{0},\delta ,C_{1},C_{2}$, and $C_{3}$.

Aided with the key result in (\ref{eq:thm l1 abs}) above, the analysis of Theorem \ref{thm:l2 testing} is straightforward. Indeed we have
\begin{eqnarray*}
&&\left\vert \left( \sum_{t=1}^{k}n^{(t)}\hat{\omega}_{2,2}^{(t)}\hat{\omega}%
_{1,1}^{(t)}\left( T_{n,k,1,2}^{(t)}-J_{n,k,1,2}^{(t)}\right) ^{2}\right)
^{1/2}-U_{n,k,1,2}^{\ast }\right\vert  \\
&\leq &\left[ \sum_{t=1}^{k}\left( \sqrt{n^{(t)}\hat{\omega}_{2,2}^{(t)}\hat{%
\omega}_{1,1}^{(t)}}\left( T_{n,k,1,2}^{(t)}-J_{n,k,1,2}^{(t)}\right)
-V_{n,k,1,2}^{\ast (t)}\right) ^{2}\right] ^{1/2} \\
&\leq &\sum_{t=1}^{k}\left\vert \sqrt{n^{(t)}\hat{\omega}_{2,2}^{(t)}\hat{%
\omega}_{1,1}^{(t)}}\left( T_{n,k,1,2}^{(t)}-J_{n,k,1,2}^{(t)}\right)
-V_{n,k,1,2}^{\ast (t)}\right\vert  \\
&\leq &Cs\frac{k+(\log p)}{\sqrt{n^{(0)}}},
\end{eqnarray*}%
where the last inequality is due to  (\ref{eq:thm l1 abs}). The remaining part of the proof for Theorem \ref{thm:l2 testing} follows easily.

Note that the chi distribution $U_{n,k,1,2}^{\ast }$ always has constant level standard deviation. Hence Proposition \ref{cor:l2 testing} follows from the fact
that the error bound of $\vert U_{n,k,1,2}-U_{n,k,1,2}^{\ast
}\vert $ is $o(1)$ with significant probability under the sample size assumption, which completes the proofs.

\subsection{Proofs of Theorem \ref{thm:l1 testing} and Proposition \ref{cor:l1 testing}} \label{SecA.2}
Theorem \ref{thm:l1 testing} is an immediate consequence of  (\ref{eq:thm l1 abs}) established in Section \ref{SecA.1}, since the left-hand side of (\ref{eq:thm l1 abs}) is an upper bound of the left-hand side of  (\ref{eq:thm l1}) regardless of what sign vector is picked.

Note that $V_{n,k,1,2}^{\ast }(\xi)$ follows distribution $N(0,k)$. The error bound of $\vert V_{n,k,1,2}(\xi)-V_{n,k,1,2}^{\ast
}(\xi)\vert $ is negligible compared to the standard deviation of $V_{n,k,1,2}^{\ast }(\xi)$ with significant probability under the sample size
assumption, that is, $s(k+(\log p))/\sqrt{n^{(0)}}=o(k^{1/2})$, which concludes the proofs of both Theorem \ref{thm:l1 testing} and Proposition \ref{cor:l1 testing}.

\subsection{Proof of Theorem \ref{thm:lower rate}} \label{SecA.3}

The first part of the analysis serves as a general tool for both the lower bound arguments in Theorem %
\ref{thm:lower rate} and the proof of
Theorem \ref{thm:lower sample size}. It suffices to assume without loss of generality that the sample sizes of all $k$ graphs are identical,
that is, $n^{(1)}=\cdots =n^{(k)}=n^{(0)}$, noting that Condition \ref{CondA2} is
valid under this setting. Consider a least favorable finite subset $\mathcal{G}=\{\Omega
_{1}^{0},\cdots ,\Omega _{m}^{0}\}\subset \mathcal{A}$ in the alternative
sets, where $\mathcal{A=A%
}^{l2}(s,c'\sqrt{k^{1/2}/n^{(0)}})$ for Theorem \ref{thm:lower rate} (1),  $\mathcal{A=A}^{l1}(s,c'\sqrt{k/n^{(0)}},\xi)$ for Theorem \ref%
{thm:lower rate} (2), and $\mathcal{A=A}^{l1}(s,c\sqrt{k/n^{(0)}},\xi)$ for Theorem \ref{thm:lower sample size}. In
addition, we consider one element in $\Omega _{0}^{0}\in \mathcal{N}(s)$. \
The choice of $\mathcal{G}$ and $\Omega _{0}^{0}$ will be determined later.

Recall that each index denotes each of the $k$ graphs, that is, $\Omega _{h}^{0}=\{\Omega
_{h}^{(t)}\}_{t=1}^{k}$ for $h=0,\cdots ,m$. Let $\mathbb{P}_{h}\equiv\mathbb{P}%
_{\Omega _{h}^{0}}$ denote the joint distribution of the observations when
the true parameter is $\Omega _{h}^{0}$. In other words, $\mathbb{P}_{h}$ is
the joint distribution of $n^{(0)}$ copies of  $k$ graphs $%
\prod\nolimits_{t=1}^{k}g_{h}^{(t)}(x_{t})$, where $g_{h}^{(t)}(\cdot )$ is
the density of $N(0,(\Omega _{h}^{(t)})^{-1})$ for $t\in \lbrack k]$. We use
$\mathbb{E}_{v}$ and $f_{h}$ to denote the expectation under $\mathbb{P}_{v}$
and the density function under $\mathbb{P}_{h}$, respectively. Moreover, let $\mathbb{\bar{P}}=\frac{1%
}{m}\sum_{h=1}^{m}\mathbb{P}_{h}$ be the average measure of these joint
distributions indexed by elements in $\mathcal{G}$. For any test $\psi _{0}$%
, we have
\begin{eqnarray*}
\sup_{v\in \mathcal{G}}\left( \mathbb{E}_{0}\psi _{0}+\mathbb{E}_{v}(1-\psi
_{0})\right) &\geq &\inf_{\psi }\left( \sup_{v\in \mathcal{G}}\mathbb{E}%
_{0}\psi +\mathbb{E}_{v}(1-\psi )\right) \\
&\geq &\inf_{\psi }\left( \mathbb{E}_{0}\psi +\mathbb{\bar{E}}(1-\psi
)\right) \\
&=&\left\Vert \mathbb{P}_{0}\wedge \mathbb{\bar{P}}\right\Vert ,
\end{eqnarray*}%
where $\Vert \mathbb{P}_{0}\wedge \mathbb{\bar{P}}\Vert $ is the
total variation affinity between two measures. Therefore, if $\psi _{0}$ has
significance level $\alpha $ it holds that
\begin{equation}
\inf_{v\in \mathcal{A}}\mathbb{P}_{v}(\psi _{0}\mbox{
\rm  rejects }H_{0,12})\leq \inf_{v\in \mathcal{G}}\mathbb{E}_{v}(\psi
_{0})\leq 1+\alpha -\left\Vert \mathbb{P}_{0}\wedge \mathbb{\bar{P}}%
\right\Vert \mbox{\rm .}  \label{eq:proof lower 1}
\end{equation}

To show that for any given $\beta >\alpha $ and some constant $c>0$, no test
of significance level $\alpha $ satisfies  (\ref{eq:test upper}), it
is sufficient to prove that $\Vert \mathbb{P}_{0}\wedge \mathbb{\bar{P}}%
\Vert >1-(\beta -\alpha )/2$, which together with (\ref%
{eq:proof lower 1}) implies that \[ \inf_{v\in \mathcal{A}}\mathbb{P}_{v}(\psi
_{0}\mbox{ \rm  rejects }H_{0,12})\leq \beta -(\beta -\alpha )/2. \]
We will use this fact in the lower bound arguments in Theorem \ref{thm:lower rate} and the proof of Theorem \ref{thm:lower sample size} with different
constructions of $\mathcal{G}$ and $\Omega _{0}^{0}$, and constant $c>0$.

\subsubsection{Proof of Theorem \ref{thm:lower rate} (1)} \label{SecA.3.1}

To show that $\epsilon _{n}=\sqrt{k^{1/2}/n^{(0)}}$ is the separating rate, we
first establish the lower bound (\ref{eq:test lower}) and then prove that
our test $\phi _{2}$ satisfies  (\ref{eq:test upper}) with $\mathcal{A}=\mathcal{A}^{l2}(s,c\sqrt{k^{1/2}/n^{(0)}})$. With the aid of  (\ref{eq:proof lower 1}), it suffices to show that for
fixed $\beta >\alpha $, there exists some constant $c^{\prime }>0$ such that
$\Vert \mathbb{P}_{0}\wedge \mathbb{\bar{P}}\Vert >1-%
(\beta -\alpha )/2$ with appropriate choices of $\mathcal{G}\subset \mathcal{A=A}%
^{l2}(s,c^{\prime }\sqrt{k^{1/2}/n^{(0)}})$ and $\Omega _{0}^{0}\in \mathcal{%
N}(s)$.

We define
\begin{equation} \label{neweq010}
\Omega _{0}^{0}=\{\Omega _{0}^{(t)}\}_{t=1}^{k} \text{ \ such
that \ } \Omega _{0}^{(1)}=\cdots =\Omega _{0}^{(k)}=I.
\end{equation}
For simplicity,
assume that $\tau \sqrt{k}$ is an integer with some small constant $\tau >0$ to
be determined later. Otherwise, $\tau \sqrt{k}$ can be replaced by its floor
function $\lfloor \tau \sqrt{k}\rfloor $ in the analysis below.
Then we construct a subset
\begin{eqnarray}
\mathcal{G} &=&\Big\{\Omega ^{0}=\{\Omega ^{(t)}\}_{t=1}^{k}:%
\mbox{ there exists some }T\subset [k] \mbox{ with }\left\vert T\right\vert =\tau
\sqrt{k}\mbox{ such that } \notag\\ \label{eq:ConstructionG_l2}
&& \quad \Omega ^{(t)}=I\mbox{ for }t\notin T  \mbox{ and } (\Omega _{0}^{(k)})^{-1} =I+(n^{(0)})^{-1/2}e_{12}\mbox{ for }t\in T\Big\},
\end{eqnarray}%
where $e_{12}$ is the matrix with the $(1,2)$th and $(2,1)$th entries being
one and all other entries being zero. Therefore, there are $\binom{k}{\tau
\sqrt{k}}$ distinct elements in $\mathcal{G}$ and thus $m=\binom{k}{\tau
\sqrt{k}}$. It is easy to check that $\Omega _{0}^{0}\in \mathcal{N}(s)$
and $\mathcal{G\subset A}^{l2}(s,c^{\prime }\sqrt{k^{1/2}/n^{(0)}})$ with $%
c^{\prime }\equiv 2\sqrt{\tau }$, by noting that for each element in $\mathcal{G}$, $%
\Vert \omega _{h,12}^{0}\Vert =\frac{1}{1-1/n^{(0)}}\sqrt{\tau
k^{1/2}/n^{(0)}}$. Hence we omit the details here. Lemma \ref{lem:lower l2} in Section \ref{SecB.3} helps
us finish the proof of the lower bound, that is,  (\ref{eq:test lower}).

It remains to show that the proposed chi-based test $\phi _{2}$ satisfies
 (\ref{eq:test upper}), that is, with a sufficiently large $%
c>0$, $\mathcal{A(}c\mathcal{)=A}^{l2}(s,c%
\sqrt{k^{1/2}/n^{(0)}})$, and $n^{(0)}$, we have
\begin{equation}
\inf_{v\in \mathcal{A(}c\mathcal{)}}\mathbb{P}%
_{v}\left(U_{n,k,1,2}>z_{k}^{l2}(1-\alpha )\right)\geq \beta .
\label{eq:upper l2 alter1}
\end{equation}%
We show this fact in three steps. During the first two steps, we reduce the
goal in (\ref{eq:upper l2 alter1}) to a relatively simple one so that
during the third step we are able to apply Chebyshev's inequality to finish
our proof. Hereafter we use $C>0$ to denote a generic constant. Before
proceeding, note that under the assumptions of Proposition \ref{cor:l2
testing}, including $\delta >1$ and $\delta _{1}=o(1)$, the last inequality
of Lemma \ref{lem:diagnoal} and Condition \ref{CondA1} entail that with probability $%
1-o(1)$,%
\begin{eqnarray}
\max_{t\in \lbrack k],j=1,2}\left\{ \left\vert \omega _{j,j}^{(t)}\left( \hat{%
\omega}_{j,j}^{(t)}\right) ^{-1}-1\right\vert \right\}  &\leq &C\left( s\frac{%
\left( k+\log p\right) }{n^{(0)}}+\sqrt{\frac{\log (k/\delta _{1})}{n^{(0)}}}%
\right) ,  \label{eq:upper l2 alter2} \\
J_{n,k,1,2}^{(t)}/\left( \omega _{1,2}^{(t)}/\left( \omega _{1,1}^{(t)}\omega
_{2,2}^{(t)}\right) \right)  &\in &\left( -1.1,-0.9\right) ,
\label{eq:upper l2 alter2.5}
\end{eqnarray}%
where the second expression (\ref{eq:upper l2 alter2.5}) follows from (\ref%
{eq:upper l2 alter2}) and the definition of $J_{n,k,1,2}^{(t)}$ in (\ref%
{eq:J(t)}).

Define $\bar{U}_{n,k,1,2}^{2}\equiv\sum_{t=1}^{k}n^{(t)}\omega _{2,2}^{(t)}\omega
_{1,1}^{(t)}( T_{n,k,1,2}^{(t)}) ^{2}$. Comparing $\bar{U}%
_{n,k,1,2}^{2}$ with the definition of $U_{n,k,1,2}^{2}$ in (\ref{eq:test
stat chi}), we obtain that with probability $1-o(1)$,
\begin{equation*}
\frac{\bar{U}_{n,k,1,2}^{2}}{U_{n,k,1,2}^{2}}\leq \max_{t\in \lbrack k]}%
\frac{\omega _{1,1}^{(t)}}{\hat{\omega}_{1,1}^{(t)}}\frac{\omega _{2,2}^{(t)}}{%
\hat{\omega}_{2,2}^{(t)}}\leq 1+C\left( s\frac{\left( k+\log p\right) }{%
n^{(0)}}+\sqrt{\frac{\log (k/\delta _{1})}{n^{(0)}}}\right) \equiv\left( 1+\eta
_{1}^{l2}\right) ^{2},
\end{equation*}%
where the second inequality follows from  (\ref{eq:upper l2 alter2}%
). Note that according to our assumptions, it holds that $\eta _{1}^{l2}\leq
C( s\frac{\left( k+\log p\right) }{n^{(0)}}+\sqrt{\frac{\log (k/\delta
_{1})}{n^{(0)}}}) =o(1)$. Therefore, due to the union bound argument,
to prove (\ref{eq:upper l2 alter1}) it is sufficient to show
\begin{equation}
\inf_{v\in \mathcal{A(}c\mathcal{)}}\mathbb{P}_{v}\left(\bar{U}%
_{n,k,1,2}>\left( 1+\eta _{1}^{l2}\right) \cdot z_{k}^{l2}(1-\alpha )\right)>\beta
.  \label{eq:upper l2 alter3}
\end{equation}

We further reduce (\ref{eq:upper l2 alter3}) in the second step. Denote by  $%
\bar{V}_{n,k,1,2}^{\ast (t)}=\sqrt{\frac{\omega _{2,2}^{(t)}\omega _{1,1}^{(t)}%
}{n^{(t)}}}\sum_{i=1}^{n^{(t)}}( E_{i,1}^{(t)}E_{i,2}^{(t)}-\mathbb{E}%
E_{i,1}^{(t)}E_{i,2}^{(t)}) $ with $\mathbb{E}\bar{V}_{n,k,1,2}^{\ast
(t)}=0$. Lemma \ref{lem:lin comb} implies that with probability $1-o(1)$,
\begin{eqnarray*}
&&\left\vert \bar{U}_{n,k,1,2}-\left( \sum_{t=1}^{k}\left[ \sqrt{%
n^{(t)}\omega _{2,2}^{(t)}\omega _{1,1}^{(t)}}J_{n,k,1,2}^{(t)}+\bar{V}%
_{n,k,1,2}^{\ast (t)}\right] ^{2}\right) ^{1/2}\right\vert  \\
&\leq &\sum_{t=1}^{k}\sqrt{n^{(t)}\omega _{2,2}^{(t)}\omega _{1,1}^{(t)}}%
\left\vert T_{n,k,1,2}^{(t)}-J_{n,k,1,2}^{(t)}-\frac{1}{n^{(t)}}%
\sum_{i=1}^{n^{(t)}}\left( E_{i,1}^{(t)}E_{i,2}^{(t)}-\mathbb{E}%
E_{i,1}^{(t)}E_{i,2}^{(t)}\right) \right\vert  \\
&\leq &C\left( s\frac{k+(\log p)}{n^{(0)}}\right) \equiv \eta _{2}^{l2}\text{.}
\end{eqnarray*}%
Therefore, by the union bound argument again, to show (\ref{eq:upper l2
alter3}) it is sufficient to prove that
\begin{equation*}
\inf_{v\in \mathcal{A(}c\mathcal{)}}\mathbb{P}_{v}\left(
\sum_{t=1}^{k}\left[ \sqrt{n^{(t)}\omega _{2,2}^{(t)}\omega _{1,1}^{(t)}}%
J_{n,k,1,2}^{(t)}+\bar{V}_{n,k,1,2}^{\ast (t)}\right] ^{2}>\left[ \left(
1+\eta _{1}^{l2}\right) \cdot z_{k}^{l2}(1-\alpha )+\eta _{2}^{l2}\right]
^{2}\right) >\beta .
\end{equation*}%
We denote $\Xi _{t} \equiv ( \sqrt{n^{(t)}\omega _{2,2}^{(t)}\omega _{1,1}^{(t)}%
}J_{n,k,1,2}^{(t)}+\bar{V}_{n,k,1,2}^{\ast (t)}) ^{2}$, $t\in \lbrack
k]$ to simplify our notation. Then it suffices to show
\begin{equation}
\inf_{v\in \mathcal{A(}c\mathcal{)}}\mathbb{P}_{v}\left(
\sum_{t=1}^{k}\left( \Xi _{t}-\mathbb{E}\Xi _{t}\right) >\left[ \left(
1+\eta _{1}^{l2}\right) \cdot z_{k}^{l2}(1-\alpha )+\eta _{2}^{l2}\right]
^{2}-\sum_{t=1}^{k}\mathbb{E}\Xi _{t}\right) >\beta .
\label{eq:upper l2 alter4}
\end{equation}

In the third step, we need a careful analysis of both sides of (\ref{eq:upper
l2 alter4}). We first calculate the right-hand side term. According to the
third result in Lemma \ref{prop:chi} in Section \ref{SecC} with $z=\sqrt{2\log (1/\alpha )/k}
$, it holds that $z_{k}^{l2}(1-\alpha )\leq \sqrt{k}(1+\sqrt{2\log (1/\alpha
)/k})$. By our sample size assumption $s^{2}\left( k+\log p\right)
^{2}=o(n^{(0)})$ and the definitions of $\eta _{1}^{l2}$ and $\eta _{2}^{l2}$%
, we deduce that $s\frac{\left( k+\log p\right) }{n^{(0)}}\leq C\left(
n^{(0)}\right) ^{-1/2}$, which further yields%
\begin{eqnarray}
&&\left[ \left( 1+\eta _{1}^{l2}\right) \cdot z_{k}^{l2}(1-\alpha )+\eta
_{2}^{l2}\right] ^{2}  \notag \\
&\leq &\left( \sqrt{k}(1+\sqrt{2\log (1/\alpha )/k})\left( 1+C\sqrt{\frac{%
\log (k/\delta _{1})}{n^{(0)}}}\right) +C\left( n^{(0)}\right)
^{-1/2}\right) ^{2}  \notag \\
&\leq &\left( \sqrt{k}(1+\sqrt{2\log (1/\alpha )/k})\left( 1+C\sqrt{\frac{%
\log (k/\delta _{1})}{n^{(0)}}}\right) \right) ^{2}+C\sqrt{\frac{k}{n^{(0)}}}
\notag \\
&\leq &\left( k+3\sqrt{2k\log (1/\alpha )}\right) \left( 1+C\sqrt{\frac{\log
(k/\delta _{1})}{n^{(0)}}}\right)   \notag \\
&\leq &k+4\sqrt{2k\log (1/\alpha )}\text{.}  \label{eq:upper l2 alter5}
\end{eqnarray}%
Next we calculate a lower bound of $\sum_{t=1}^{k}\mathbb{E}\Xi _{t}$. By
the definition of $\bar{V}_{n,k,1,2}^{\ast (t)}$ and the joint Gaussianity of $%
E_{i,1}^{(t)}$ and $E_{i,2}^{(t)}$, we have $\mathbb{E}( \bar{V}%
_{n,k,1,2}^{\ast (t)}) ^{2}=1+( \omega _{1,2}^{(t)})
^{2}/( \omega _{2,2}^{(t)}\omega _{1,1}^{(t)}) $. This fact
together with (\ref{eq:upper l2 alter2.5}) results in
\begin{eqnarray}
\sum_{t=1}^{k}\mathbb{E}\Xi _{t} &=&\sum_{t=1}^{k}\mathbb{E}\left[ \sqrt{%
n^{(t)}\omega _{2,2}^{(t)}\omega _{1,1}^{(t)}}J_{n,k,1,2}^{(t)}+\bar{V}%
_{n,k,1,2}^{\ast (t)}\right] ^{2}  \notag \\
&\geq &\sum_{t=1}^{k}\mathbb{E}\left( \bar{V}_{n,k,1,2}^{\ast (t)}\right)
^{2}+Cn^{(0)}\sum_{t=1}^{k}\frac{\left( \omega _{1,2}^{(t)}\right) ^{2}}{%
\omega _{2,2}^{(t)}\omega _{1,1}^{(t)}}  \notag \\
&\geq &k+Cn^{(0)}\left\Vert \omega _{1,2}^{0}\right\Vert ^{2}.
\label{eq:upper l2 alter6}
\end{eqnarray}%

We can further upper bound the variance of $\sum_{t=1}^{k}\left( \Xi _{t}-%
\mathbb{E}\Xi _{t}\right) $ by the joint Gaussianity of $%
E_{i,1}^{(t)}$ and $E_{i,2}^{(t)}$,
\begin{equation}
\mathrm{var}\left( \sum_{t=1}^{k}\left( \Xi _{t}-\mathbb{E}\Xi _{t}\right)
\right) \leq C\left(k+n^{(0)}\left\Vert \omega _{1,2}^{0}\right\Vert ^{2}\right).  \label{eq:upper l2 alter7}
\end{equation}%
Expressions (\ref{eq:upper l2 alter5}) and (\ref{eq:upper l2 alter6}) imply
that under alternative $\mathcal{A(}c\mathcal{)=A}%
^{l2}(s,c\sqrt{k^{1/2}/n^{(0)}})$ with a sufficiently large $%
c>0$, the right-hand side of (\ref{eq:upper l2 alter4}) is
negative, that is,
\begin{eqnarray}
&&\left[ \left( 1+\eta _{1}^{l2}\right) \cdot z_{k}^{l2}(1-\alpha )+\eta
_{2}^{l2}\right] ^{2}-\sum_{t=1}^{k}\mathbb{E}\Xi _{t}  \notag \\
&<&-Cn^{(0)}\left\Vert \omega _{1,2}^{0}\right\Vert ^{2}+4\sqrt{2k\log
(1/\alpha )}  \notag \\
&\leq &-cC\sqrt{k}+4\sqrt{2k\log (1/\alpha )}<0.
\label{eq:upper l2 alter8}
\end{eqnarray}%
Therefore, by Chebyshev's inequality we obtain that for any $v\in \mathcal{A(}%
c\mathcal{)}$,
\begin{eqnarray*}
&&\mathbb{P}_{v}\left( \sum_{t=1}^{k}\left( \Xi _{t}-\mathbb{E}\Xi
_{t}\right) \leq \left[ \left( 1+\eta _{1}^{l2}\right) \cdot
z_{k}^{l2}(1-\alpha )+\eta _{2}^{l2}\right] ^{2}-\sum_{t=1}^{k}\mathbb{E}\Xi
_{t}\right)  \\
&\leq &\mathrm{var}\left( \sum_{t=1}^{k}\left( \Xi _{t}-\mathbb{E}\Xi
_{t}\right) \right) /\left( Cn^{(0)}\left\Vert \omega _{1,2}^{0}\right\Vert ^{2}\right) ^{2}<1-\beta,
\end{eqnarray*}%
where the first inequality follows from (\ref{eq:upper l2 alter8}) and the
last inequality follows from (\ref{eq:upper l2 alter7}) and a large constant  $%
c>0.$ Thus  (\ref{eq:upper l2 alter4}) is an
immediate consequence, which completes the proof for the first part of Theorem \ref{thm:lower rate}.

\subsubsection{Proof of Theorem \ref{thm:lower rate} (2)} \label{SecA.3.2}

To prove that $\epsilon _{n}=\sqrt{k/n^{(0)}}$ is the separating rate, we first
show the lower bound (\ref{eq:test lower}) and then establish that the proposed linear functional-based test $\phi _{1}$ satisfies  (\ref{eq:test upper}). Without loss of generality, assume that the sign vector $\xi=(1,\cdots,1)'$ and denote by $\mathcal{A}^{l1}(s,c'\sqrt{k/n^{(0)}})\equiv\mathcal{A}^{l1}(s,c'\sqrt{%
k/n^{(0)}},\xi)$ for short. Facilitated with (\ref{eq:proof lower 1}), it suffices to show that for
fixed $\beta >\alpha $, there exists some constant $c^{\prime }>0$ such that
$\Vert \mathbb{P}_{0}\wedge \mathbb{\bar{P}}\Vert >1-%
(\beta -\alpha )/2$ with appropriate choices of $\mathcal{G\subset A=A}%
^{l1}(s,c^{\prime }\sqrt{k/n^{(0)}})$ and $\Omega _{0}^{0}\in \mathcal{N}(s)$%
.

The constructions of $\mathcal{G}$ and $\Omega _{0}^{0}$ are straightforward. There
is only one element in $\mathcal{G}$, that is, $m=1$ and $\mathbb{\bar{P}=P}_{1}
$. We define $\Omega _{0}^{0}=\{\Omega _{0}^{(t)}\}_{t=1}^{k}$ such that $%
\Omega _{0}^{(1)}=\cdots =$ $\Omega _{0}^{(k)}=I$ and set $\Omega
_{1}^{0}=\{\Omega _{1}^{(t)}\}_{t=1}^{k}$ such that $(\Omega
_{0}^{(1)})^{-1}=\cdots =(\Omega _{0}^{(k)})^{-1}=I+( \tau /\sqrt{n^{(0)}k}%
) e_{12}$, where\ $\tau >0$ is some small constant to be determined later and
$e_{12}$ is the matrix with all but two entries being zero and the $(1,2)$th and $(2,1)$th entries being one. It is easy to see that $\Omega _{0}^{0}\in
\mathcal{N}(s)$. In addition, it is easy to check that all eigenvalues of $\Omega
_{1}^{0}$ are in $[M^{-1},M]$, and thus $\Omega _{1}^{0}\in \mathcal{F}(s)$
since $\tau /\sqrt{n^{(0)}k}=o(1)$. Note that $\left\Vert \omega
_{1,12}^{0}\right\Vert _{1}=\frac{\tau }{1-\tau ^{2}/(n^{(0)}k)}\sqrt{%
k/n^{(0)}}$. Therefore, we have shown that $\Omega _{1}^{0}\in \mathcal{A}%
^{l1}(s,c^{\prime }\sqrt{k/n^{(0)}})$ with $c^{\prime }\equiv 2\tau $, where we have used $\tau ^{2}/(n^{(0)}k)<1/2$.

To finish the lower bound (\ref%
{eq:test lower}), it remains to prove $\Vert \mathbb{P}_{0}\wedge
\mathbb{P}_{1}\Vert >1-(\beta -\alpha )/2$. A similar
argument to that in the proof of Lemma \ref{lem:lower sample} in Section \ref{SecB.4} (see expression (%
\ref{eq:proof lower 2})) implies that it is sufficient to show that the $\chi ^{2}
$ divergence between $\mathbb{P}_{0}$ and $\mathbb{P}_{1}$ is small enough,
that is, $\Delta =\int f_{1}^{2}/f_{0}-1<(\beta -\alpha )^{2}$. By the simple
constructions of $\Omega _{0}^{0}$ and $\Omega _{1}^{0}$, together with the $%
\chi ^{2}$ divergence of two Gaussian distributions (see expression (\ref%
{eq:Chi dist Gaussian})), it can be easily checked that $\Delta =(1-\tau
^{2}/(n^{(0)}k))^{-n^{(0)}k}-1$. Since $\tau ^{2}/(n^{(0)}k)<1/2$, we can
further bound the $\chi ^{2}$ divergence as
\[ \Delta \leq (1+2\tau
^{2}/(n^{(0)}k))^{n^{(0)}k}-1\leq \exp (2\tau ^{2})-1. \]
Therefore, by picking
$\tau $ small enough we deduce that $\Delta <(\beta -\alpha )^{2}$ and thus
$\Vert \mathbb{P}_{0}\wedge \mathbb{P}_{1}\Vert >1-(\beta -\alpha )/2$, which finishes the proof of (\ref{eq:test lower}).

It remains to show that the proposed linear functional-based test $\phi _{1}$
satisfies  (\ref{eq:test upper}), that is, with a sufficiently large
$c>0$, $\mathcal{A(}c\mathcal{)=A}^{l1}(s,c%
\sqrt{k/n^{(0)}})$, and $n^{(0)}$, it holds that
\begin{equation*}
\inf_{v\in \mathcal{A(}c\mathcal{)}}\mathbb{P}_{v}\left( \frac{%
V_{n,k,1,2}(\xi )}{\sqrt{k}}<z(\alpha )\right) \geq \beta \text{.}
\end{equation*}
Observe that under the assumptions of Proposition \ref{cor:l1 testing}, including
$\delta >1$ and $\delta _{1}=o(1)$, the last three inequalities of Lemma \ref{lem:diagnoal} and Condition \ref{CondA1} lead to the following two facts: (i) $\omega
_{1,1}^{(t)}( \hat{\omega}_{1,1}^{(t)}) ^{-1}=1+o(1)$ and $\omega
_{2,2}^{(t)}( \hat{\omega}_{2,2}^{(t)}) ^{-1}=1+o(1)$ uniformly over
$t\in [k],$ and (ii) $\sum_{t=1}^{k}\vert (\omega
_{1,1}^{(t)})^{1/2}-(\tilde{\omega}_{1,1}^{(t)})^{1/2}\vert =o(1)$ with
probability $1-o(1)$, which will be used later in our analysis.

With bound (\ref{eq:thm l1}) in Theorem \ref{thm:l1 testing} and the
definition of $V_{n,k,1,2}(\xi )$ in (\ref{eq:test stat l1}), along with
a union bound argument, we see that it suffices to prove that as $%
n^{(0)}\rightarrow \infty $,
\begin{equation}
\inf_{v\in \mathcal{A(}c\mathcal{)}}\mathbb{P}_{v}\left( \frac{%
V_{n,k,1,2}^{\ast }}{\sqrt{k}}<z(\alpha )-\eta _{1}^{l1}-\Psi \right) >\beta
,  \label{eq: proof l1 alter 1}
\end{equation}%
where $\Psi \equiv \sum_{t=1}^{k}\xi _{t}(n^{(t)}\hat{\omega}_{2,2}^{(t)}\hat{%
\omega}_{1,1}^{(t)})^{1/2}J_{n,k,1,2}^{(t)}/\sqrt{k}$ and $\eta _{1}^{l1}\equiv Cs\left(
k+\log p\right) /\sqrt{n^{(0)}k}$. To deal with the bias issue of $%
V_{n,k,1,2}^{\ast }$, we define $\bar{V}%
_{n,k,1,2}^{\ast }=\sum_{t=1}^{k}\xi _{t} (\frac{\omega
_{2,2}^{(t)}\omega _{1,1}^{(t)}}{n^{(t)}})^{1/2}\sum_{i=1}^{n^{(t)}}(
E_{i,1}^{(t)}E_{i,2}^{(t)}-\mathbb{E}E_{i,1}^{(t)}E_{i,2}^{(t)})$ and
reduce the problem of showing (\ref{eq: proof l1 alter 1}) to that of
showing
\begin{equation}
\inf_{v\in \mathcal{A(}c\mathcal{)}}\mathbb{P}_{v}\left( \frac{%
\bar{V}_{n,k,1,2}^{\ast }}{\sqrt{k}}<z(\alpha )-\eta _{1}^{l1}-\eta
_{2}^{l1}-\Psi \right) >\beta ,  \label{eq: proofl1 alter 1.5}
\end{equation}%
where $\eta _{2}^{l1}\equiv( V_{n,k,1,2}^{\ast }-\bar{V}_{n,k,1,2}^{\ast
}) /\sqrt{k}$.

We claim that $\eta _{1}^{l1}+\eta _{2}^{l1}=o_{P}(1)$
and  $z(\alpha )-\Psi <0$ under alternative $v\in \mathcal{A(}c%
\mathcal{)}$ with a sufficiently large constant $c>0$. Note that by
definition $\mathbb{E}\bar{V}_{n,k,1,2}^{\ast }=0$. Hence according to
Chebyshev's inequality and the union bound argument, it suffices to prove that \[
\mathrm{var}(\bar{V}_{n,k,1,2}^{\ast }/\sqrt{k})/\left\vert z(\alpha )-\Psi
\right\vert ^{2}<\left( 1-\beta \right) /2 \]
under alternative $v\in \mathcal{%
A(}c\mathcal{)}$. We finish the proof by showing that $\eta
_{1}^{l1}+\eta _{2}^{l1}=o_{P}(1)$, $\mathrm{var}(\bar{V}_{n,k,1,2}^{\ast }/%
\sqrt{k})\leq 2$ and that $\Psi <0$ can be arbitrarily small under
alternative $v\in \mathcal{A(}c\mathcal{)}$ by picking a
sufficiently large constant $c>0$, respectively. Indeed, assuming that
the latter two facts hold, $\mathrm{var}(\bar{V}_{n,k,1,2}^{\ast }/\sqrt{k}%
)/\left\vert z(\alpha )-\Psi \right\vert ^{2}\\<\left( 1-\beta \right) /2$ follows as
an immediate consequence, which will finish our proof.

In particular, fact (i) above entails that $J_{n,k,1,2}^{(t)}=(-1+o(1))%
\omega _{1,2}^{(t)}/( \omega _{1,1}^{(t)}\omega _{2,2}^{(t)}) $
uniformly over $t\in \lbrack k]$, following from the definition of $%
J_{n,k,1,2}^{(t)}$ in (\ref{eq:J(t)}). Since the sign vector of $\omega
_{1,2}^{0}$ is encoded in $\xi $, the boundedness of $\omega
_{1,1}^{(t)}\omega _{2,2}^{(t)}$ and $(\hat{\omega}_{2,2}^{(t)}\hat{\omega}%
_{1,1}^{(t)})^{1/2}$ for $t\in [p]$ (due to Condition \ref{CondA1} and fact (i) above)
further implies that with some constant $C>0$,
\begin{equation*}
\Psi \leq -C\sqrt{\frac{n^{(0)}}{k}}\left\Vert \omega _{1,2}^{0}\right\Vert
_{1}\leq -Cc,
\end{equation*}%
under alternative $\mathcal{A(}c\mathcal{)=A}^{l1}(s,c%
\sqrt{k/n^{(0)}})$. Therefore, with a sufficiently large constant $c>0$, $\Psi <0$ is smaller than any pre-determined negative constant.

Note that by the independence and joint Gaussianity of $E_{1,1}^{(t)}$ and $E_{1,2}^{(t)}
$, we have $\mathrm{var}(\bar{V}_{n,k,1,2}^{\ast }/\sqrt{k}%
)=k^{-1}\sum_{t=1}^{k}\mathrm{var}(E_{1,1}^{(t)}E_{1,2}^{(t)})\omega
_{2,2}^{(t)}\omega _{1,1}^{(t)}\leq 2$. Thus it remains to show that $\eta
_{1}^{l1}+\eta _{2}^{l1}=o_{P}(1)$. It is easy to see that $\eta
_{1}^{l1}=Cs\left( k+\log p\right) /\sqrt{n^{(0)}k}=o(1)$ by our sample size
assumption. In addition, we have with probability at least $1-2\delta
_{1}^{-10}$,%
\begin{eqnarray}
\left\vert \eta _{2}^{l1}\right\vert  &=&\left\vert \sum_{t=1}^{k}\frac{\xi
_{t}}{\sqrt{k}}\cdot \sqrt{\frac{\omega _{2,2}^{(t)}}{n^{(t)}}}%
\sum_{i=1}^{n^{(t)}}\left( E_{i,1}^{(t)}E_{i,2}^{(t)}-\mathbb{E}%
E_{i,1}^{(t)}E_{i,2}^{(t)}\right) \left( \sqrt{\omega _{1,1}^{(t)}}-\sqrt{%
\tilde{\omega}_{1,1}^{(t)}}\right) \right\vert   \notag \\
&\leq &\frac{1}{\sqrt{k}}\max_{t\in \lbrack k]}\left\vert \sqrt{\frac{\omega
_{2,2}^{(t)}}{n^{(t)}}}\sum_{i=1}^{n^{(t)}}\left( E_{i,1}^{(t)}E_{i,2}^{(t)}-%
\mathbb{E}E_{i,1}^{(t)}E_{i,2}^{(t)}\right) \right\vert \cdot
\sum_{t=1}^{k}\left\vert \sqrt{\omega _{1,1}^{(t)}}-\sqrt{\tilde{\omega}%
_{1,1}^{(t)}}\right\vert   \notag \\
&<&C\sqrt{\frac{\log (k/\delta _{1})}{k}}\cdot \sum_{t=1}^{k}\left\vert
\sqrt{\omega _{1,1}^{(t)}}-\sqrt{\tilde{\omega}_{1,1}^{(t)}}\right\vert ,
\label{eq: proofl1 alter 2}
\end{eqnarray}%
where the first inequality is due to H\"{o}lder's inequality and the second
one follows from Bernstein's inequality (see, e.g., Proposition 5.16,
\cite{vershynin2010introduction}). It follows from fact (ii) above and inequality (\ref%
{eq: proofl1 alter 2}) that $\eta _{2}^{l1}=o_{P}(1)$,
in view of $\delta _{1}=o(1)$. Therefore, we have shown  (\ref{eq:
proofl1 alter 1.5}), which further entails that $\phi _{1}$ satisfies (\ref{eq:test upper}) with a sufficiently large constant $c>0$. This concludes the proof for the second part of Theorem \ref{thm:lower rate}.

\subsection{Proof of Theorem \ref{thm:lower sample size}} \label{SecA.4}

The general tool established in (\ref{eq:proof lower 1}) of Section \ref{SecA.3} plays a key role in our analysis. We need to show that for any fixed $\beta >\alpha $ and $c>0$, there is no
test of significance level $\alpha $ satisfying  (\ref{eq:test upper}) with $\mathcal{A}=\mathcal{A}^{l1}(s,c\sqrt{k/n^{(0)}},\xi)$.
In light of (\ref{eq:proof lower 1}), it is sufficient to show that as
long as $s^{2}k^{-1}(k+\log p)>Cn^{(0)}$ for some sufficiently large positive constant $C$
depending on $M_{1},\mu $, and $c$, we have
\[ \Vert \mathbb{P}_{0}\wedge
\mathbb{\bar{P}}\Vert >1-(\beta -\alpha )/2 \]
with appropriate
choices of $\mathcal{G\subset A}^{l1}(s,c\sqrt{k/n^{(0)}},\xi)$ and $\Omega
_{0}^{0}\in \mathcal{N}(s)$. Since the lower bound does not depend on the choice of the sign vector $\xi$, hereafter we assume $\xi=(1,\cdots,1)'$ without loss of generality.

To construct $\mathcal{G}$ and $\Omega _{0}^{0}$, it suffices to assume that  the $k$
precision matrices are identical for each $\Omega _{h}^{0}$, $h=0,\cdots ,m$%
, that is, $\Omega _{h}^{(1)}=\cdots =$ $\Omega _{h}^{(k)}$. Therefore, we
only need to construct $\Omega _{h}^{(1)}$ for each $h$. The element in null
is defined as $\Omega _{0}^{(1)}=I$ which gives
\begin{equation} \label{neweq012}
\Omega _{0}^{0}=\{\Omega _{0}^{(t)}\}_{t=1}^{k} \text{ \ with \ } \Omega _{0}^{(1)}=\cdots =\Omega _{0}^{(k)}=I.
\end{equation}
Besides, we construct a subset
\begin{equation} \label{neweq01,1}
\mathcal{G}=\left\{\Omega ^{0}=\{\Omega ^{(t)}\}_{t=1}^{k}:\Omega ^{(1)}=\cdots
=\Omega ^{(k)}=(I+aH)^{-1}\mbox{\rm  for some }H\in \mathcal{H}\right\}
\end{equation}
with $a=\sqrt{\tau \frac{1+(\log p)/k}{n^{(0)}}}$and $\tau >0$ some
small constant to be determined later. Here $\mathcal{H}$ is the set containing
the collection of all $p\times p$ symmetric matrices with exactly $s-1$
elements equal to $1$ between the third and the last elements of the first
and second\textbf{\ }rows (and hence columns by symmetry) and the rest all
zeros. We also assume that for each $H\in \mathcal{H}$, the supports of the first
row and the second row are identical. Clearly, there are $\binom{p-2}{s-1}$
distinct elements in $\mathcal{G}$ and thus $m=\binom{p-2}{s-1}$. To finish
the proof, we need to show two claims: (i) $\mathcal{G}\subset \mathcal{A}^{l1}(s,c\sqrt{%
k/n^{(0)}},\xi)$ and $\Omega _{0}^{0}\in \mathcal{N}(s)$ and (ii) $\Vert
\mathbb{P}_{0}\wedge \mathbb{\bar{P}}\Vert >1-(\beta
-\alpha )/2$.

The desired result in claim (ii) is established in Lemma \ref{lem:lower sample} in Section \ref{SecB.4}. Thus it remains to prove the desired result in claim (i). It is
easy to see that $\Omega _{0}^{0}\in \mathcal{N}(s)$ since all $k$
precision matrices are identity matrices and particularly $\omega
_{0,12}^{0}=\mathbf{0}$. For each $\Omega _{h}^{0}\in \mathcal{G}$, we can  check that $\Omega _{h}^{0}$ satisfies the sparsity assumption $%
\max_{a}\sum_{b\neq a}1\{\omega _{h,ab}^{0}\neq \mathbf{0}\}\leq s$.
Moreover, the largest and smallest eigenvalues of $\Omega _{h}^{(1)}$ are%
\begin{equation*}
\lambda _{\max }(\Omega _{h}^{(1)})=\frac{1+\sqrt{2(s-1)a^{2}}}{1-2(s-1)a^{2}%
},\lambda _{\min }(\Omega _{h}^{(1)})=\frac{1-\sqrt{2(s-1)a^{2}}}{%
1-2(s-1)a^{2}},
\end{equation*}%
respectively, with all remaining eigenvalues being ones. Under the assumption that $%
s(1+(\log p)/k)/n^{(0)}\\=o(1)$, we see that $2(s-1)a^{2}$ is sufficiently
small and hence all eigenvalues are bounded between $1/M$ and $M$, which satisfies
Condition \ref{CondA1}. Therefore, we have shown that $\mathcal{G\subset }$ $\mathcal{F}(s)$.

Finally, some elementary algebra implies that for each $\Omega _{h}^{0}\in
\mathcal{G}$, we always have $\omega _{h,12}^{(1)}=\frac{(s-1)a^{2}}{%
1-2(s-1)a^{2}}$. As a result, it holds that
\begin{equation*}
\left\Vert \omega _{h,12}^{0}\right\Vert _{1}=\frac{k(s-1)a^{2}}{%
1-2(s-1)a^{2}}\geq 2k(s-1)\tau \left(\frac{1+(\log p)/k}{n^{(0)}}\right)>c\sqrt{\frac{k%
}{n^{(0)}}},
\end{equation*}%
where the first inequality follows from $2(s-1)a^{2}<1/2$ and the last
inequality is due to the main assumption of Theorem \ref{thm:lower sample
size}, that is, $s^{2}k^{-1}(k+\log p)^{2}>Cn^{(0)}$ with $C\equiv(c/\tau )^{2}$%
. Therefore, we have shown $\mathcal{G}\subset \mathcal{A}^{l1}(s,c\sqrt{%
k/n^{(0)}},\xi)$, which completes the proof.

\subsection{Proof of Theorem \ref{thm:GSRLH}} \label{SecA.5}

Without loss of generality, we only prove the results for the case of $j=1$. This is because by
symmetry, the results remain valid for any $j\in [p]$. Hereafter, we
follow the same notation for any vector $u\in \mathbb{R}^{(p-1)k}$ as
defined for $C_{1}^{0}$, that is, $u^{(t)}$ denotes its subvector
corresponding to the $t$th class and $u_{(l)}$ represents its subvector
corresponding to the $l$th group. The purpose of normalization diagonal
matrices $\bar{D}_{1}^{(t)}$ for our method HGSL defined in (\ref{eq:GSRLH}) is to obtain a tight universal regularization
parameter $\lambda $ by normalizing each column of $\mathbf{X}_{\ast
,-1}^{(t)}$ such that its $\ell_2$ norm is $\sqrt{n^{(t)}}$, that is,  $\mathbf{%
\bar{X}}_{\ast ,-1}^{(t)}=\mathbf{X}_{\ast ,-1}^{(t)}( \bar{D}%
_{1}^{(t)}) ^{-1/2}$.

Define $\bar{C}_{1}^{(t)}=( \bar{D}%
_{1}^{(t)}) ^{1/2}C_{1}^{(t)}$ and $\hat{\bar{C}}_{1}^{(t)}=(
\bar{D}_{1}^{(t)}) ^{1/2}\hat{C}_{1}^{(t)}$, and correspondingly $\bar{C}%
_{1}^{0}$ and $\hat{\bar{C}}_{1}^{0}$. Then the right-hand side of (\ref%
{eq:HeterGroupLasso}) becomes $\mathbf{\bar{X}}_{\ast ,-1}^{0}\bar{C}%
_{1}^{0}+E_{\ast ,1}^{0}$ and the method HGSL in (\ref{eq:GSRLH}) becomes%
\begin{equation*}
\hat{\bar{C}}_{1}^{0}={\arg\min}_{\beta ^{0}\in \mathbb{R}^{k(p-1)}}%
\left\{\sum_{t=1}^{k}\bar{Q}_{t}^{1/2}(\beta ^{(t)})+\lambda
\sum_{l=2}^{p}\left\Vert \beta _{(l)}^{0}\right\Vert \right\}
\end{equation*}%
with $\bar{Q}_{t}(\beta ^{(t)})=\frac{1}{n^{(0)}}\Vert X_{\ast
,1}^{(t)}-\mathbf{\bar{X}}_{\ast ,-1}^{(t)}\beta ^{(t)}\Vert ^{2}$.
Our main results involve the difference $\Delta =\hat{C}_{1}^{0}-C_{1}^{0}$.
In what follows, we establish all results in terms of $\bar{\Delta}=\hat{\bar{%
C}}_{1}^{0}-\bar{C}_{1}^{0}=\left( \bar{D}_{1}^{0}\right) ^{1/2}\Delta $.
It is worth mentioning that this does not affect our results much. Indeed, our Condition \ref{CondA1} and the
fact of $\mathbf{X}_{\ast ,l}^{(t)\prime }\mathbf{X}_{\ast ,l}^{(t)}/\sigma
_{ll}^{(t)}\sim \chi ^{2}(n^{(t)})$, together with an application of
Lemma \ref{prop:chi} and the union bound, entail that with probability at least $%
1-2pk\exp (-n^{(0)}/32),$ all diagonal entries of $\bar{D}_{1}^{0}$ are
bounded from below by $M/2$ and from above by $3M/2$ simultaneously. Therefore, $\Delta $ and $\bar{\Delta}$ are of the same order
componentwise and globally. To make it rigorous, define an event \[ \mathcal{E}_{scale}=\left\{ \mathbf{X}%
_{\ast ,l}^{(t)\prime }\mathbf{X}_{\ast ,l}^{(t)}/n^{(t)}\in [
1/(2M),3M/2]\mbox{\rm\ \ for all }t\in [k],l\in [p]\right\} \]
and it holds that $\mathbb{P}\{\mathcal{E}_{scale}\}\geq 1-2pk\exp (-n^{(0)}/32)$.


We begin with introducing the group-wise restricted eigenvalue
(gRE) condition proposed by \cite{nardi2008asymptotic} and \cite%
{lounici2011oracle}, which is needed to establish our main results. Recall that
the true coefficient vector $C_{1}^{0}$ is a group sparse vector. Denote by $T=\{l:%
\bar{C}_{1(l)}^{0}\neq \mathbf{0}\}$. By the definition of the maximum node
degree given in (\ref{eq:sparsity}) and the relationship between $\bar{C}_{1}^{(t)}$ and $%
\Omega ^{(t)}$, we deduce that $\left\vert T\right\vert \leq s$, where $%
\left\vert \cdot \right\vert $ stands for the cardinality of a set.

\begin{definition} \label{def:gRE}
\label{def:comp} The group-wise restricted eigenvalue (gRE) condition holds on the design matrix $\mathbf{\bar{X}}_{\ast
,-1}^{0}$ if
\begin{equation*}
gRE(\xi ,T)\equiv\inf_{u\neq 0}\left\{\frac{\left\Vert \mathbf{\bar{X}}_{\ast
,-1}^{0}u\right\Vert }{\sqrt{n^{(0)}}\left\Vert u\right\Vert }:u\in \Psi
(\xi ,T)\right\}>0,
\end{equation*}%
where $\Psi (L,T)=\{u\in \mathbb{R}^{(p-1)k}:\sum_{j\in
T^{c}}\Vert u_{(j)}\Vert \leq L\sum_{j\in T}\Vert
u_{(j)}\Vert \}$ is a cone.
\end{definition}

The above gRE 
condition is an extension of  the restricted
eigenvalue (RE) condition for the regular Lasso proposed in \cite{bickel2009simultaneous},
in which the $\ell_1$ norm is replaced by the group-wise $\ell_1$ norm. It was
also assumed in \cite{lounici2011oracle} to tackle the usual group Lasso as
a direct condition. \cite{nardi2008asymptotic} derived the gRE condition based
on some incoherence condition. However, to the best of our knowledge, there is no existing result for the random design matrix satisfying the gRE condition in the literature. In this paper, we first establish that the gRE condition is satisfied with large probability as a consequence of our assumptions in Lemma \ref{lem:GRE} presented in Section \ref{SecB.5}.

We would like to mention that other commonly used conditions on the design matrix $\mathbf{\bar{X}}_{\ast
,-1}^{0}$, including the group-wise compatibility condition \citep{bunea2014group}
and the group-wise cone invertibility factor condition \citep{mitra2014benefit}, can also be applied here. In fact, the group-wise compatibility condition $%
\kappa (\xi ,T)>0$ is a natural consequence of the gRE condition thanks to the Cauchy-Schwarz
inequality, since
\begin{eqnarray}
\kappa (\xi ,T) &\equiv&\inf_{u\neq \mathbf{0}}\left\{\frac{\sqrt{\left\vert T\right\vert}\left\Vert
\mathbf{\bar{X}}_{\ast ,-1}^{0}u\right\Vert }{\sqrt{n^{(0)}}\sum_{l\in
T}\left\Vert u_{(l)}\right\Vert }:u\in \Psi (\xi ,T)\right\}  \notag \\
&\geq &\inf_{u\neq \mathbf{0}}\left\{\frac{\left\Vert \mathbf{\bar{X}}_{\ast
,-1}^{0}u\right\Vert }{\sqrt{n^{(0)}}\left( \sum_{l\in T}\left\Vert
u_{(l)}\right\Vert ^{2}\right) ^{1/2}}:u\in \Psi (\xi ,T)\right\}  \notag \\
&\geq &\inf_{u\neq 0}\left\{\frac{\left\Vert \mathbf{\bar{X}}_{\ast
,-1}^{0}u\right\Vert }{\sqrt{n^{(0)}}\left\Vert u\right\Vert }:u\in \Psi
(\xi ,T)\right\}=gRE(\xi ,T).  \label{eq: GCC}
\end{eqnarray}
In particular, on the event $\mathcal{E}_{1,gRE}$ defined in Lemma \ref{lem:GRE} it holds that $$\kappa (\xi ,T)>\min_{l,t}\{(n^{(t)}/%
\mathbf{X}_{\ast ,l}^{(t)\prime }\mathbf{X}_{\ast
,l}^{(t)})^{1/2}\}/(2M)^{1/2}.$$

As discussed in Section \ref{sec:initial_1}, the analysis of Theorem \ref{thm:GSRLH} relies critically on the event $\mathcal{B}_{1}$ defined in (\ref{eq: lambda condition}), which guides us to pick a
sharp parameter $\lambda $. Lemma \ref{lem:key event} in Section \ref{SecB.6}  implies that our explicit
choice of $\lambda $ is indeed feasible. Thus with the aid of Lemmas \ref{lem:GRE} and \ref{lem:key event}, we are now ready
to establish our main results in the following two steps.

\medskip

\textbf{Step 1.} It follows from the definition that
\begin{eqnarray}
\sum_{t=1}^{k}\left(\bar{Q}_{t}^{1/2}(\hat{\bar{C}}_{1}^{(t)})-\bar{Q}_{t}^{1/2}(%
\bar{C}_{1}^{(t)})\right) &\leq &\lambda \sum_{l=2}^{p}\left(\left\Vert \bar{C}%
_{1(l)}^{0}\right\Vert -\left\Vert \hat{\bar{C}}_{1(l)}^{0}\right\Vert \right)
\notag \\
&\leq &\lambda \left(\sum_{l\in T}\left\Vert \bar{\Delta}_{(l)}\right\Vert
-\sum_{l\in T^{c}}\left\Vert \bar{\Delta}_{(l)}\right\Vert \right).
\label{eq: by definition}
\end{eqnarray}%
Observe that $\frac{\partial \bar{Q}_{t}^{1/2}(\bar{C}_{1}^{(t)})}{\partial
\beta ^{(t)}}=\frac{-1}{\sqrt{n^{(0)}}}\frac{ \mathbf{\bar{X}}_{\ast
,-1}^{(t)\prime}E_{\ast ,1}^{(t)}}{\Vert E_{\ast
,1}^{(t)}\Vert }$. By the convexity of $\bar{Q}_{t}^{1/2}(\cdot )$, we have%
\begin{eqnarray}
\sum_{t=1}^{k}\left(\bar{Q}_{t}^{1/2}(\hat{\bar{C}}_{1}^{(t)})-\bar{Q}_{t}^{1/2}(%
\bar{C}_{1}^{(t)})\right) &\geq &-\frac{1}{\sqrt{n^{(0)}}}\sum_{t=1}^{k}\frac{\bar{%
\Delta}^{(t)\prime }\mathbf{\bar{X}}_{\ast
,-1}^{(t)\prime}E_{\ast ,1}^{(t)}}{\left\Vert E_{\ast ,1}^{(t)}\right\Vert }  \notag \\
&\geq &-\left( \sum_{l=2}^{p}\left\Vert \bar{\Delta}_{(l)}\right\Vert
\right) \cdot \max_{2\leq l\leq p}\frac{\left\Vert \bar{D}_{E1}^{-1/2}
\mathbf{\bar{X}}_{\ast ,(l)}^{0\prime} E_{\ast ,1}^{0}\right\Vert
}{\sqrt{n^{(0)}}}  \notag \\
&\geq &-\lambda \frac{\xi -1}{\xi +1}\sum_{l=2}^{p}\left\Vert \bar{\Delta}%
_{(l)}\right\Vert ,  \label{eq: convexity}
\end{eqnarray}%
where the last inequality follows from Lemma \ref{lem:key event}. Combining
inequalities (\ref{eq: by definition}) and (\ref{eq: convexity}), we obtain
\begin{equation*}
-\lambda \frac{\xi -1}{\xi +1}\sum_{l=2}^{p}\left\Vert \bar{\Delta}%
_{(l)}\right\Vert \leq \lambda \left(\sum_{l\in T}\left\Vert \bar{\Delta}%
_{(l)}\right\Vert -\sum_{l\in T^{c}}\left\Vert \bar{\Delta}_{(l)}\right\Vert
\right),
\end{equation*}%
which entails that%
\begin{equation*}
\sum_{l\in T^{c}}\left\Vert \bar{\Delta}_{(l)}\right\Vert \leq \xi
\sum_{l\in T}\left\Vert \bar{\Delta}_{(l)}\right\Vert \mbox{\rm .}
\end{equation*}%
Hence, we have shown that $\bar{\Delta}\in \Psi (\xi ,T)$.

\medskip

\textbf{Step 2.} We will make use of the following facts with $\zeta _{t}=\bar{Q}%
_{t}^{1/2}(\hat{\bar{C}}_{1}^{(t)})+\bar{Q}_{t}^{1/2}(\bar{C}_{1}^{(t)})$
\begin{eqnarray}
\bar{Q}_{t}(\hat{\bar{C}}_{1}^{(t)})-\bar{Q}_{t}(\bar{C}_{1}^{(t)}) &=&\frac{%
\left\Vert \mathbf{\bar{X}}_{\ast ,-1}^{(t)}\bar{\Delta}^{(t)}\right\Vert
^{2}}{n^{(0)}}-\frac{2\bar{\Delta}^{(t)\prime }\mathbf{\bar{X}}_{\ast
,-1}^{(t)\prime }E_{\ast ,1}^{(t)}}{n^{(0)}},  \label{eq:step2_1}
\\
\bar{Q}_{t}(\hat{\bar{C}}_{1}^{(t)})-\bar{Q}_{t}(\bar{C}_{1}^{(t)}) &=&\left(\bar{%
Q}_{t}^{1/2}(\hat{\bar{C}}_{1}^{(t)})-\bar{Q}_{t}^{1/2}(\bar{C}%
_{1}^{(t)})\right)\cdot \zeta _{t},  \label{eq:step2_2} \\
\sum_{l\in T}\left\Vert \bar{\Delta}_{(l)}\right\Vert &\leq &\frac{\sqrt{s}%
\left\Vert \mathbf{\bar{X}}_{\ast ,-1}^{0}\bar{\Delta}\right\Vert }{\sqrt{%
n^{(0)}}\kappa (\xi ,T)},  \label{eq:step2_3} \\
\sum_{t=1}^{k}\frac{\bar{\Delta}^{(t)\prime }\mathbf{\bar{X}}_{\ast
,-1}^{(t)\prime }E_{\ast ,1}^{(t)}}{n^{(0)}\zeta _{t}} &\leq
&\left( \sum_{l=2}^{p}\left\Vert \bar{\Delta}_{(l)}\right\Vert \right)
\max_{2\leq l\leq p}\frac{\left\Vert \bar{D}_{E1}^{-1/2} \mathbf{%
\bar{X}}_{\ast ,(l)}^{0\prime }E_{\ast ,1}^{0}\right\Vert }{\sqrt{%
n^{(0)}}}\cdot \max_{t\in [k]}\frac{\left\Vert E_{\ast
,1}^{(t)}\right\Vert }{\zeta _{t}\sqrt{n^{(0)}}},  \label{eq:step2_4}
\end{eqnarray}%
where the first two facts are due to some elementary algebra and the third one
follows from the definition of $\kappa (\xi ,T)$ in (\ref{eq: GCC}) and the fact of $\bar{\Delta}\in
\Psi (\xi ,T)$ proved in Step 1. It follows from (\ref{eq:step2_1}) and  (\ref{eq:step2_2}%
) that
\begin{equation*}
\sum_{t=1}^{k}(\bar{Q}_{t}^{1/2}(\hat{\bar{C}}_{1}^{(t)})-\bar{Q}_{t}^{1/2}(%
\bar{C}_{1}^{(t)}))=\sum_{t=1}^{k}\left(\frac{\left\Vert \mathbf{\bar{X}}_{\ast
,-1}^{(t)}\bar{\Delta}^{(t)}\right\Vert ^{2}}{n^{(0)}\zeta _{t}}-\frac{2\bar{%
\Delta}^{(t)\prime }\mathbf{\bar{X}}_{\ast ,-1}^{(t)\prime
}E_{\ast ,1}^{(t)}}{n^{(0)}\zeta _{t}}\right).
\end{equation*}%
Therefore, by (\ref{eq:step2_4}), Lemma \ref{lem:key event}, and the
fact of $\max_{t\in [k]}(\frac{\Vert E_{\ast ,1}^{(t)}\Vert }{%
\zeta _{t}\sqrt{n^{(0)}}})\leq 1,$ we further deduce that
\begin{eqnarray}
\sum_{t=1}^{k}\frac{\left\Vert \mathbf{\bar{X}}_{\ast ,-1}^{(t)}\bar{\Delta}%
^{(t)}\right\Vert ^{2}}{n^{(0)}\zeta _{t}} &\leq &\sum_{t=1}^{k}\left(\bar{Q}%
_{t}^{1/2}(\hat{\bar{C}}_{1}^{(t)})-\bar{Q}_{t}^{1/2}(\bar{C}%
_{1}^{(t)})\right)+2\lambda \frac{\xi -1}{\xi +1}\left( \sum_{l=2}^{p}\left\Vert
\bar{\Delta}_{(l)}\right\Vert \right)  \notag \\
&\leq &\lambda \left(\sum_{l\in T}\left\Vert \bar{\Delta}_{(l)}\right\Vert
-\sum_{l\in T^{c}}\left\Vert \bar{\Delta}_{(l)}\right\Vert \right)+2\lambda \frac{%
\xi -1}{\xi +1}\left( \sum_{l=2}^{p}\left\Vert \bar{\Delta}_{(l)}\right\Vert
\right)  \notag \\
&=&\lambda \left(\frac{3\xi -1}{\xi +1}\sum_{l\in T}\left\Vert \bar{\Delta}%
_{(l)}\right\Vert +\frac{\xi -3}{\xi +1}\sum_{l\in T^{c}}\left\Vert \bar{%
\Delta}_{(l)}\right\Vert \right)  \notag \\
&\leq &\lambda \left( \frac{3\xi -1}{\xi +1}+\xi \frac{\left( \xi -3\right)
_{+}}{\xi +1}\right) \sum_{l\in T}\left\Vert \bar{\Delta}_{(l)}\right\Vert
\notag \\
&\leq &\frac{\sqrt{s}\left\Vert \mathbf{\bar{X}}_{\ast ,-1}^{0}\bar{\Delta}%
\right\Vert }{\sqrt{n^{(0)}}\kappa (\xi ,T)}\lambda \left( \frac{3\xi -1}{%
\xi +1}+\xi \frac{\left( \xi -3\right) _{+}}{\xi +1}\right) ,
\label{eq: step2 predict}
\end{eqnarray}%
where the second inequality is due to (\ref{eq: by definition}) and the last
one follows from the definition of $\kappa (\xi ,T)$ in (\ref{eq: GCC}%
).

Lemma \ref{lem:constant upper} presented in Section \ref{SecB.7} provides a natural constant level upper bound for the
fitted prediction error. Then we can lower bound the left-hand side of (\ref{eq: step2 predict})
according to Lemma \ref{lem:constant upper} on the event $\mathcal{E}%
_{1,up}$ as
\begin{equation*}
\sum_{t=1}^{k}\frac{\left\Vert \mathbf{\bar{X}}_{\ast ,-1}^{(t)}\bar{\Delta}%
^{(t)}\right\Vert ^{2}}{n^{(0)}\zeta _{t}}\geq \frac{1}{\sqrt{6MM_{0}}}%
\sum_{t=1}^{k}\frac{\left\Vert \mathbf{\bar{X}}_{\ast ,-1}^{(t)}\bar{\Delta}%
^{(t)}\right\Vert ^{2}}{n^{(0)}}.
\end{equation*}%
Thus combining (\ref{eq: step2 predict}) with the above inequality leads to
\begin{equation*}
\frac{\left\Vert \mathbf{\bar{X}}_{\ast ,-1}^{0}\bar{\Delta}\right\Vert }{%
\sqrt{n^{(0)}}}\leq \frac{\sqrt{s}}{\kappa (\xi ,T)}\lambda \left( \frac{%
3\xi -1}{\xi +1}+\xi \frac{\left( \xi -3\right) _{+}}{\xi +1}\right) \sqrt{%
6MM_{0}}.
\end{equation*}%

In summary, by (\ref{eq: GCC}) and with our well specified $\lambda $%
, on the event $\mathcal{E}_{scale}\cap \mathcal{E}_{1,up}\cap \mathcal{B}%
_{1}\cap \mathcal{E}_{1,gRE}$ there exists some constant $C>0$
such that
\begin{equation*}
\sum_{t=1}^{k}\frac{\left\Vert \mathbf{X}_{\ast ,-1}^{(t)}\left( \hat{C}%
_{1}^{(t)}-C_{1}^{(t)}\right) \right\Vert ^{2}}{n^{(0)}}=\sum_{t=1}^{k}\frac{%
\left\Vert \mathbf{\bar{X}}_{\ast ,-1}^{0}\left( \hat{\bar{C}}_{1}^{(t)}-%
\bar{C}_{1}^{(t)}\right) \right\Vert ^{2}}{n^{(0)}}\leq Cs\frac{k+\log p}{%
n^{(0)}}.
\end{equation*}%
Moreover, since $\hat{\bar{C}}_{1}^{0}-\bar{C}_{1}^{0}=\bar{\Delta}\in \Psi
(\xi ,T)$, by the definitions of $\kappa (\xi ,T)$ in (\ref{eq: GCC}) and the gRE condition in
Definition \ref{def:comp} we can derive the following two inequalities from the expression above
\begin{eqnarray*}
\sum_{l=2}^{p}\left\Vert \hat{C}_{1(l)}^{0}-C_{1(l)}^{0}\right\Vert  &\leq &%
\sqrt{2M}\sum_{l=2}^{p}\left\Vert \hat{\bar{C}}_{1(l)}^{0}-\bar{C}%
_{1(l)}^{0}\right\Vert \leq Cs\left(\frac{k+\log p}{n^{(0)}}\right)^{1/2}, \\
\left\Vert \hat{C}_{1}^{0}-C_{1}^{0}\right\Vert  &\leq &\sqrt{2M}\left\Vert
\hat{\bar{C}}_{1}^{0}-\bar{C}_{1}^{0}\right\Vert \leq C\left(s\frac{k+\log p}{%
n^{(0)}}\right)^{1/2},
\end{eqnarray*}%
noting that conditional on the event $\mathcal{E}_{scale}$, $%
\Delta $ is less than or equal to$\sqrt{2M}\bar{\Delta}$ componentwise. Finally we conclude the proof by an application of the union bound argument using
Lemmas \ref{lem:GRE}--\ref{lem:constant upper}.

\subsection{Proof of Theorem \ref{com theorem}} \label{SecA.6}
The main idea of the proof consists of two parts. First we prove that our suggested algorithm in Section \ref{Sec3.2} has a unique guaranteed point of convergence  $\beta^{*}$. Then we show that such a point is the global optimum of the HGSL optimization problem \eqref{eq: group square-root Lasso1}.
%

\medskip

{\bf Step 1: Convergence of $\beta(m)$.}
Let us denote by
\begin{equation} \label{neweq006}
F(\beta)= (n^{(0)})^{-1/2} \sum_{t=1}^{k}\Vert Y^{(t)}-\bX^{(t)}\beta ^{(t)}\Vert+\lambda \sum_{l=1}^{p}\Vert \beta_{(l)}\Vert
\end{equation}
the objective function in (\ref{eq: group square-root Lasso2}) which is a reformulation of \eqref{eq: group square-root Lasso1} in simplified notation. To prove the desired result, we first construct a surrogate function and show that the updating rule optimizes the surrogate function. Then we characterize the relationship between the objective function and the surrogate function, which entails that the limit of $\beta(m)$ from the $m$th iteration of the algorithm is in fact optimal for our objective function.

We begin with introducing a surrogate function
\begin{align}\label{sur}
G(\beta ,\gamma )=& \sum_{t=1}^{k}\frac{\left\Vert Y^{(t)}-\bX^{(t)}\beta ^{(t)}\right\Vert}{\sqrt{n^{(0)}}}+\frac{1}{2}\sum_{t=1}^{k}\frac{1}{\sqrt{n^{(0)}}\left\Vert Y^{(t)}-\bX^{(t)}\beta ^{(t)}\right\Vert} \left \| \gamma -\beta  \right \|^2  +\lambda \sum_{l=1}^{p}\left\Vert \gamma_{(l)}\right\Vert \notag\\  & +\sum_{t=1}^{k}\frac{1}{\sqrt{n^{(0)}}\left\Vert Y^{(t)}-\bX^{(t)}\beta ^{(t)}\right\Vert}(\gamma ^{(t)}-\beta^{(t)})' (\bX^{(t)})'(\bX^{(t)}\beta^{(t)}-Y^{(t)}),
\end{align}
where $\gamma^{(t)}$ and $\gamma_{(l)}$ are the subvectors of $\gamma$ defined similarly as $\beta^{(t)}$ and $\beta_{(l)}$, respectively. It is easy to see that
\begin{equation} \label{neweq003}
F(\beta) = G(\beta, \beta).
\end{equation}
Denote by $ R^{(t)}= (n^{(0)})^{-1/2} (\bX^{(t)})'(\bX^{(t)}\beta^{(t)}-Y^{(t)})/\Vert Y^{(t)}-\bX^{(t)}\beta ^{(t)}\Vert$ and $R = ((R^{(1)})', \cdots, (R^{(k)})')'$. Then we can rewrite the last term in \eqref{sur} as
$$\sum_{t=1}^{k}\frac{1}{\sqrt{n^{(0)}}\left\Vert Y^{(t)}-\bX^{(t)}\beta ^{(t)}\right\Vert}(\gamma ^{(t)}-\beta^{(t)})' (\bX^{(t)})'(\bX^{(t)}\beta^{(t)}-Y^{(t)})=(\gamma-\beta)'R.$$
Thus given a fixed $\beta$, minimizing the above surrogate function $G$ over $\gamma$ is equivalent to minimizing the following objective function formed by the last three terms of $G$ in \eqref{sur} with respect to $\gamma$
\[
\frac{1}{2}A \left \| \gamma -\beta  \right \|^2  +\lambda \sum_{l=1}^{p}\left\Vert \gamma_{(l)}\right\Vert + (\gamma-\beta)'R, 
\]
where we denote by $A = \sum_{t=1}^{k}(n^{(0)})^{-1/2}\Vert Y^{(t)}-\bX^{(t)}\beta ^{(t)}\Vert^{-1}$. The optimization problem above is further equivalent to minimizing the following objective function with respect to $\gamma$
\begin{equation}\label{ming}
\frac{1}{2}\left \| \gamma-\beta +\frac{R}{A}\right \|^2+ \frac{\lambda}{A} \sum_{l=1}^{p}\left\Vert \gamma_{(l)}\right\Vert.
\end{equation}
Combining the above results yields that for any given $\beta$, the minimizer of the objective function $G(\beta, \gamma)$ defined in \eqref{sur} with respect to $\gamma$ is the same as that of the objective function given in  \eqref{ming}.

We now set $\beta = \beta(m)$ and correspondingly define the vector $R(m)$ and the scalar $A(m)$ similarly as $R$ and $A$, respectively, with $\beta(m)$ in place of $\beta$. We update $\beta(m+1)$ as the minimizer of the objective function \eqref{ming} with respect to $\gamma$ given $\beta = \beta(m)$. Thus  $\beta(m+1)$ is also the minimizer of $G(\beta(m),\gamma)$ with respect to $\gamma$. Since the optimization problem in \eqref{ming} is separable, it can be rewritten in the following form
\begin{align}\label{eq:separate}
\sum_{l=1}^p \left\{ \frac{1}{2}\Big\|\beta_{(l)} - \frac{R_{(l)}}{A} - \gamma_{(l)}\Big\|^2 + \frac{\lambda}{A} \left\|\gamma_{(l)}\right\|\right\}
.\end{align}
In view of \eqref{eq:separate}, the optimization problem in \eqref{ming} can be solved componentwise by minimizing each of the $p$ summands above. In particular, the resulting solution admits an explicit form and we obtain by Lemmas 1 and 2 in \cite{she2012} that $\beta(m+1)$ is given by
\begin{equation}\label{upd}
\beta (m+1)_{(l)}=\overrightarrow{ \Theta }\left(\beta (m)_{(l)}-\frac{R(m)_{(l)}}{A(m)};\frac{\lambda }{A(m)}\right),  \qquad l \in[p],
\end{equation}
where $R(m)_{(l)}$ is a subvector of $R(m)$ defined in a similar way to $\beta_{(l)}$ as a subvector of $\beta$ and $\overrightarrow{ \Theta }(\cdot; \cdot)$ is the multivariate soft-thresholding operator introduced in (\ref{neweq001}). Thus, it follows from (\ref{neweq003}) that
\begin{equation} \label{neweq002}
G(\beta(m),\beta(m+1))\leq G(\beta(m),\beta(m))=F(\beta(m)).
\end{equation}

Let us consider the function $(n^{(0)})^{-1/2}\left\Vert Y^{(t)}-\bX^{(t)}\gamma ^{(t)}\right\Vert$ with respect to $\gamma ^{(t)}$. Some routine calculations show that its gradient is given by
\begin{equation} \label{neweq005}
(n^{(0)})^{-1/2}\left\Vert Y^{(t)}-\bX^{(t)}\gamma ^{(t)}\right\Vert^{-1} (\bX^{(t)})' (\bX^{(t)}\gamma ^{(t)} - Y^{(t)})
\end{equation}
and its Hessian matrix is
\begin{align} \label{neweq004}
(n^{(0)})^{-1/2} & \left\Vert Y^{(t)}-\bX^{(t)}\gamma ^{(t)}\right\Vert^{-1} (\bX^{(t)})' \bX^{(t)} - (n^{(0)})^{-1/2}\left\Vert Y^{(t)}-\bX^{(t)}\gamma ^{(t)}\right\Vert^{-3} \nonumber \\
& \quad \cdot (\bX^{(t)})' (\bX^{(t)}\gamma ^{(t)} - Y^{(t)}) (\bX^{(t)}\gamma ^{(t)} - Y^{(t)})'\bX^{(t)} \nonumber \\
& \leq (n^{(0)})^{-1/2} \left\Vert Y^{(t)}-\bX^{(t)}\gamma ^{(t)}\right\Vert^{-1} (\bX^{(t)})' \bX^{(t)},
\end{align}
where $\leq$ means that the difference between the matrices on the right-hand side and the left-hand side of the inequality is positive semidefinite. Thus for any given $\beta$ and $\gamma$, an application of the Taylor expansion of the function $(n^{(0)})^{-1/2}\left\Vert Y^{(t)}-\bX^{(t)}\gamma ^{(t)}\right\Vert$ at the point $\beta^{(t)}$ to the first order with the Lagrange remainder, together with (\ref{neweq005})--(\ref{neweq004}), results in
\begin{align}\label{tay}
\notag \sum_{t=1}^{k}\frac{\left\Vert Y^{(t)}-\bX^{(t)}\beta^{(t)}\right\Vert}{\sqrt{n^{(0)}}}&+\sum_{t=1}^{k}\frac{1}{\sqrt{n^{(0)}} \left\| \bX^{(t)}\beta^{(t)}-Y^{(t)} \right \|}(\gamma ^{(t)}-\beta^{(t)})' (\bX^{(t)})'(\bX^{(t)}\beta^{(t)}-Y^{(t)})\\
\notag&\qquad -\sum_{t=1}^{k}\frac{\left\Vert Y^{(t)}-\bX^{(t)}\gamma ^{(t)}\right\Vert}{\sqrt{n^{(0)}}} \\
&\geq \sum_{t=1}^{k}-\frac{(\gamma ^{(t)}-\beta^{(t)})' (\bX^{(t)})'\bX^{(t)}(\gamma ^{(t)}-\beta^{(t)})}{2\sqrt{n^{(0)}}\left \| \bX^{(t)}\xi^{(t)}-Y^{(t)} \right \|},
\end{align}
where $\xi^{(t)}$ lies on the line segment connecting $\beta^{(t)}$ and $\gamma^{(t)}$ for each $t\in[k]$.

For now set $\beta = \beta(m)$ and $\gamma = \beta(m+1)$. Then it follows from (\ref{sur})  and (\ref{tay}) that
\begin{align}\label{ine}
\notag F(\beta(m)) & - F(\beta(m+1))
\geq G(\beta(m),\beta(m+1))-F(\beta(m+1))\\ \notag
&\geq \sum_{t=1}^{k}-\frac{(\beta(m+1)^{(t)}-\beta(m)^{(t)})' (\bX^{(t)})'\bX^{(t)}(\beta(m+1) ^{(t)}-\beta(m)^{(t)})}{2\sqrt{n^{(0)}}\left \| \bX^{(t)}\xi^{(t)}-Y^{(t)} \right \|}\\ \notag
&\qquad +\frac{1}{2}A(m)\left \| \beta(m+1)-\beta(m) \right \|^2 \\ \notag
&=\sum_{t=1}^{k} (\beta(m+1)^{(t)}-\beta(m)^{(t)})'\left(\frac{A(m)}{2}I-\frac{(\bX^{(t)})'\bX^{(t)}}{2\sqrt{n^{(0)}}\left \| \bX^{(t)}\xi^{(t)}-Y^{(t)}\right \|}\right) \\
& \qquad \cdot (\beta(m+1)^{(t)}-\beta(m)^{(t)} ) \nonumber\\ 
&\geq \sum_{t=1}^{k}\frac{1}{2\sqrt{n^{(0)}}} \left(\frac{1}{\left \| \bX^{(t)}\beta(m)^{(t)}-Y^{(t)}\right \|}-\frac{\left \| \bX^{(t)} \right \|_{\ell_2}^2}{2\left \| \bX^{(t)}\xi^{(t)}-Y^{(t)}\right \|}\right) \nonumber \\
& \qquad \cdot
\left \| \beta(m+1)^{(t)}-\beta(m)^{(t)} \right \|^2,
\end{align}
where $I$ stands for the identity matrix and $\left \| \bX \right \|_{\ell_2}$ denotes the spectral norm of matrix $\bX$.

To show the descent property of our algorithm and thus the convergence of the sequence $\beta(m)$ due to the nonnegativity of the objective function $F(\beta)$ in (\ref{neweq006}), we need to prove that the right-hand side of (\ref{ine}) is positive. At the initial step $m=0$, it is easy to see that this can be achieved by picking a large enough scalar $K_0 > 0$ in the scaling step \eqref{eq: scale} as long as $\| \bX^{(t)}\xi^{(t)}-Y^{(t)} \| \neq 0$. This fact and the regularity condition assumed in Theorem \ref{com theorem} can guarantee that $F(\beta(m))$ is monotonically decreasing. To see this, set $B_0=(n^{(0)})^{1/2}F(\beta(0))$ and recall that $\| \bX^{(t)}\xi^{(t)}-Y^{(t)} \|>c_0$ by assumption. It suffices to show that $\| \bX^{(t)}  \|_{\ell_2}^2<c_0/B_0$. From the definition of $B_0$, this claim is equivalent to
\begin{equation} \label{neweq007}
\| \bX^{(t)}  \|_{\ell_2}^2F(\beta(0))<(n^{(0)})^{-1/2}c_0.
\end{equation}
In light of the rescaling step for $Y^{(t)}$, $\bX^{(t)}$, and $\lambda$ in (\ref{eq: scale}), we see that the term on the left-hand side of (\ref{neweq007}) scales down with a factor of $K_0^{-3}$. This entails that as long as $K_0 >0$ is chosen large enough, inequality (\ref{neweq007}) can be easily satisfied and thus the above claim $\| \bX^{(t)}  \|_{\ell_2}^2<c_0/B_0$ holds.

Moreover, we can use the induction later to prove
\begin{equation}
\label{eq: inducation}
\left \| \bX^{(t)}\beta(m)^{(t)}-Y^{(t)}\right \| \leq B_0
\quad \text{ and } \quad  F(\beta(m))\leq F(\beta(0))
\end{equation}
for all $t$ and $m$. Combining the above inequalities (\ref{eq: inducation}), $\| \bX^{(t)}  \|_{\ell_2}^2<c_0/B_0$, and $\| \bX^{(t)}\xi^{(t)}-Y^{(t)} \|>c_0$ results in
\begin{align*}
\frac{1}{\left \| \bX^{(t)}\beta(m)^{(t)}-Y^{(t)}\right \|}-\frac{\left \| \bX^{(t)} \right \|_{\ell_2}^2}{2\left \| \bX^{(t)}\xi^{(t)}-Y^{(t)}\right \|} \geq \frac{1}{2B_0},
\end{align*}
which along with \eqref{ine} entails that
\begin{equation} \label{neweq008}
F(\beta(m)) - F(\beta(m+1))
\geq \frac{1}{4} (n^{(0)})^{-1/2} B_0^{-1} \sum_{t=1}^{k} \left \| \beta(m+1)^{(t)}-\beta(m)^{(t)} \right \|^2.
\end{equation}
This shows that $F(\beta(m)) \geq F(\beta(m+1))$. Since $F(\beta(m))$ is always bounded from below by zero, it follows that $\lim_{m\rightarrow \infty}F(\beta(m))$ exists and $\lim_{m\rightarrow \infty} |F(\beta(m+1)) - F(\beta(m))| = 0$. Thus in view of \eqref{neweq008}, we have
\begin{equation}
 \label{eq: beta-conv}
  \lim_{m\rightarrow \infty}\left \| \beta(m+1)-\beta(m) \right \| = 0.
\end{equation}
Observe that for each $m \geq 0$,
$$\left \|\beta(m)\right \| \leq \sum_{l=1}^{p}\left\Vert \beta(m)_{(l)}\right\Vert \leq \frac{F(\beta(m))}{\lambda}\leq \frac{F(\beta(0))}{\lambda},$$
which means that all $\beta(m)$ lie in a compact subset of $ \mathbb{R}^{kp}$. This fact entails that the sequence $\beta(m)$ has at least one point of convergence. Furthermore, \eqref{eq: beta-conv} ensures that $\beta(m)$ has a unique limit point $\beta^{*}$, which is a fixed point of the soft-thresholding rule given in \eqref{upd}.

It now remains to establish the results in \eqref{eq: inducation} using induction. When $m=0$, it is easy to verify that $ \| \bX^{(t)}\beta(m)^{(t)}-Y^{(t)} \| \leq B_0$ and $F(\beta(m))\leq F(\beta(0))$. Let us assume that the inequalities $ \| \bX^{(t)}\beta(m)^{(t)}-Y^{(t)} \| \leq B_0$ and $F(\beta(m))\leq F(\beta(0))$ in \eqref{eq: inducation} hold for all $ m\leq T$. Then it follows that $$\frac{1}{\left \| \bX^{(t)}\beta(T)^{(t)}-Y^{(t)}\right \|}-\frac{\left \| \bX^{(t)} \right \|_{\ell_2}^2}{2\left \| \bX^{(t)}\xi^{(t)}-Y^{(t)}\right \|}\geq\frac{1}{2B_0},$$
which together with \eqref{ine} leads to
\[ F(\beta(T+1))\leq F(\beta(T))\leq F(\beta(0)). \]
We can also obtain $ \| \bX^{(t)}\beta(T+1)^{(t)}-Y^{(t)} \| \leq (n^{(0)})^{1/2}F(\beta(T+1))\leq (n^{(0)})^{1/2}F(\beta(0))=B_0 $. Thus \eqref{eq: inducation} also holds for $m=T+1$. This completes the proof of \eqref{eq: inducation} for all $m$ and $t$ and also concludes the proof of the first step.

\medskip

{\bf Step 2: Global optimality.} To conclude the proof, we need to show that the unique point of convergence $\beta^*$ of our algorithm established in Step 1 is the global optimum of the HGSL optimization problem \eqref{eq: group square-root Lasso1}. Since $F(\beta)$ defined in (\ref{neweq006}) is the sum of two convex functions of $\beta$, it follows that $F(\beta)$ is also a convex function. Thus a vector $\beta$ is a global minimizer of the objective function $F(\cdot)$ if and only if it satisfies the  Karush-Kuhn-Tucker (KKT) conditions
\begin{align}
\label{kkt}& \frac{((\bX^{(t)})'(\bX^{(t)}\beta^{(t)}-Y^{(t)}))_{l}}{\sqrt{n^{(0)}}\left \| \bX^{(t)}\beta^{(t)}-Y^{(t)} \right \|}=-\lambda \frac{\beta _{l}^{(t)}}{\left \|\ \beta _{(l)} \right \|} \qquad \text{ for } \beta _{(l)} \neq \mathbf{0},\\
&
\label{kktt} \frac{\left|((\bX^{(t)})'(\bX^{(t)}\beta^{(t)}-Y^{(t)}))_{l}\right|}{\sqrt{n^{(0)}}\left \| \bX^{(t)}\beta^{(t)}-Y^{(t)} \right \|}\leq\lambda \qquad \text{ for } \beta _{(l)} = \mathbf{0},
\end{align}
where the subscript $l$ in both expressions represents the $l$th component of a vector.

Recall that we have shown in Step 1 that $\beta^{*}$ is the fixed point of the soft-thresholding rule in \eqref{upd}, that is,
$$
\beta ^{*}_{(l)} =\overrightarrow{ \Theta }\Big(\beta ^{*}_{(l)} -\frac{R^{*}_{(l)}}{A^{*}};\frac{\lambda }{A^{*}}\Big), \qquad   l\in[p],
$$
where $R^*_{(l)}$ and $A^*$ are defined similarly as $R(m)_{(l)}$ and $A(m)$ in \eqref{upd} with $\beta(m)$ replaced by $\beta^*$. Let us first consider the case when $\beta_{(l)}^* =\mathbf{0}$. Then by the definition of the soft-thresholding rule, we have $ \|R^{*}_{(l)}/A^{*}\| \leq \lambda /A^{*}$,
which entails that $\|R^{*}_{(l)}\| \leq \lambda $. Thus it holds that
\begin{equation}\label{kkt-beta-star1}
\frac{\left|((\bX^{(t)})'(\bX^{(t)}\beta^{*(t)}-Y^{(t)}))_{l}\right|}{\sqrt{n^{(0)}}\left \| \bX^{(t)}\beta^{*(t)}-Y^{(t)} \right \|}=|R^{*(t)}_{l}|\leq \|R^{*}_{(l)}\| \leq \lambda
\end{equation}
for $\beta_{(l)}^* =\mathbf{0}$, which verifies the second KKT condition \eqref{kktt} for the fixed point $\beta^*$.

We next consider the case when $\beta_{(l)}^* \neq \mathbf{0}$. It follows from the soft-thresholding rule that
\begin{equation} \label{kkt1}
\beta ^{*}_{(l)} =\frac{\left\|\beta ^{*}_{(l)} -\frac{R^{*}_{(l)}}{A^{*}}\right\|-\frac{\lambda }{A^{*}}}{\left\|\beta ^{*}_{(l)} -\frac{R^{*}_{(l)}}{A^{*}}\right\|}\left(\beta ^{*}_{(l)} -\frac{R^{*}_{(l)}}{A^{*}}\right).
\end{equation}
Taking the $\ell_2$ norm on both sides of the above equation leads to  $\|\beta ^{*}_{(l)}\| =\|\beta ^{*}_{(l)} -R^{*}_{(l)}/A^{*}\|-\lambda /A^{*}$. Moreover, equation \eqref{kkt1} can be rewritten as
$$-\frac{\lambda }{A^{*}}\left(\beta ^{*}_{(l)} -\frac{R^{*}_{(l)}}{A^{*}}\right)=\frac{R^{*}_{(l)}}{A^{*}}\left\|\beta ^{*}_{(l)} -\frac{R^{*}_{(l)}}{A^{*}}\right\|,$$
which along with the above fact results in
\begin{equation} \label{neweq009}
\lambda\beta ^{*}_{(l)}= R^{*}_{(l)}\left(\left\|\beta ^{*}_{(l)} -\frac{R^{*}_{(l)}}{A^{*}}\right\|-\frac{\lambda}{A^{*}}\right)= R^{*}_{(l)}\left\|\beta ^{*}_{(l)}\right\|.
\end{equation}
The representation in (\ref{neweq009}) further entails that
\begin{equation}\label{kkt-beta-star2}
R^{*(t)}_{l}=\frac{\big((\bX^{(t)})'(\bX^{(t)}\beta^{*(t)}-Y^{(t)})\big)_{l}}{\sqrt{n^{(0)}}\left \| \bX^{(t)}\beta^{*(t)}-Y^{(t)} \right \|}=-\lambda \frac{\beta _{l}^{(t)}}{\left \|\ \beta _{(l)} \right \|}
\end{equation}
for $\beta_{(l)}^* \neq \mathbf{0}$, which establishes the first KKT condition \eqref{kkt} for the fixed point $\beta^*$. Combining \eqref{kkt-beta-star1} and \eqref{kkt-beta-star2}, we conclude that $\beta^{(*)}$ is indeed a global minimizer of the HGSL optimization problem \eqref{eq: group square-root Lasso1}, which completes the proof of Theorem \ref{com theorem}.

\subsection{Proof of Proposition \ref{prop:suppreco}} \label{SecA.7}
The support recovery property of our THP estimator 
$\mathcal{\hat{E}}$ given in (\ref{neweq015}) follows from the proofs of Theorems \ref{thm:l2 testing} and \ref{thm:lower rate} (1) in Sections \ref{SecA.1} and \ref{SecA.3.1}, in view of the conditions of Proposition \ref{cor:l2 testing} and the assumption that the minimum signal strength $\min_{(a,b)\in \mathcal{E}%
}\Vert \omega _{a,b}^{0}\Vert$ is above $C\sqrt{[(k\log p)^{1/2}+\log p]/n^{(0)}}$. Specifically, we need a 
refined technical analysis in the proof of Theorem \ref{thm:lower rate} (1) in Section \ref{SecA.3.1} through replacing Chebyshev's inequality used in the third step by an accurate coupling inequality such as Proposition KMT 
in \cite{mason2012quantile}, which was also used in Theorem 2 (iii) of \cite{ren2013asymptotic} for support recovery in the setting of a single Gaussian graphical model. We omit the details here for simplicity.

\section{Key lemmas and their proofs} \label{SecB}

\subsection{Lemma \ref{lem:diagnoal} and its proof} \label{SecB.1}

\medskip

\begin{lemma}
\label{lem:diagnoal} Assume that Conditions \ref{CondA1}--\ref{CondA2} hold and $\max \{\log p,\log k\}=o(n^{(0)})$. Let $\hat{C}%
_{j}^{0}=( \hat{C}_{j}^{(1)\prime },\cdots ,\\\hat{C}_{j}^{(k)\prime
}) ^{\prime }$ be any estimator satisfying working assumptions (\ref%
{eq:assumption1})--(\ref{eq:assumption3}) for a fixed $j\in \lbrack p]$. Then there exists some positive
constant $C$ depending on constants $M$,$\delta $, $C_{1}$, and $C_{3}$ such that
\begin{eqnarray*}
\mathbb{P}\left( \frac{1}{k}\sum_{t=1}^{k}\left\vert \left( \hat{\omega}%
_{j,j}^{(t)}\right) ^{-1}-\frac{1}{n^{(t)}}\sum_{i=1}^{n^{(t)}}\left(
E_{i,j}^{(t)}\right) ^{2}\right\vert \geq Cs\frac{1+(\log p)/k}{n^{(0)}}%
\right)  &\leq &3p^{1-\delta }, \\
\mathbb{P}\left( \frac{1}{k}\sum_{t=1}^{k}\left\vert \left( \hat{\omega}%
_{j,j}^{(t)}\right) ^{-1}-\left( \omega _{j,j}^{(t)}\right) ^{-1}\right\vert
\geq C\left( \sqrt{\frac{\log (k/\delta _{1})}{n^{(0)}}}+s\frac{1+(\log p)/k%
}{n^{(0)}}\right) \right)  &\leq &3p^{1-\delta }+\delta _{1}
\end{eqnarray*}%
as long as $\log (\delta _{1}^{-1})=o(n^{(0)})$. Moreover, whenever $\max \{%
\sqrt{\frac{\log (k/\delta _{1})}{n^{(0)}}},s\frac{\left( k+\log p\right) }{%
n^{(0)}}\}=o(1)$, there exists some positive constant $C^{\prime }$ depending on $M$,$%
\delta $, $C_{1}$, and $C_{3}$ such that%
\begin{eqnarray*}
\mathbb{P}\left( \frac{1}{k}\sum_{t=1}^{k}\left\vert \hat{\omega}%
_{j,j}^{(t)}-\left( \frac{1}{n^{(t)}}\sum_{i=1}^{n^{(t)}}\left(
E_{i,j}^{(t)}\right) ^{2}\right) ^{-1}\right\vert \geq C^{\prime }s\frac{%
\left( 1+(\log p)/k\right) }{n^{(0)}}\right)  &\leq &3p^{1-\delta }, \\
\mathbb{P}\left( \frac{1}{k}\sum_{t=1}^{k}\left\vert \hat{\omega}%
_{j,j}^{(t)}-\omega _{j,j}^{(t)}\right\vert \geq C^{\prime }\left( \sqrt{\frac{%
\log (k/\delta _{1})}{n^{(0)}}}+s\frac{\left( 1+(\log p)/k\right) }{n^{(0)}}%
\right) \right)  &\leq &3p^{1-\delta }+\delta _{1}, \\
\mathbb{P}\left( \max_{t\in \lbrack k]}\left\vert \hat{\omega}%
_{j,j}^{(t)}-\omega _{j,j}^{(t)}\right\vert \geq C^{\prime }\left( \sqrt{\frac{%
\log (k/\delta _{1})}{n^{(0)}}}+s\frac{\left( k+\log p\right) }{n^{(0)}}%
\right) \right)  &\leq &3p^{1-\delta }+\delta _{1}.
\end{eqnarray*}
\end{lemma}

\medskip

\textit{Proof.} Observe that $\frac{1}{n^{(t)}}\sum_{i=1}^{n^{(t)}}( \hat{E}_{i,j}^{(t)}) ^{2}=(
\hat{\omega}_{j,j}^{(t)}) ^{-1}$. For each $j\in \lbrack p]$, in view of  $\hat{E}%
_{i,j}^{(t)}=E_{i,j}^{(t)}+X_{i,-j}^{(t)\prime }( C_{j}^{(t)}-\hat{C}%
_{j}^{(t)})$ we deduce that
\begin{eqnarray}
\frac{1}{n^{(t)}}\sum_{i=1}^{n^{(t)}}\left( \hat{E}_{i,j}^{(t)}\right) ^{2}
&=&\frac{1}{n^{(t)}}\bigg\{\sum_{i=1}^{n^{(t)}}\left( E_{i,j}^{(t)}\right)
^{2}+2E_{\ast ,j}^{(t)\prime }\mathbf{X}_{\ast ,-j}^{(t)}(C_{j}^{(t)}-\hat{C}%
_{j}^{(t)})  \notag \\
&&+(C_{j}^{(t)}-\hat{C}_{j}^{(t)})^{\prime}\mathbf{X}_{\ast ,-j}^{(t)\prime }%
\mathbf{X}_{\ast ,-j}^{(t)}(C_{j}^{(t)}-\hat{C}_{j}^{(t)})\bigg\}.
\label{eq: decomp of noise est}
\end{eqnarray}%
Thus we have%
\begin{eqnarray}
&&\frac{1}{k}\sum_{t=1}^{k}\left\vert \left( \hat{\omega}_{j,j}^{(t)}\right)
^{-1}-\sum_{i=1}^{n^{(t)}}\left( E_{i,j}^{(t)}\right) ^{2}/n^{(t)}\right\vert
\notag \\
&\leq &\frac{1}{k}\sum_{t=1}^{k}\frac{1}{n^{(t)}}\left( 2\left\vert E_{\ast
,j}^{(t)\prime }\mathbf{X}_{\ast ,-j}^{(t)}(C_{j}^{(t)}-\hat{C}%
_{j}^{(t)})\right\vert +\left\Vert \mathbf{X}_{\ast ,-j}^{(t)}(C_{j}^{(t)}-%
\hat{C}_{j}^{(t)})\right\Vert ^{2}\right)  \label{eq:decomp of noise est 1}
\notag\\
&\equiv&T_{1}+T_{2}.
\end{eqnarray}
We will consider the above two terms $T_{1}$ and $T_{2}$ separately.

For the second term $T_{2}$, we can bound it by our working assumption (\ref{eq:assumption3}) as
\begin{equation}
T_{2}=\frac{1}{k}\sum_{t=1}^{k}\frac{1}{n^{(t)}}\left\Vert \mathbf{X}_{\ast
,-j}^{(t)}(C_{j}^{(t)}-\hat{C}_{j}^{(t)})\right\Vert ^{2}\leq C_{3}s\frac{%
1+(\log p)/k}{n^{(0)}}.  \label{eq:T1}
\end{equation}%
The first term $T_{1}$ can be bounded with probability at least $%
1-3p^{1-\delta }$ as
\begin{eqnarray}
T_{1} &\leq &\frac{2}{k}\sum_{l\neq j}\sum_{t=1}^{k}\left\vert \frac{E_{\ast
,j}^{(t)\prime }X_{\ast ,l}^{(t)}}{n^{(t)}}\right\vert \cdot \left\vert
C_{j,l}^{(t)}-\hat{C}_{j,l}^{(t)}\right\vert   \notag \\
&\leq &\sum_{l\neq j}\left( \frac{1}{k}\sum_{t=1}^{k}(\frac{E_{\ast
,j}^{(t)\prime }X_{\ast ,l}^{(t)}}{n^{(t)}})^{2}\right) ^{1/2}\left( \frac{1%
}{k}\sum_{t=1}^{k}(C_{j,l}^{(t)}-\hat{C}_{j,l}^{(t)})^{2}\right) ^{1/2}
\notag \\
&\leq &\max_{l\neq j}\left( \frac{1}{k}\sum_{t=1}^{k}(\frac{E_{\ast
,j}^{(t)\prime }X_{\ast ,l}^{(t)}}{n^{(t)}})^{2}\right) ^{1/2}\sum_{l\neq j}%
\frac{1}{\sqrt{k}}\left\Vert \Delta _{j(l)}\right\Vert   \notag \\
&\leq &c_{\delta }\left( \frac{1+(\log p)/k}{n^{(0)}}\right) ^{1/2}s\left(
\frac{1+(\log p)/k}{n^{(0)}}\right) ^{1/2},  \label{eq:T2}
\end{eqnarray}%
where the last inequality is due to working assumption (\ref{eq:assumption2}%
) and Lemma \ref{prop:Bern} in Section \ref{SecC} with $c_{\delta }$ some positive constant depending only on $\delta $,
$M$, and $C_{1}$. Thus we have shown the first desired result.

Let us further bound the difference between the oracle estimator $%
\sum_{i=1}^{n^{(t)}}( E_{i,j}^{(t)}) ^{2}/n^{(t)}$ and its mean $%
( \omega _{j,j}^{(t)}) ^{-1}$. Indeed, it holds that $%
\sum_{i=1}^{n^{(t)}}( E_{i,j}^{(t)}) ^{2}(\omega _{j,j}^{(t)})\sim
\chi ^{2}(n^{(t)})$. This representation entails that as long as $\log (\delta
_{1}^{-1})=o(n^{(0)})$, by Lemma \ref{prop:chi} and $n^{(0)}\leq n^{(t)}$ we have
\begin{equation}
\left\vert \frac{1}{n^{(t)}}\sum_{i=1}^{n^{(t)}}\left( E_{i,j}^{(t)}\right)
^{2}-1/\omega _{j,j}^{(t)}\right\vert =\frac{1}{n^{(t)}}\left\vert
\sum_{i=1}^{n^{(t)}}\left( \left( E_{i,j}^{(t)}\right) ^{2}-\mathbb{E}\left(
E_{i,j}^{(t)}\right) ^{2}\right) \right\vert \leq c_{M}\sqrt{\frac{\log
(k/\delta _{1})}{n^{(0)}}}  \label{eq:decomp of noise est 2}
\end{equation}%
with probability at least $1-\delta _{1}/k$, where $c_{M}$ is some positive  constant depending only on $M$. Combining inequalities (\ref{eq:decomp of noise est 1})--(\ref{eq:decomp of
noise est 2}) with the union bound argument, we obtain the second desired  result that with
probability at least $1-3p^{1-\delta }-\delta _{1},$
\begin{equation*}
\frac{1}{k}\sum_{t=1}^{k}\left\vert \left( \hat{\omega}_{j,j}^{(t)}\right)
^{-1}-\left( \omega _{j,j}^{(t)}\right) ^{-1}\right\vert \leq C\left( \sqrt{%
\frac{\log (k/\delta _{1})}{n^{(0)}}}+s\frac{1+(\log p)/k}{n^{(0)}}\right),
\end{equation*}%
where $C$ is some positive constant that depends on $M$, $\delta $, $C_{1}$, and $C_{3}$.

Note that whenever $\max \{\sqrt{\frac{\log (k/\delta _{1})}{n^{(0)}}},s%
\frac{\left( k+\log p\right) }{n^{(0)}}\}=o(1)$, it follows from inequalities (\ref{eq:decomp of noise est 1})--(\ref{eq:decomp of
noise est 2}) and the union bound argument that with probability at least $1-3p^{1-\delta }-\delta _{1}$,
\begin{equation}
\max_{t}\left\vert 1/\hat{\omega}_{j,j}^{(t)}-1/\omega_{j,j}^{(t)}\right\vert
\leq C\left( \sqrt{\frac{\log (k/\delta _{1})}{n^{(0)}}}+s\frac{k+\log p}{n^{(0)}}\right),  \label{eq:decomp of noise est 3}
\end{equation}
which is sufficiently small for large $n^{(0)}$. Consequently, we see that $\hat{\omega}_{j,j}^{(t)}$ is
uniformly bounded from above by some positive constant for all $t\in \lbrack k]$, since $%
\omega _{j,j}^{(t)}$ is bounded from above by $M$ by Condition \ref{CondA1}. Therefore, in light of $\vert \hat{\omega}%
_{j,j}^{(t)}-\omega _{j,j}^{(t)}\vert =\vert 1/\hat{\omega}%
_{j,j}^{(t)}-1/\omega _{j,j}^{(t)}\vert \omega _{j,j}^{(t)}\hat{\omega}%
_{j,j}^{(t)}$ the last three desired
inequalities follow from the first two established above and inequality (\ref{eq:decomp of noise est 3}), which concludes the proof.

\subsection{Lemma \ref{lem:lin comb} and its proof} \label{SecB.2}

\medskip

\begin{lemma}
\label{lem:lin comb} Assume that Conditions \ref{CondA1}--\ref{CondA2} hold, working assumptions (%
\ref{eq:assumption1})--(\ref{eq:assumption3}) are valid for $j=1,2$, and
$\max \{\log p,\log k\}=o(n^{(0)})$. Then there exists some positive constant $C$
depending only on constants $M,\delta ,C_{1},C_{2}$, and $C_{3}$ such that%
\begin{align}
\frac{1}{k}&\sum_{t=1}^{k}\left\vert T_{n,k,1,2}^{(t)}-J_{n,k,1,2}^{(t)}-%
\frac{1}{n^{(t)}}\sum_{i=1}^{n^{(t)}}\left( E_{i,1}^{(t)}E_{i,2}^{(t)}-%
\mathbb{E}E_{i,1}^{(t)}E_{i,2}^{(t)}\right) \right\vert   \notag \label{eq:Normal_a}
\\
&\leq C^{\prime \prime }\left( s\frac{1+(\log p)/k}{n^{(0)}}(1+\sqrt{ks%
\frac{1+(\log p)/k}{n^{(0)}}})\right)
\end{align}%
holds with probability at least $1-6p^{1-\delta }$.
\end{lemma}

\textit{Proof.} At a high level, the first term $\frac{1}{k}\sum_{t=1}^{k}\vert
\sum_{i=1}^{n^{(t)}}\hat{E}_{i,1}^{(t)}\hat{E}_{i,2}^{(t)}/n^{(t)}\vert $ in $%
T_{n,k,1,2}$ is constructed to approximate $\frac{1}{k}\sum_{t=1}^{k}%
\vert \sum_{i=1}^{n^{(t)}}E_{i,1}^{(t)}E_{i,2}^{(t)}/n^{(t)}%
\vert $, but some bias appears in the approximation. The remaining two terms $%
\sum_{i=1}^{n^{(t)}}( \hat{E}_{i,1}^{(t)}) ^{2}%
\hat{C}_{2,1}/n^{(t)}$ and $\sum_{i=1}^{n^{(t)}}( \hat{E}%
_{i,2}^{(t)}) ^{2}\hat{C}_{1,2}/n^{(t)}$ in each $T_{n,k,1,2}^{(t)}$ serve as
the remedy to correct the bias when the null $\omega _{1,2}^{0}=%
\mathbf{0}$ is true. In view of  $\hat{E}_{i,j}^{(t)}=E_{i,j}^{(t)}+X_{i,-j}^{(t)%
\prime }( C_{j}^{(t)}-\hat{C}_{j}^{(t)}) ,$ we can deduce
\begin{eqnarray}
&&\frac{1}{k}\sum_{t=1}^{k}\left\vert \frac{1}{n^{(t)}}\sum_{i=1}^{n^{(t)}}%
\hat{E}_{i,1}^{(t)}\hat{E}_{i,2}^{(t)}\right\vert   \notag \\
&=&\frac{1}{k}\sum_{t=1}^{k}\left\vert \frac{1}{n^{(t)}}E_{\ast
,1}^{(t)\prime }E_{\ast ,2}^{(t)}+\frac{1}{n^{(t)}}E_{\ast ,1}^{(t)\prime }%
\mathbf{X}_{\ast ,-2}^{(t)}(C_{2}^{(t)}-\hat{C}_{2}^{(t)})\right.   \notag \\
&&+\frac{1}{n^{(t)}}E_{\ast ,2}^{(t)\prime }\mathbf{X}_{\ast
,-1}^{(t)}(C_{1}^{(t)}-\hat{C}_{1}^{(t)})  \notag \\
&&\left. +\frac{1}{n^{(t)}}(C_{1}^{(t)}-\hat{C}_{1}^{(t)})^{T}\mathbf{X}%
_{\ast ,-1}^{(t)\prime }\mathbf{X}_{\ast ,-2}^{(t)}(C_{2}^{(t)}-\hat{C}%
_{2}^{(t)})\right\vert   \notag \\
&=&\frac{1}{k}\sum_{t=1}^{k}\left\vert
H_{1}^{(t)}+H_{2}^{(t)}+H_{3}^{(t)}+H_{4}^{(t)}\right\vert .
\label{eq: decomp of test stat}
\end{eqnarray}

The main term $H_{1}^{(t)}$ above enjoys the following property
\begin{eqnarray}
H_{1}^{(t)} &=&\frac{1}{n^{(t)}}%
\sum_{i=1}^{n^{(t)}}E_{i,1}^{(t)}E_{i,2}^{(t)}=\mathbb{E}%
E_{1,1}^{(t)}E_{1,2}^{(t)}+\frac{1}{n^{(t)}}\sum_{i=1}^{n^{(t)}}\left(
E_{i,1}^{(t)}E_{i,2}^{(t)}-\mathbb{E}E_{i,1}^{(t)}E_{i,2}^{(t)}\right)
\notag \\
&=&\frac{\omega _{1,2}^{(t)}}{\omega _{1,1}^{(t)}\omega _{2,2}^{(t)}}+\frac{1}{%
n^{(t)}}\sum_{i=1}^{n^{(t)}}\left( E_{i,1}^{(t)}E_{i,2}^{(t)}-\mathbb{E}%
E_{i,1}^{(t)}E_{i,2}^{(t)}\right) .  \label{eq:oracle average}
\end{eqnarray}
We can bound the last term $\sum_{t=1}^{k}\vert H_{4}^{(t)}\vert/k $ in (\ref{eq: decomp of test stat}) as
\begin{eqnarray*}
\frac{1}{k}\sum_{t=1}^{k}\left\vert H_{4}^{(t)}\right\vert  &\leq &\frac{1}{k%
}\sum_{t=1}^{k}\frac{1}{n^{(t)}}\left\Vert \mathbf{X}_{\ast
,-2}^{(t)}(C_{2}^{(t)}-\hat{C}_{2}^{(t)})\right\Vert \left\Vert \mathbf{X}%
_{\ast ,-1}^{(t)}(C_{1}^{(t)}-\hat{C}_{1}^{(t)})\right\Vert  \\
&\leq &\frac{1}{2k}\sum_{t=1}^{k}\frac{1}{n^{(t)}}\left( \left\Vert \mathbf{X%
}_{\ast ,-2}^{(t)}(C_{2}^{(t)}-\hat{C}_{2}^{(t)})\right\Vert ^{2}+\left\Vert
\mathbf{X}_{\ast ,-1}^{(t)}(C_{1}^{(t)}-\hat{C}_{1}^{(t)})\right\Vert
^{2}\right)  \\
&\leq &C_{3}s\frac{1+(\log p)/k}{n^{(0)}},
\end{eqnarray*}%
where the last inequality follows from our working assumption (\ref%
{eq:assumption3}).

The second term $H_{2}^{(t)}$ in (\ref{eq: decomp of test stat}) can be further decomposed as%
\begin{eqnarray} \label{neweq014}
H_{2}^{(t)} &=&\frac{1}{n^{(t)}}\left( E_{\ast ,1}^{(t)\prime }X_{\ast
,1}^{(t)}(C_{2,1}^{(t)}-\hat{C}_{2,1}^{(t)})+E_{\ast ,1}^{(t)\prime }\mathbf{%
X}_{\ast ,\{1,2\}^{c}}^{(t)}(C_{2,-1}^{(t)}-\hat{C}_{2,-1}^{(t)})\right)  \notag \\
&\equiv&H_{2,0}^{(t)}+H_{2,1}^{(t)}.
\end{eqnarray}%
We can bound $\sum_{t=1}^{k}\vert H_{2,1}^{(t)}\vert/k $
such that with probability at least $1-3p^{1-\delta }$,%
\begin{eqnarray*}
\frac{1}{k}\sum_{t=1}^{k}\left\vert H_{2,1}^{(t)}\right\vert  &\leq &\frac{1%
}{k}\sum_{j=3}^{p}\sum_{t=1}^{k}\left\vert \frac{E_{\ast ,1}^{(t)\prime
}X_{\ast ,j}^{(t)}}{n^{(t)}}\right\vert \cdot \left\vert C_{2,j}^{(t)}-\hat{C%
}_{2,j}^{(t)}\right\vert  \\
&\leq &\sum_{j=3}^{p}\left( \frac{1}{k}\sum_{t=1}^{k}(\frac{E_{\ast
,1}^{(t)\prime }X_{\ast ,j}^{(t)}}{n^{(t)}})^{2}\right) ^{1/2}\left( \frac{1%
}{k}\sum_{t=1}^{k}(C_{2,j}^{(t)}-\hat{C}_{2,j}^{(t)})^{2}\right) ^{1/2} \\
&\leq &\max_{j}\left( \frac{1}{k}\sum_{t=1}^{k}(\frac{E_{\ast ,1}^{(t)\prime
}X_{\ast ,j}^{(t)}}{n^{(t)}})^{2}\right) ^{1/2}\sum_{j=3}^{p}\frac{1}{\sqrt{k%
}}\left\Vert \Delta _{2(j)}\right\Vert  \\
&\leq &C\left( \frac{1+(\log p)/k}{n^{(0)}}\right) ^{1/2}s\left( \frac{%
1+(\log p)/k}{n^{(0)}}\right) ^{1/2},
\end{eqnarray*}%
where the last inequality is due to working assumption (\ref{eq:assumption2}) and
Lemma \ref{prop:Bern}. Observe that similar decomposition, notation, and analysis
apply to term $H_{3}^{(t)}$ as well. Hence, it holds that with probability at least $%
1-3p^{1-\delta },$%
\begin{equation}
\frac{1}{k}\sum_{t=1}^{k}\left\vert \frac{1}{n^{(t)}}\sum_{i=1}^{n^{(t)}}%
\hat{E}_{i,1}^{(t)}\hat{E}_{i,2}^{(t)}-\left(
H_{1}^{(t)}+H_{2,0}^{(t)}+H_{3,0}^{(t)}\right) \right\vert \leq C\left(
\frac{s}{n^{(0)}}(1+(\log p)/k)\right) \mbox{\rm .}
\label{eq: test decomp 1}
\end{equation}

Let us decompose term $H_{2,0}^{(t)}$ in (\ref{neweq014}) as
\begin{eqnarray}
H_{2,0}^{(t)} &=&\frac{1}{n^{(t)}}\Big\{\hat{E}_{\ast ,1}^{(t)\prime }\hat{E}%
_{\ast ,1}^{(t)}(C_{2,1}^{(t)}-\hat{C}_{2,1}^{(t)})+%
\sum_{i=1}^{n^{(t)}}E_{i,1}^{(t)}X_{i,-1}^{(t)\prime
}C_{1}^{(t)}(C_{2,1}^{(t)}-\hat{C}_{2,1}^{(t)}) \notag \\
&&+(E_{\ast ,1}^{(t)\prime }E_{\ast ,1}^{(t)}-\hat{E}_{\ast ,1}^{(t)\prime }%
\hat{E}_{\ast ,1}^{(t)})(C_{2,1}^{(t)}-\hat{C}_{2,1}^{(t)})\Big\}  \notag \\
&\equiv&H_{2,0,0}^{(t)}+H_{2,0,1}^{(t)}+H_{2,0,2}^{(t)}.  \label{eq:H_2_0}
\end{eqnarray}%
Now we control the two terms $\sum_{t=1}^{k}\vert
H_{2,0,1}^{(t)}\vert/k $ and $\sum_{t=1}^{k}\vert
H_{2,0,2}^{(t)}\vert/k $ separately, and leave $H_{2,0,0}^{(t)}$ as the
main term. By Lemma \ref{prop:Bern} and working assumption (\ref%
{eq:assumption2}), we obtain that with probability at least $1-3p^{-\delta }$%
,
\begin{eqnarray}
\frac{1}{k}\sum_{t=1}^{k}\left\vert H_{2,0,1}^{(t)}\right\vert  &\leq
&\left( \frac{1}{k}\sum_{t=1}^{k}(\frac{E_{\ast ,1}^{(t)\prime }\mathbf{X}%
_{\ast ,-1}^{(t)}C_{1}^{(t)}}{n^{(t)}})^{2}\right) ^{1/2}\left( \frac{1}{k}%
\sum_{t=1}^{k}(C_{2,1}^{(t)}-\hat{C}_{2,1}^{(t)})^{2}\right) ^{1/2}  \notag
\\
&\leq &C\left( \frac{1+(\log p)/k}{n^{(0)}}\right) ^{1/2}\left( s\frac{%
1+(\log p)/k}{n^{(0)}}\right) ^{1/2}.  \label{eq:H_2_0_1}
\end{eqnarray}%

As for the term $H_{2,0,2}^{(t)}$ in (\ref{eq:H_2_0}), we can show that with probability at least $%
1-3p^{1-\delta },$%
\begin{eqnarray}
\frac{1}{k}\sum_{t=1}^{k}\left\vert H_{2,0,2}^{(t)}\right\vert  &=&\frac{1}{k%
}\sum_{t=1}^{k}\left\vert (E_{\ast ,1}^{(t)\prime }E_{\ast ,1}^{(t)}-\hat{E}%
_{\ast ,1}^{(t)\prime }\hat{E}_{\ast ,1}^{(t)})(C_{2,1}^{(t)}-\hat{C}%
_{2,1}^{(t)})\right\vert   \notag \\
&\leq &\frac{1}{k}\sum_{t=1}^{k}\left\vert \frac{1}{n^{(t)}}%
(\sum_{i=1}^{n^{(t)}}\left( \hat{E}_{i,1}^{(t)}\right)
^{2}-\sum_{i=1}^{n}\left( E_{i,1}^{(t)}\right) ^{2})\right\vert
\max_{t}\left\vert C_{2,1}^{(t)}-\hat{C}_{2,1}^{(t)}\right\vert   \notag \\
&\leq &Cs\frac{1+(\log p)/k}{n^{(0)}}\cdot \max_{t}\left\Vert \Delta
_{1(t)}\right\Vert   \notag \\
&\leq &Cs\frac{1+(\log p)/k}{n^{(0)}}\cdot \left( ks\frac{1+(\log p)/k}{%
n^{(0)}}\right) ^{1/2},  \label{eq:H_2_0_2}
\end{eqnarray}%
where the second inequality follows from expressions (\ref{eq: decomp of noise
est})--(\ref{eq:T2}) in the earlier proof of Lemma \ref{lem:diagnoal} in Section \ref{SecB.1} and the last
inequality follows from our working assumption (\ref{eq:assumption1}). Note that similar decomposition, notation, and analysis also apply to term $H_{3,0}^{(t)}$. Thus combining the above expressions (\ref{eq: test decomp 1})--(\ref{eq:H_2_0_2}) yields that
with probability at least $1-3p^{-\delta }-3p^{1-\delta }$,
\begin{eqnarray}
&&\frac{1}{k}\sum_{t=1}^{k}\left\vert \frac{1}{n^{(t)}}\sum_{i=1}^{n^{(t)}}%
\hat{E}_{i,1}^{(t)}\hat{E}_{i,2}^{(t)}-\left(
H_{1}^{(t)}+H_{2,0,0}^{(t)}+H_{3,0,0}^{(t)}\right) \right\vert   \notag \\
&\leq &C\left( \frac{s}{n^{(0)}}(1+(\log p)/k)\right) \left(1+(ks\frac{1+(\log
p)/k}{n^{(0)}})^{1/2}\right).  \label{eq:final decom of inner}
\end{eqnarray}

We finally correct the bias in $H_{2,0,0}^{(t)}$ and $H_{3,0,0}^{(t)}$ induced from $%
\hat{C}_{2,1}$. To this end, we take the sum of $\hat{E}_{\ast ,1}^{(t)\prime }\hat{E}_{\ast ,2}^{(t)}/n^{(t)}$ and two terms $\hat{E}_{\ast ,1}^{(t)\prime }\hat{E}_{\ast ,1}^{(t)}\hat{C}%
_{2,1}/n^{(t)}$, $\hat{E}_{\ast ,1}^{(t)\prime }\hat{E}_{\ast
,1}^{(t)}\hat{C}_{1,2}/n^{(t)}$ out of $H_{2,0,0}^{(t)}$ and $H_{3,0,0}^{(t)}$ as
the statistic $T_{n,k,1,2}^{(t)}$. The remaining terms in $H_{2,0,0}^{(t)}$
and $H_{3,0,0}^{(t)}$ together with the first term of decomposition of $%
H_{1}^{(t)}$ in (\ref{eq:oracle average}) form $J_{n,k,1,2}^{(t)}$ defined
in (\ref{eq:J(t)}), in light of $C_{2,1}^{(t)}=-\omega _{1,2}^{(t)}/\omega
_{2,2}^{(t)}$ and $C_{1,2}^{(t)}=-\omega _{1,2}^{(t)}/\omega _{1,1}^{(t)}$.
Therefore, the desired result follows from (\ref{eq:final decom of inner}),
that is, with probability at least $1-3p^{-\delta }-3p^{1-\delta },$%
\begin{eqnarray*}
&&\frac{1}{k}\sum_{t=1}^{k}\left\vert T_{n,k,1,2}^{(t)}-J_{n,k,1,2}^{(t)}-%
\frac{1}{n^{(t)}}\sum_{i=1}^{n^{(t)}}\left( E_{i,1}^{(t)}E_{i,2}^{(t)}-%
\mathbb{E}E_{i,1}^{(t)}E_{i,2}^{(t)}\right) \right\vert  \\
&\leq &C^{\prime \prime }\left( \frac{s}{n^{(0)}}(1+(\log p)/k)\right) \left(1+(ks%
\frac{1+(\log p)/k}{n^{(0)}})^{1/2}\right)
\end{eqnarray*}%
with $C^{\prime
\prime }$ some positive constant. Keeping track of all relevant constants, we see that the positive constant $C^{\prime
\prime }$ depends only on $M,\delta ,C_{1},C_{2}$, and $C_{3}$, which completes the proof.

\subsection{Lemma \ref{lem:lower l2} and its proof} \label{SecB.3}

\medskip

\begin{lemma}
\label{lem:lower l2} With $\mathcal{G}$ and $\Omega _{0}^{0}$ chosen as
in (\ref{eq:ConstructionG_l2}) and (\ref{neweq010}), we have $\Vert \mathbb{P}_{0}\wedge \mathbb{\bar{P}}\Vert
>1-\frac{1}{2}(\beta -\alpha )$ with some sufficiently small constant $\tau >0$
depending only on $\beta -\alpha $.
\end{lemma}

\textit{Proof.} A similar argument to that used in the later proof of Lemma \ref{lem:lower sample} in Section \ref{SecB.4} (see
inequality (\ref{eq:proof lower 2})) entails that it is sufficient to show that the
$\chi ^{2}$ divergence between $\mathbb{P}_{0}$ and $\mathbb{\bar{P}}$ is
small enough, that is,
\[ \Delta =\int \left( \frac{1}{m}\sum_{h=1}^{m}f_{h}%
\right) ^{2}/f_{0}-1=\sum_{h_{1},h_{2}=1}^{m}( \int (%
\frac{f_{h_{1}}f_{h_{2}}}{f_{0}})-1)/(m)^{2} <(\beta -\alpha )^{2}. \]
Recall that $g_{h}^{(t)}$ denotes the density of $N(0,( \Omega
_{h}^{(t)}) ^{-1})$ for $h=0,\cdots ,m$. By our construction
of $\Omega _{0}^{0}$ and $\Omega _{1}^{0}$, together with the $\chi ^{2}$
divergence of two Gaussian distributions in (\ref{eq:Chi dist
Gaussian}), we can deduce that for any $h_{1},h_{2}\in \lbrack
m] $,
\begin{eqnarray*}
\int \frac{f_{h_{1}}f_{h_{2}}}{f_{0}} &=&\left( \int
\prod\nolimits_{t=1}^{h}g_{h_{1}}^{(t)}g_{h_{2}}^{(t)}/g_{0}^{(t)}\right)
^{n^{(0)}}=\left( 1-1/n^{(0)}\right) ^{-\dot{J}(h_{1},h_{2})n^{(0)}} \\
&\leq &\left( 1+2/n^{(0)}\right) ^{\dot{J}(h_{1},h_{2})n^{(0)}}\leq \exp (2%
\dot{J}(h_{1},h_{2})),
\end{eqnarray*}%
where we have used $1/n^{(0)}<1/2$ in the second to last inequality and  $\dot{%
J}=\dot{J}(h_{1},h_{2})$ is the cardinality of $T_{h1}\cap T_{h_{2}}$ with  the index sets $T_{h_{i}}\subset \lbrack k]$ denoting those graphs with
non-identity precision matrices in (\ref{eq:ConstructionG_l2}) for $i=1,2$. In
other words, $\dot{J}(h_{1},h_{2})$ is the number of overlapping
non-identity precision matrices between two sets of $k$ precision matrices
indexed by $\Omega _{h_{1}}^{0}$ and $\Omega _{h_{2}}^{0}$. It is easy to
see that integer $\dot{J}=\dot{J}(h_{1},h_{2})\in \lbrack 0,\cdots ,\tau
\sqrt{k}]$.

Recall that $m=\binom{k}{\tau \sqrt{k}}$. Thus we have%
\begin{eqnarray*}
\Delta &=&\frac{1}{(m)^{2}}\sum_{0\leq j\leq \tau \sqrt{k}}\sum_{\dot{J}%
(h_{1},h_{2})=j}\left( \exp (2\dot{J}(h_{1},h_{2}))-1\right) \\
&\leq &\frac{1}{(m)^{2}}\sum_{1\leq j\leq \tau \sqrt{k}}\binom{k}{\tau \sqrt{%
k}}\binom{\tau \sqrt{k}}{j}\binom{k-j}{\tau \sqrt{k}-j}\exp (2j) \\
&=&\sum_{1\leq j\leq \tau \sqrt{k}}\binom{\tau \sqrt{k}}{j}\binom{k-j}{\tau
\sqrt{k}-j}/\binom{k}{\tau \sqrt{k}}\cdot \exp (2j) \\
&\leq &\sum_{1\leq j\leq \tau \sqrt{k}}\frac{1}{j!}\left( \frac{\tau
^{2}k\exp (2)}{k-\tau \sqrt{k}}\right) ^{j} \\
&\leq &\exp (\lambda )\mathbb{P}(Z>0)=\exp (\lambda )-1,
\end{eqnarray*}%
where in the last inequality we bounded the sum using a Poisson random
variable $Z$ with parameter $\lambda =\tau ^{2}k\exp (2)/(k-\tau \sqrt{k})$.
Finally, we can conclude the proof by picking a small enough constant $\tau $
depending on $\beta -\alpha $ to obtain $\Delta \leq (\beta -\alpha
)^{2}$.

\subsection{Lemma \ref{lem:lower sample} and its proof} \label{SecB.4}

\medskip

\begin{lemma}
\label{lem:lower sample} With $\mathcal{G}$ and $\Omega _{0}^{0}$
specified in (\ref{neweq01,1}) and (\ref{neweq012}), it holds that $\Vert \mathbb{P}_{0}\wedge \mathbb{\bar{P}}\Vert
>1-\frac{1}{2}(\beta -\alpha )$ with some sufficiently small constant $\tau >0$
depending only on $M_{1}$ and $\mu $.
\end{lemma}

\textit{Proof.} Recall that the densities of distributions $\mathbb{P}_{h}$ and $N(0,( \Omega
_{h}^{(1)}) ^{-1})$ are denoted as $f_{h}$ and $g_{h}$, respectively,
for each $0\leq h\leq m$. By Jensen's inequality we have
\[ \Vert \mathbb{P}%
_{0}\wedge \mathbb{\bar{P}}\Vert =\int ( f_{0}\wedge \bar{f}%
) \geq 1-\frac{1}{2}( \int \frac{\bar{f}^{2}}{f_{0}}-1)
^{1/2}=1-\sqrt{\Delta }/2. \]
Thus it suffices to show that the $\chi ^{2}$
divergence is small enough
\begin{equation}
\Delta =\int \frac{\left( \frac{1}{m}\sum_{h=1}^{m}f_{h}\right) ^{2}}{f_{0}}%
-1=\frac{1}{m^{2}}\sum_{h_{1},h_{2}=1}^{m}\left( \int (\frac{%
f_{h_{1}}f_{h_{2}}}{f_{0}})-1\right) <(\beta -\alpha )^{2},
\label{eq:proof lower 2}
\end{equation}%
which yields the desired bound $\Vert \mathbb{P}_{0}\wedge \mathbb{\bar{%
P}}\Vert >1-\frac{1}{2}(\beta -\alpha )$.

The following representation of the $\chi ^{2}$ divergence of two Gaussian
distributions
\begin{equation}
\int \frac{g_{1}g_{2}}{g_{0}}=[\det (I-\Sigma _{0}^{-1}(\Sigma _{1}-\Sigma
_{0})\Sigma _{0}^{-1}(\Sigma _{2}-\Sigma _{0}))]^{-1/2},
\label{eq:Chi dist Gaussian}
\end{equation}
with $g_{i}$ the density of $%
N(0,\Sigma _{i})$ for $i=0,1,2$, is helpful to our analysis. By our construction of $\mathbb{P}_{h}$ and (\ref{eq:Chi dist
Gaussian}), some algebra results in $$\int \frac{f_{h_{1}}f_{h_{2}}}{f_{0}}%
=\left( \int
\prod\nolimits_{t=1}^{h}g_{h_{1}}^{(t)}g_{h_{2}}^{(t)}/g_{0}^{(t)}\right)
^{n^{(0)}}=\left( 1-2Ja^{2}\right) ^{-n^{(0)}k},$$ where $J=J(h_{1},h_{2})$
is the number of overlapping $a$ between the first rows of $( \Omega
_{h_{1}}^{(1)}) ^{-1}$ and $( \Omega _{h_{2}}^{(1)}) ^{-1}$%
. Hence it follows that
\begin{eqnarray*}
\Delta &=&\frac{1}{m^{2}}\sum_{0\leq j\leq
s-1}\sum_{J(h_{1},h_{2})=j}\left( \left( 1-2ja^{2}\right)
^{-n^{(0)}k}-1\right) \\
&=&\frac{1}{m^{2}}\sum_{1\leq j\leq s-1}\binom{p-1}{s-1}\binom{s-1}{j}%
\binom{p-s}{s-1-j}\left( \left( 1-2ja^{2}\right) ^{-n^{(0)}k}-1\right) .
\end{eqnarray*}%

Observe that since $2ja^{2}\leq $ $2(s-1)a^{2}<1/2$ and $k\leq M_{1}\log p$%
, we have
\begin{align*}
\left( 1-2ja^{2}\right) ^{-n^{(0)}k} & \leq \left( 1+4ja^{2}\right)
^{n^{(0)}k}\leq \exp (4ja^{2}n^{(0)}k) =\exp (4j\tau (k+\log p)) \\
& \leq
(p)^{4(1+M_{1})\tau j}.
\end{align*}%
Moreover, it can be checked that with $m=$ $\binom{p-1}{s-1}$,%
\begin{equation*}
\frac{1}{m^{2}}\binom{p-1}{s-1}\binom{s-1}{j}\binom{p-s}{s-1-j}\leq \left(
\frac{s^{2}}{p-s}\right) ^{j}.
\end{equation*}%
Therefore, combining the three expressions above we can complete the proof by noting that
\begin{equation*}
\Delta \leq \sum_{1\leq j\leq s-1}\left( \frac{s^{2}p^{4(1+M_{1})\tau }}{p-s}%
\right) ^{j}\rightarrow 0,
\end{equation*}%
where we have used $p>s^{\mu }$ for some $\mu >2$ and picked a small enough constant $\tau
$ depending on $\mu $ and $M_{1}$.

\subsection{Lemma \ref{lem:GRE} and its proof} \label{SecB.5}

\medskip

\begin{lemma}
\label{lem:GRE} For any fixed $\xi $, under
Conditions \ref{CondA1}--\ref{CondA2} and the assumption of $s<C_{\xi }n^{(0)}/\log p$ with some sufficiently
small constant $C_{\xi }>0$ depending on $\xi $, $M$, and $M_{0}$, we have $%
\mathbb{P}\{\mathcal{E}_{1,gRE}\}>1-2k\exp (-cn^{(0)})$, where $\mathcal{E}_{1,gRE}=\{gRE(\xi ,T)>\min_{l,t}\{(n^{(t)}/%
\mathbf{X}_{\ast ,l}^{(t)\prime }\mathbf{X}_{\ast
,l}^{(t)})^{1/2}\}/(2M)^{1/2}\}$ and $c>0$ is some constant depending
on $\xi $, $M$, and $M_{0}$.
\end{lemma}

\textit{Proof.} The proof of the group-wise restricted eigenvalue (gRE) condition follows from a
similar reduction principle to that developed in \cite%
{rudelson2013reconstruction} and \cite{loh2012high} for dealing with the regular
restricted eigenvalue (RE) condition. First of all, due to the normalization constant, that is,  $\mathbf{\bar{X}}_{\ast
,-1}^{0}=\mathbf{X}_{\ast ,-1}^{0}( \bar{D}_{1}) ^{-1/2}$, it
suffices to show that with probability at least $1-2k\exp (-cn^{(0)})$,
\begin{equation}
\inf_{u\neq 0}\left\{\frac{\left\Vert \mathbf{X}_{\ast ,-1}^{0}u\right\Vert }{%
\sqrt{n^{(0)}}\left\Vert u\right\Vert }:u\in \Psi (\xi ,T)\right\}\geq \left(
2M\right) ^{-1/2}.  \label{eq:reduction1}
\end{equation}%
To further reduce the condition in (\ref{eq:reduction1}), we note that
\begin{equation*}
\frac{u^{\prime }\mathbf{X}_{\ast ,-1}^{0\prime }\mathbf{X}_{\ast ,-1}^{0}u}{%
n^{(0)}\left\Vert u\right\Vert ^{2}}=\frac{u^{\prime }\mathbb{E}\left(
\mathbf{X}_{\ast ,-1}^{0\prime }\mathbf{X}_{\ast ,-1}^{0}\right) u}{%
n^{(0)}\left\Vert u\right\Vert ^{2}}+\frac{u^{\prime }\left( \mathbf{X}%
_{\ast ,-1}^{0\prime }\mathbf{X}_{\ast ,-1}^{0}-\mathbb{E}\left( \mathbf{X}%
_{\ast ,-1}^{0\prime }\mathbf{X}_{\ast ,-1}^{0}\right) \right) u}{%
n^{(0)}\left\Vert u\right\Vert ^{2}}
\end{equation*}%
and the first term above is lower bounded by $M^{-1}$, that is,
\begin{equation*}
\frac{u^{\prime }\mathbb{E}\left( \mathbf{X}_{\ast ,-1}^{0\prime }\mathbf{X}%
_{\ast ,-1}^{0}\right) u}{n^{(0)}\left\Vert u\right\Vert ^{2}}=\sum_{t=1}^{k}%
\frac{u^{(t)\prime }\Sigma _{-1,-1}^{(t)}u^{(t)}}{\left\Vert
u^{(t)}\right\Vert ^{2}}\cdot \frac{n^{(t)}}{n^{(0)}}\geq \frac{1}{M},
\end{equation*}%
where the last inequality follows from Conditions \ref{CondA1}--\ref{CondA2}. Thus it
remains to prove that with probability at least $1-2k\exp (-cn^{(0)})$,
\begin{equation}
\left\vert \frac{u^{\prime }\left( \mathbf{X}_{\ast ,-1}^{0\prime }\mathbf{X}%
_{\ast ,-1}^{0}-\mathbb{E}\left( \mathbf{X}_{\ast ,-1}^{0\prime }\mathbf{X}%
_{\ast ,-1}^{0}\right) \right) u}{n^{(0)}\left\Vert u\right\Vert ^{2}}%
\right\vert \leq \frac{1}{2M}\mbox{\quad for all }u\in \Psi (\xi ,T).
\label{eq:reduction2}
\end{equation}

Before proceeding, let us introduce some notation. Let
\[ \mathbb{K}(m)=\{u\in
\mathbb{R}^{k(p-1)}: \sum_{l=2}^{p}1\{u_{(l)}\neq 0\}\leq m\} \]
be the
group-wise $m$-sparse set. The proof of (\ref{eq:reduction2}) is comprised of two steps. In the first step we prove that the following inequality
holds with probability at least $1-2k\exp (-cn^{(0)})$ for all $u\in \mathbb{%
K}(2s)$,%
\begin{eqnarray}
&&\left\vert \frac{u^{\prime }\left( \mathbf{X}_{\ast ,-1}^{0\prime }\mathbf{%
X}_{\ast ,-1}^{0}-\mathbb{E}\left( \mathbf{X}_{\ast ,-1}^{0\prime }\mathbf{X}%
_{\ast ,-1}^{0}\right) \right) u}{n^{(0)}\left\Vert u\right\Vert ^{2}}%
\right\vert \notag  \label{eq:reduction 3} \\
&=&\left\vert \sum_{t=1}^{k}\frac{u^{(t)\prime }(\mathbf{X}_{\ast
,-1}^{(t)\prime }\mathbf{X}_{\ast ,-1}^{(t)}/n^{(t)}-\Sigma
_{-1,-1}^{(t)})u^{(t)}}{\left\Vert u^{(t)}\right\Vert ^{2}}\cdot \frac{%
n^{(t)}}{n^{(0)}}\right\vert   \notag \\
&\leq &\frac{1}{6(2+\xi )^{2}M},
\end{eqnarray}%
while the second step shows that (\ref{eq:reduction 3}) entails (\ref%
{eq:reduction2}) deterministically.

The inequality (\ref{eq:reduction 3}) can be established by the
standard $\delta $-net argument for each of the design matrices $\mathbf{X}%
_{\ast ,-1}^{(t)}$ and a union bound argument. Denote by
\[ \mathbb{K}%
^{(t)}(m)=\left\{u^{(t)}\in \mathbb{R}^{(p-1)}: \sum_{l=2}^{p}1\{u_{l}^{(t)}\neq
0\}\leq m\right\}. \]
Then an application of Lemma 15 in \cite{loh2012high} implies that there exists
some absolute
constant $c_{0}>0$ such that
\begin{eqnarray*}
&&\mathbb{P}\left( \sup_{u^{(t)}\in \mathbb{K}^{(t)}(2s)}\left\vert \frac{%
u^{(t)\prime }\left( \mathbf{X}_{\ast ,-1}^{(t)\prime }\mathbf{X}_{\ast
,-1}^{(t)}/n^{(t)}-\Sigma _{-1,-1}^{(t)}\right) u^{(t)}}{\left\Vert
u^{(t)}\right\Vert ^{2}}\right\vert >x\right)  \\
&\leq &2\exp (-c_{0}n^{(t)}\min \{x^{2}/M^{2},x/M\}+4s\log
p).
\end{eqnarray*}%
Note that $n^{(t)}/n^{(0)}\leq M_{0}$ from Condition \ref{CondA2}.
Therefore, the union bound of the above inequality for all $t\in \lbrack k]$,
together with the choice $x=(6(2+\xi )^{2}MM_{0})^{-1}$ and our assumption $s<C_{\xi
}n^{(0)}/\log p$ with some sufficiently small constant $C_{\xi }>0$ depending
on $\xi $, $M$, and $M_{0},$ yields that (\ref{eq:reduction 3})
holds with probability at least $1-2k\exp (-cn^{(0)})$ for some positive constant $c$
depending on $\xi $, $M$, and $M_{0}$.

It remains to show that (\ref{eq:reduction 3}) in fact implies the desired
result in (\ref{eq:reduction2}). From now on, denote by
\[ \mathbf{\Gamma } = ( \mathbf{X}%
_{\ast ,-1}^{0\prime }\mathbf{X}_{\ast ,-1}^{0}-\mathbb{E}( \mathbf{X}%
_{\ast ,-1}^{0\prime }\mathbf{X}_{\ast ,-1}^{0}) ) /n^{(0)}. \]
In order to show (\ref{eq:reduction2}), by the scaling
property it suffices to establish
\begin{equation}
\left\vert u^{\prime }\mathbf{\Gamma }u\right\vert \leq \frac{1}{2M}%
\mbox{\quad
for all }u\in \Psi (\xi ,T)\cap B_{2}(1),  \label{eq:reduction4}
\end{equation}%
where $B_{2}(1)$ is the unit $\ell_2$ ball in $\mathbb{R}^{k(p-1)}$. To
finish our proof, given (\ref{eq:reduction 3}) we show that  $\left\vert
u^{\prime }\mathbf{\Gamma }u\right\vert \leq \frac{1}{2M}$ for any $u\in
\mathrm{cl}(\mathrm{conv}\{\mathbb{K}(s)\cap B_{2}(2+\xi) \})$, the closure
of the convex hull covering $\mathbb{K}(2s)\cap B_{2}(2+\xi)$, followed by
an argument showing that $\Psi (\xi ,T)\cap B_{2}(1)\subset $ $\mathrm{cl}(%
\mathrm{conv}\{\mathbb{K}(s)\cap B_{2}(2+\xi) \})$.

For any $u\in \mathrm{cl}(%
\mathrm{conv}\{\mathbb{K}(s)\cap B_{2}(2+\xi) \})$, we can write $%
u=\sum_{i}\alpha _{i}u_{i}$, where $u_{i}\in \mathbb{K}(s)$, $\left\Vert
u_{i}\right\Vert \leq 2+\xi $, $\alpha _{i}>0$, and $\sum_{i}\alpha _{i}=1$.
Thus it follows from (\ref{eq:reduction 3}) and the fact of $%
u_{i}+u_{j}\in $ $\mathbb{K}(2s)$ for any $i$ and $j$ that
\begin{eqnarray*}
\left\vert u^{\prime }\mathbf{\Gamma }u\right\vert &=&\left\vert
(\sum_{i}\alpha _{i}u_{i})^{\prime }\mathbf{\Gamma }(\sum_{i}\alpha
_{i}u_{i})\right\vert \leq \sum_{i,j}\alpha _{i}\alpha _{j}\left\vert
u_{i}{}^{\prime }\mathbf{\Gamma }u_{j}\right\vert \\
&=&\frac{1}{2}\sum_{i,j}\alpha _{i}\alpha _{j}\left\vert
(u_{i}{}+u_{j})^{\prime }\mathbf{\Gamma }(u_{i}{}+u_{j})-u_{i}{}^{\prime }%
\mathbf{\Gamma }u_{i}-u_{j}{}^{\prime }\mathbf{\Gamma }u_{j}\right\vert \\
&\leq &\frac{1}{2}\frac{1}{6(2+\xi )^{2}M}\sum_{i,j}\alpha _{i}\alpha
_{j}\left(4(2+\xi )^{2}+(2+\xi )^{2}+(2+\xi )^{2}\right) \\
&\leq &\frac{1}{2M}\sum_{i,j}\alpha _{i}\alpha _{j}=\frac{1}{2M}\mbox{\rm ,}
\end{eqnarray*}%
where (\ref{eq:reduction 3}) has been applied in the second inequality. It remains
to show that
\[ \Psi (\xi ,T)\cap B_{2}(1)\subset \mathrm{cl}(\mathrm{conv}\{%
\mathbb{K}(s)\cap B_{2}(2+\xi) \}). \]
We exploit a similar analysis to that designed
for the regular sparse set (see Lemma 1,1 of \cite{loh2012high}). To show that a set $A$
belongs to a convex set $B$, it suffices to prove
\[ \phi _{A}(z)\leq \phi
_{B}(z) \text{\quad for all } z\in \mathbb{R}^{k(p-1)}, \]
where $\phi _{A}(z)=\sup_{u\in
A}\left\langle u,z\right\rangle $; see, e.g., Theorem 2.3.1 of \cite%
{hug2010course}.

Hereafter we denote by $A=\Psi (\xi ,T)\cap B_{2}(1)$ and $B=%
\mathrm{cl}(\mathrm{conv}\{\mathbb{K}(s)\cap B_{2}(2+\xi \})$. For any $z\in
\mathbb{R}^{k(p-1)}$, let the index set $S$ consist of the top $s$ groups of $z$
in terms of the $\ell_2$ norm. Consequently, for any $l\in S^{c}$ we have $%
\Vert z_{(l)}\Vert \leq (\sum_{l\in S}\Vert
z_{(l)}\Vert ^{2})^{1/2}/\sqrt{s}$. Now we upper bound $\phi
_{A}(z)$ by considering index sets $S$ and $S^{c}$ separately,%
\begin{eqnarray*}
\phi _{A}(z) &\leq &\sup_{u\in A}\sum_{l\in S}\left\langle
u_{(l)},z_{(l)}\right\rangle +\sup_{u\in A}\sum_{l\in S^{c}}\left\langle
u_{(l)},z_{(l)}\right\rangle \\
&\leq &(\sum_{l\in S}\left\Vert z_{(l)}\right\Vert ^{2})^{1/2}+\max_{l\in
S^{c}}\left\Vert z_{(l)}\right\Vert \cdot \sum_{l\in S^{c}}\left\Vert
u_{(l)}\right\Vert  \\
&\leq &(\sum_{l\in S}\left\Vert z_{(l)}\right\Vert ^{2})^{1/2}(1+
(1+\xi )\sqrt{s}/\sqrt{s})=(2+\xi )(\sum_{l\in S}\left\Vert
z_{(l)}\right\Vert ^{2})^{1/2},
\end{eqnarray*}%
where we have used the fact that $u$ is a unit vector and the Cauchy--Schwarz inequality
in the second inequality, and the third inequality follows from the fact that  \[
\sum_{l\in S^{c}}\Vert u_{(l)}\Vert \leq \sum_{l=2}^{p}\Vert
u_{(l)}\Vert \leq (1+\xi )\sum_{l\in T}\Vert u_{(l)}\Vert
\leq (1+\xi )\sqrt{s}\Vert u\Vert \]
in light of $u\in \Psi (\xi ,T)$.
On the other hand, since $B$ is a convex set we have%
\begin{equation*}
\phi _{B}(z)=\sup_{u\in B}\left\langle u,z\right\rangle =(2+\xi
)\max_{L:\left\vert L\right\vert =s}\sup_{u\in B_{2}(1)}\sum_{l\in
L}\left\langle u_{(l)},z_{(l)}\right\rangle =(2+\xi )(\sum_{l\in
S}\left\Vert z_{(l)}\right\Vert ^{2})^{1/2},
\end{equation*}%
where we have used the definition of the index set $S$. Clearly, it holds that $\phi
_{A}(z)\leq \phi _{B}(z)$ for all $z\in \mathbb{R}^{k(p-1)}$, which
concludes the proof.

\subsection{Lemma \ref{lem:key event} and its proof} \label{SecB.6}

\medskip

\begin{lemma}
\label{lem:key event} With the choice of regularization parameter $\lambda $ specified in Theorem \ref{thm:GSRLH}, the event $\mathcal{B}_{1}$ defined in (\ref{eq: lambda condition}) holds
with probability at least $1-3p^{-\delta +1}$.
\end{lemma}

\textit{Proof.} Throughout this proof we condition on $\mathbf{X}_{\ast ,-1}^{0}$. For any fixed $l\in \lbrack k]$, we have
\begin{equation*}
\bar{D}_{1(l)}^{-1/2}\mathbf{X}_{\ast ,(l)}^{0\prime }E_{\ast ,1}^{0}\overset%
{d}{\sim } \left(N(0,n^{(1)}/\omega _{1,1}^{(1)}),\cdots ,N(0,n^{(k)}/\omega
_{1,1}^{(k)})\right)^{\prime },
\end{equation*}%
where $\overset{d}{\sim }$ denotes equivalence in distribution and the $k$ components on the right-hand side are independent of each other. By the definition of $\bar{D}_{E1}$, we can further write
\begin{equation*}
\bar{D}_{E1}^{-1/2}\bar{D}_{1(l)}^{-1/2}\mathbf{X}_{\ast ,(l)}^{0\prime
}E_{\ast ,1}^{0}\overset{d}{\sim } \left(T^{(1)}Z^{(1)},\cdots
,T^{(k)}Z^{(k)}\right)^{\prime },
\end{equation*}%
where $Z^{(t)}$, $t\in \lbrack k]$, are i.i.d. standard Gaussian and $(
T^{(t)}) ^{-2}\overset{d}{\sim } \chi ^{2}(n^{(t)})/n^{(t)}$. \
Consequently, we obtain
\begin{equation}
\mathbb{P}\left( \left\Vert \bar{D}_{E1}^{-1/2}\bar{D}_{1(l)}^{-1/2}\mathbf{X%
}_{\ast ,(l)}^{0\prime }E_{\ast ,1}^{0}\right\Vert ^{2}>z\right) \leq
\mathbb{P}\left( \max_{t\in \lbrack k]}\left( T^{(t)}\right) ^{2}\chi
^{2}(k)>z\right).  \label{eq:transfer_chi}
\end{equation}

To control the term $T^{(t)}$, we apply Lemma \ref{prop:chi} with $x=\tau
=( 8(\delta \log p+\log k)/n^{(0)}) ^{1/2}=o(1)$ to deduce that
\begin{equation}
\mathbb{P}\left( \left( T^{(t)}\right) ^{2}>\frac{1}{1-\tau }\right) \leq
2k^{-1}p^{-\delta },  \label{eq:control T}
\end{equation}%
where we have used the fact of $n^{(0)}\leq n^{(t)}$. Similarly, to control the term $\chi
^{2}(k)$ an application of Lemma \ref{prop:chi} with $y=\delta \log p$ leads to
\begin{equation}
\mathbb{P}\left( \chi ^{2}(k)>k+2\delta \log p+2\sqrt{\delta k\log p}\right)
\leq p^{-\delta }.  \label{eq:control Chi_k}
\end{equation}%
Thus the union bound argument applied to inequalities (\ref{eq:control T}) over $%
t\in \lbrack k]$ and (\ref{eq:control Chi_k}) yields
\begin{equation*}
\mathbb{P}\left( \max_{t\in \lbrack k]}\left( T^{(t)}\right) ^{2}\chi
^{2}(k)>\frac{k+2\delta \log p+2\sqrt{\delta k\log p}}{1-\tau }\right) \leq
3p^{-\delta }\mbox{\rm .}
\end{equation*}%
Finally, we can apply another union bound argument over all $2\leq l\leq p$ and (\ref{eq:transfer_chi}) to obtain
\begin{equation*}
\mathbb{P}\left( \max_{2\leq l\leq p}\left\Vert \bar{D}_{E1}^{-1/2}\bar{D}%
_{1(l)}^{-1/2}\mathbf{X}_{\ast ,(l)}^{0\prime }E_{\ast ,1}^{0}\right\Vert
^{2}>\frac{k+2\delta \log p+2\sqrt{\delta k\log p}}{1-\tau }\right) \leq
3p^{-\delta +1},
\end{equation*}%
which completes the proof by noting that the above conditional probability
is free of $\mathbf{X}_{\ast ,-1}^{0}$.

\subsection{Lemma \ref{lem:constant upper} and its proof} \label{SecB.7}

\medskip

\begin{lemma}
\label{lem:constant upper} Under
Conditions \ref{CondA1}--\ref{CondA2}, for the event $\mathcal{E}_{1,up}=\{$ $\zeta _{t}\leq
$ $\sqrt{6MM_{0}}$ simultaneously for all $t\in \lbrack k]\}$ it holds that  $%
\mathbb{P\{}\mathcal{E}_{1,up}\mathbb{\}\geq }1-4k\exp (-n^{(0)}/32)$.
\end{lemma}

\textit{Proof.} Be definition, we have $\zeta _{t}=\bar{Q}_{t}^{1/2}(\hat{\bar{C}}_{1}^{(t)})+\bar{%
Q}_{t}^{1/2}(\bar{C}_{1}^{(t)})$. Since $\hat{\bar{C}}_{1}^{0}$ is the
solution to the HGSL optimization problem (\ref{eq: group square-root Lasso1}), for the vector $\check{\beta}=(%
\mathbf{0,}\, \hat{\bar{C}}_{1}^{(2)\prime }, \cdots , \hat{\bar{C}}_{1}^{(k)\prime })^{\prime }$ with $\check{\beta}%
_{(l)}=(0, \hat{\bar{C}}_{1,l}^{(2)}, \cdots ,\hat{\bar{C}}_{1,l}^{(k)})^{%
\prime }$ it holds that
\begin{equation*}
\sum_{t=1}^{k}\bar{Q}_{t}^{1/2}(\hat{\bar{C}}_{1}^{(t)})+\lambda
\sum_{l=2}^{p}\left\Vert \hat{\bar{C}}_{1(l)}^{0}\right\Vert \leq \bar{Q}%
_{1}^{1/2}(\mathbf{0})+\sum_{t\neq t_{0}}\bar{Q}_{t}^{1/2}(\hat{\bar{C}}%
_{1}^{(t)})+\lambda \sum_{l=2}^{p}\left\Vert \check{\beta}_{(l)}\right\Vert
\text{.}
\end{equation*}%
Note that $\Vert \hat{\bar{C}}_{1(l)}^{0}\Vert \geq \Vert
\check{\beta}_{(l)}\Vert $ by our choice of $\check{\beta}_{(l)}$.
Thus we deduce that
\[ \bar{Q}_{1}^{1/2}(\hat{\bar{C}}_{1}^{(1)})\leq \bar{Q}%
_{1}^{1/2}(\mathbf{0})=\Vert X_{\ast ,1}^{(1)}\Vert /(n^{(0)}%
)^{1/2}. \]
By symmetry, for all $t\in [k]$ we have with probability at
least $1-4k\exp (-n^{(0)}/32)$,
\begin{eqnarray*}
\zeta _{t} &\leq &\frac{\left\Vert X_{\ast ,1}^{(t)}\right\Vert +\left\Vert
E_{\ast ,1}^{(t)}\right\Vert }{\sqrt{n^{(0)}}}\leq \frac{\left\Vert X_{\ast
,1}^{(t)}\right\Vert +\left\Vert E_{\ast ,1}^{(t)}\right\Vert }{\sqrt{n^{(t)}%
}}\frac{\sqrt{n^{(t)}}}{\sqrt{n^{(0)}}} \\
&\leq &2\sqrt{3M/2}\cdot \sqrt{M_{0}},
\end{eqnarray*}%
where the last inequality follows from Conditions \ref{CondA1}--\ref{CondA2} and the facts of  $%
X_{\ast ,1}^{(t)\prime }X_{\ast ,1}^{(t)}/\sigma _{1,1}^{(t)}\sim \chi
^{2}(n^{(t)})$ and $E_{\ast ,1}^{(t)\prime }E_{\ast ,1}^{(t)}(\omega
_{1,1}^{(t)})\sim \chi ^{2}(n^{(t)})$. Specifically, the union bound for $%
t\in \lbrack k]$ with an application of Lemma \ref{prop:chi} using $x=1/2$
yields
\[ (\Vert X_{\ast ,1}^{(t)}\Vert +\Vert E_{\ast
,1}^{(t)}\Vert )/(n^{(t)})^{1/2}\leq (3\sigma _{1,1}^{(t)}/2)^{1/2}+(%
3/2\omega _{1,1}^{(t)})^{1/2} \]
with probability at least $1-4k\exp (-n^{(0)}/32)$, which concludes the proof.

\section{Additional technical details} \label{SecC}
The following two technical lemmas are used throughout the paper from place to place.

\begin{lemma}[\cite{laurent2000adaptive}]
\label{prop:chi} The chi-square distribution with $n$
degrees of freedom satisfies the following tail probability bounds
\begin{eqnarray*}
\mathbb{P}\left(\left\vert \chi ^{2}(n)/n-1\right\vert  >x\right)&\leq& 2\exp
(-nx(x\wedge 1)/8)\mbox{\quad for any }x>0, \\
\mathbb{P}\left(\chi ^{2}(n)/n-1 >2y/n+2\sqrt{y/n}\right)&\leq& \exp (-y)\mbox{\quad
for any }y>0, \\
\mathbb{P}\left(\sqrt{\chi ^{2}(n)/n}-1 >z\right)&\leq& \exp (-nz^{2}/2) %
\mbox{\quad for any }z>0.
\end{eqnarray*}
\end{lemma}


\medskip

\begin{lemma}
\label{prop:Bern} Assume that Conditions \ref{CondA1}--\ref{CondA2} hold and $\max \{\log p,\log k\}=o(n^{(0)})$. Then for any given constant $%
\delta >0$, there exists
some positive constant $C$ depending only on $M$ and $\delta $ such that for any
fixed $j$,%
\begin{eqnarray*}
\mathbb{P}\left(\max_{l\neq j}\frac{1}{k}\sum_{t=1}^{k}(\frac{E_{\ast
,j}^{(t)\prime }X_{\ast ,l}^{(t)}}{n^{(t)}})^{2} \geq C\frac{1+(\log p)/k}{%
n^{(0)}}\right)&\leq& 3p^{1-\delta }, \\
\mathbb{P}\left(\frac{1}{k}\sum_{t=1}^{k}(\frac{E_{\ast ,j}^{(t)\prime }\mathbf{X}%
_{\ast ,-j}^{(t)}C_{j}^{(t)}}{n^{(t)}})^{2} \geq C\frac{1+(\log p)/k}{%
n^{(0)}}\right)&\leq& 3p^{-\delta }.
\end{eqnarray*}
\end{lemma}

\textit{Proof.} Since $E_{\ast ,j}^{(t)}\sim
N(0,I\cdot (\omega _{j,j}^{(t)})^{-1})$ is independent of $\mathbf{X}%
_{\ast ,-j}^{(t)}$ for each $t\in [k]$, it holds that for each $l\neq j
$, $( E_{\ast ,j}^{(t)\prime }X_{\ast ,l}^{(t)}) (\omega
_{j,j}^{(t)})^{1/2}/\Vert X_{\ast ,l}^{(t)}\Vert \sim N(0,1)$.
In addition, these random variables are independent among different $t\in \lbrack k]$. By Lemma \ref{prop:chi}, we have%
\begin{equation}
\mathbb{P}\left(\frac{1}{k}\sum_{t=1}^{k}\omega _{j,j}^{(t)}\left(E_{\ast
,j}^{(t)\prime }X_{\ast ,l}^{(t)}/\left\Vert X_{\ast ,l}^{(t)}\right\Vert%
\right)^{2}\geq 1+2\sqrt{\frac{\delta \log p}{k}}+\frac{2\delta \log p}{k}\right)\leq
2p^{-\delta }.  \label{eq:pf of prop1}
\end{equation}%
To control the term $\Vert X_{\ast ,l}^{(t)}\Vert $, we apply Lemma %
\ref{prop:chi} with $X_{\ast ,l}^{(t)}\sim N(0,I\cdot \sigma _{l,l}^{(t)})$
to deduce that $$\mathbb{P}\left(\left\Vert X_{\ast ,l}^{(t)}\right\Vert /\sqrt{%
\sigma _{l,l}^{(t)}n^{(t)}}\geq 1+\sqrt{\frac{2(\delta \log p+\log k)}{n^{(t)}%
}}\right)\leq p^{-\delta }k^{-1},$$
where $\sigma_{l,l}^{(t)}$ stands for the variance of $X^{(t)}_l$. The union bound, together with the assumption of $\max \{\log
p, \\ \log k\}=o(n^{(0)})$, entails that
\begin{equation} \label{neweq013}
\Vert X_{\ast ,l}^{(t)}\Vert
\leq 2(\sigma _{l,l}^{(t)}n^{(t)})^{1/2}\leq (4Mn^{(t)})^{1/2}
\end{equation}
simultaneously
for all $t\in [k]$ with probability at least $1-p^{-\delta }$.

We now condition on the event given by (\ref{neweq013}). Due to Conditions \ref{CondA1}--\ref{CondA2}, we have $$\frac{1}{k}%
\sum_{t=1}^{k}\omega _{j,j}^{(t)}\left(\frac{E_{\ast ,j}^{(t)\prime }X_{\ast
,l}^{(t)}}{\left\Vert X_{\ast ,l}^{(t)}\right\Vert }\right)^{2}\geq \frac{n^{(0)}}{%
4M^{2}}\frac{1}{k}\sum_{t=1}^{k}\left(\frac{E_{\ast ,j}^{(t)\prime }X_{\ast
,l}^{(t)}}{n^{(t)}}\right)^{2},$$
which along with (\ref{eq:pf of prop1}%
) leads to
\begin{equation}
\mathbb{P}\left(\frac{1}{k}\sum_{t=1}^{k}\left(E_{\ast ,j}^{(t)\prime }X_{\ast
,l}^{(t)}/n^{(t)}\right)^{2}\geq \frac{4M^{2}}{n^{(0)}}\left(1+2\sqrt{\frac{\delta
\log p}{k}}+\frac{2\delta \log p}{k}\right)\right)\leq 3p^{-\delta }.
\label{eq:pf of prop2}
\end{equation}%
Thus we see that the first desired result follows immediately from (\ref{eq:pf of prop2})
with a union bound for all $l\neq j$ and $C=4M^{2}(2+3\delta )$, in view of  $2%
((\delta \log p)/k)^{1/2}\leq 1+(\delta \log p)/k$. Since $\mathbf{X}_{\ast ,-1}^{(t)}C_{1}^{(t)}$ has i.i.d. Gaussian
entries with bounded variance and is independent of $E_{\ast ,j}^{(t)}$, the second
desired result follows from a similar analysis as for (\ref{eq:pf of prop2}), which completes the proof.

\end{document}